\documentclass[aps,pre,superscriptaddress,showpacs,nofootinbib]{revtex4}
\usepackage{amsfonts,amssymb,amsmath,bm,bbm,mathrsfs}
\usepackage[mathcal]{euscript}
\usepackage[usenames]{color}
\usepackage{color,soul}
\usepackage[utf8]{inputenc}
\usepackage[T1]{fontenc}
\usepackage{textcomp}
\usepackage{upgreek}
\usepackage{graphicx}
\usepackage[tight,sf]{subfigure}
\usepackage{hyperref}

\newcommand{\am}{\mathrm{am}}
\newcommand{\sn}{\mathrm{sn}}
\newcommand{\cn}{\mathrm{cn}}
\newcommand{\dn}{\mathrm{dn}}
\newcommand{\arctanh}{\mathrm{arctanh}}
\newcommand{\rmd}{\mathrm{d}}
\newcommand{\rmtz}{\mathrm{t}^z}
\newcommand{\rmnz}{\mathrm{n}^z}
\newcommand{\rmbz}{\mathrm{b}^z}
\newcommand{\rmF}{\mathrm{F}}
\newcommand{\rmK}{\mathrm{K}}
\newcommand{\rmE}{\mathrm{E}}
\newcommand{\calH}{\mathcal{H}}
\newcommand{\calB}{\mathcal{B}}
\newcommand{\calM}{\mathcal{M}}
\newcommand{\bfx}{\mathbf{x}}
\newcommand{\bfY}{\mathbf{Y}}
\newcommand{\bfM}{\mathbf{M}}
\newcommand{\bfS}{\mathbf{S}}
\newcommand{\bfH}{\mathbf{H}}
\newcommand{\bft}{\mathbf{t}}
\newcommand{\bfn}{\mathbf{n}}
\newcommand{\bfb}{\mathbf{b}}
\newcommand{\bfF}{\mathbf{F}}
\newcommand{\bfD}{\mathbf{D}}
\newcommand{\hbfy}{\hat{\bf y}}
\newcommand{\hbfz}{\hat{\bf z}}

\begin{document}

\title{Paramagnetic filaments in a fast precessing field:\\
 Planar versus helical conformations}

\author{Pablo V\'{a}zquez-Montejo}
\email[]{pablovazqmont@northwestern.edu}
\affiliation{Department of Materials Science and Engineering, Northwestern University, 2220 Campus Drive, Evanston, Illinois 60208, USA}
\author{Joshua M. Dempster}
\email[]{joshuadempster2016@u.northwestern.edu}
\affiliation{Department of Physics and Astronomy, Northwestern University, 2145 Sheridan Road F165, Evanston, Illinois 60208, USA}
\author{M\'onica Olvera de la Cruz}
\email[]{m-olvera@northwestern.edu}
\affiliation{Department of Materials Science and Engineering, Northwestern University, 2220 Campus Drive, Evanston, Illinois 60208, USA}
\affiliation{Department of Physics and Astronomy, Northwestern University, 2145 Sheridan Road F165, Evanston, IL 60208, USA}

\begin{abstract}
We examine analytically equilibrium conformations of elastic chains of paramagnetic beads in the presence of a precessing magnetic field. Conformations of these filaments are determined by minimizing their total energy, given in the harmonic approximation by the sum of the bending energy, quadratic in its curvature, and the magnetic dipolar interaction energy, quadratic in the projection of the vector tangent to the filament onto the precession axis. In particular, we analyze two families of open filaments with their ends aligned along the precession axis and described by segments of planar curves and helices. These configurations are characterized in terms of two parameters encoding their features such as their length, separation between their ends, as well as their bending and magnetic moduli, the latter being proportional to the magnitude and precession angle of the magnetic field. Based on energetic arguments, we present the set of parameter values for which each of these families of curves is probable to occur.
\end{abstract}


\maketitle

\section{Introduction}
\noindent
Flexible magnetic filaments can be synthesized by joining ferromagnetic or superparamagnetic beads with elastic linkers \cite{Wang2011, Tierno2014}. The combination of their elastic and magnetic properties gives rise to interesting phenomena, which have been a subject of active research since their introduction more than a decade ago \cite{Biswal2003, Goubalt2003}. 
\\
The mechanical properties of magnetic filaments have been extensively characterized. Their Young and bending moduli have been measured directly in bending and compression experiments performed with optical traps \cite{Biswal2003} and indirectly from measurements of other quantities such as the separation between beads using Bragg diffraction \cite{Goubalt2003}, or more recently, via their thermal fluctuations \cite{Gerbal2017}. It has been found that the sole magnetic field can drive an Euler buckling instability in a free filament \cite{Cebers2005a}, whose critical value has been determined theoretically and measured experimentally with a good agreement \cite{CebersJavaitis2004, Gerbal2015}.
\\
This kind of filaments exhibit diverse morphological features \cite{Cebers2007}, for instance, depending on their length, bending rigidity and magnetic field strength, they may adopt $U$ and $S$ shapes \cite{Goubalt2003, Cebers2003} or configurations with more undulations \cite{Huang2016}. Configurations of anchored superparamagnetic filaments with a free end may fold into loops, sheets and pillars for different combinations of their bending rigidity and the strength of the magnetic field \cite{Wei2016}. In a precessing magnetic field, filaments with loads at their ends can adopt planar or helical configurations by changing the precessing angle and with a time-dependent precession free filaments can assemble into gels of diverse conformations \cite{Dempster2017}. 
\\
Due to their magnetic features, they have inspired the development of several applications. They can be used as micro-mechanical sensors used in the determination of force-extension laws at the micro-scale \cite{Biswal2003, Goubalt2003, Koenig2005}. Furthermore, since they possess the interesting feature that their stiffness is tunable \cite{Cebers2005b}, and conformational changes can be controlled through the magnetic field \cite{Cebers2007, Huang2016} or the temperature \cite{Cerda2016}, they can also be used as actuators \cite{Dempster2017} or grabbers \cite{Martinez2016}. 
\\
Their dynamics have also been studied in detail. In a magnetic field rotating on a plane, free filaments rotate rigidly synchronously or asynchronously depending on their length and whether the precession frequency is smaller or bigger than a critical value \cite{Biswal2004, Cebers2004}. In the presence of an alternating magnetic field, magnetic filaments oscillate and displace, so they have been used to design self-propelled swimmers of controllable velocity and displacement direction \cite{Dreyfus2005, Belovs2006, Roper2008, Cebers2013}. 
\\
Further information about the properties and applications of these magnetic filaments can be found in the reviews \cite{Cebers2005b, Cebers2015}.
\\
In general, most of the previous works consider magnetic filaments with one or both free ends and in precessing fields at a fixed angle, typically precessing on a plane. In this paper we examine open superparamagnetic filaments in a magnetic field precessing at different constant angles, and with their ends held along the precession axis at a certain distance. Molecular dynamics simulations suggest that under these conditions filaments exhibit different behavior depending on the value of the precessing angle relative to a critical angle: if it is smaller, filaments bend but remain on a plane, whereas if it is larger, filaments explore the ambient space adopting helical structures \cite{Dempster2017}. Here we present a detailed analytic description of these two families of filaments. We determine their equilibrium configurations by minimizing their ascribed total energy, which at quadratic order has two contributions, the bending energy and the magnetic energy due to dipolar interactions between the beads \cite{Cebers2003, Goubalt2003}. In principle, the behavior of these magnetic filaments depends on their intrinsic properties: length and bending modulus; as well as of their extrinsic properties: separation between their ends and magnetic modulus, which depends on the magnetic field parameters (magnitude and precession angle) and can be positive, negative or even vanish. However, it is possible to characterize their equilibrium configurations in terms of just two parameters capturing all of their characteristics: the boundary separation and the ratio of the magnetic to bending moduli, both scaled with powers of the total length of the filament so as to adimensionalize them. We discuss the forces required to hold the boundaries of  the filaments and the behavior of their total energy as a function of these two parameters. Although both families are critical points of the total energy regardless of the precession angle, by comparing their total energies we investigate their plausibility in each precession regime, determining the parameter values for which each family is more likely to take place.
\\
This paper is organized as follows. We begin in Sec. \ref{sec:EnergyAndStresses} with the framework that we employ to describe superparamagnetic filaments, to this end we define their energy and we express the corresponding Euler-Lagrange (EL) equations, that their equilibrium configurations must satisfy, in terms of the stresses on the filaments. In Sect. \ref{sec:PlanarCuves} we specialize this framework to the case of planar curves, which in Sec. \ref{sec:vetfils} is applied to examine the family of vertical planar filaments (their ends are fixed and aligned with the precession axis) in the perturbative and non-linear regimes. In Sec. \ref{sec:Helices} we do the respective analysis of the family of helices. In Sec. \ref{sec:Hcomp} we compare the total energy of both families of filaments with same parameters to assess their possible physical realization. We close with our conclusions and discussion of future work in Sec. \ref{sec:Conc}. Some derivations and calculations used or discussed in the main text are presented in the appendices.

\section{Energy and stresses} \label{sec:EnergyAndStresses}
\begin{figure}[htbp]
\begin{center}
\includegraphics[height=5cm]{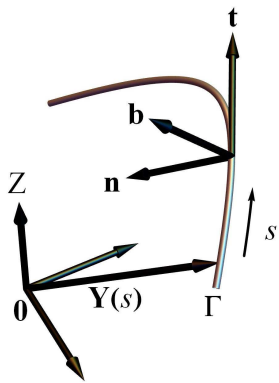}
\end{center}
\caption{The filament is described by a space curve $\Gamma: s \rightarrow \bfY(s)$ parametrized by arc length $s$. The adapted FS basis is formed by the tangent vector $\bft$, the principal normal $\bfn$ and the binormal $\bfb$.}\label{Fig:filament}
\end{figure}

\begin{figure}[htbp]
\centering
\begin{tabular}{cc}
\subfigure[]{\includegraphics[height=4cm]{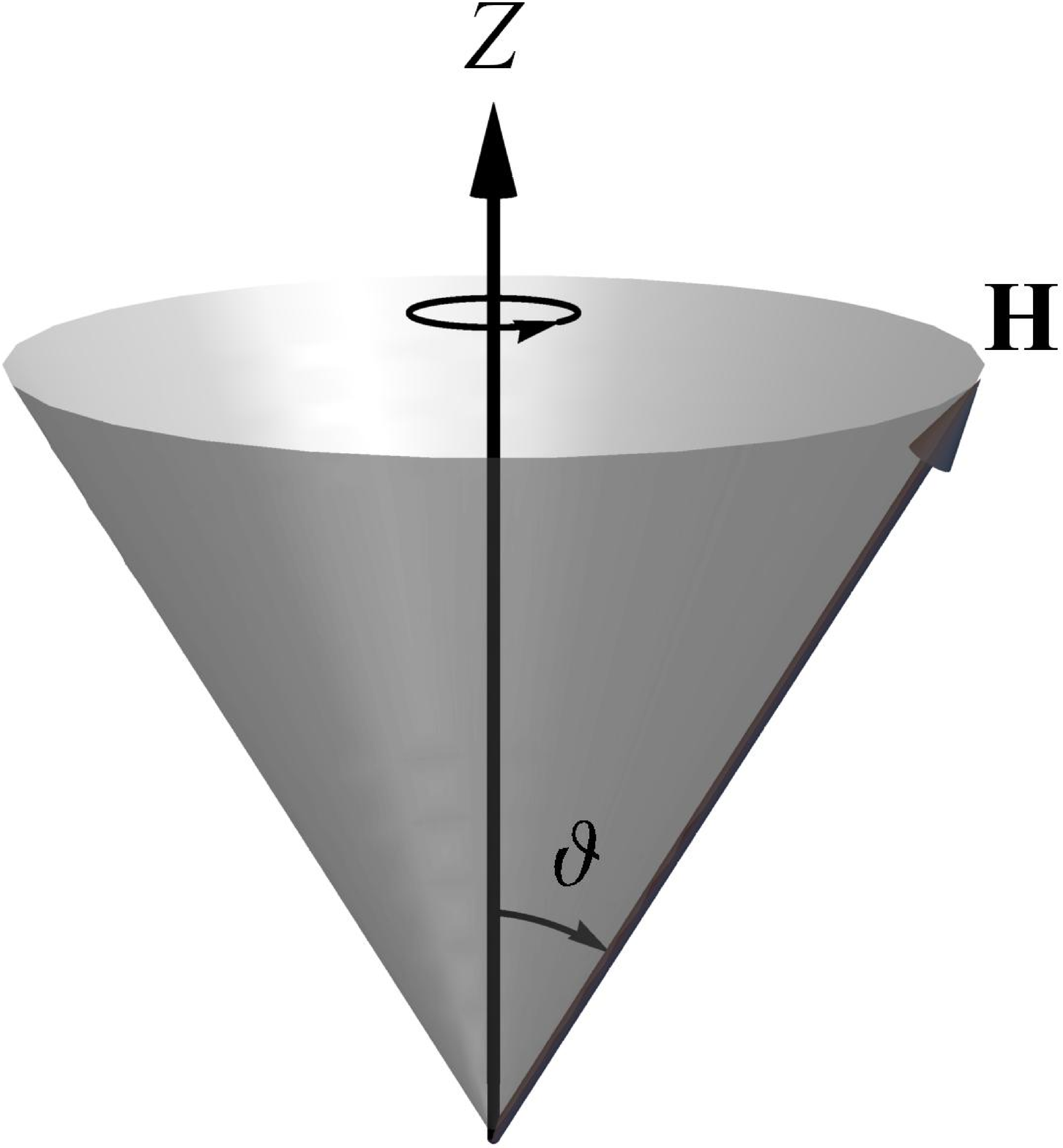}} &
\subfigure[]{\includegraphics[height=4cm]{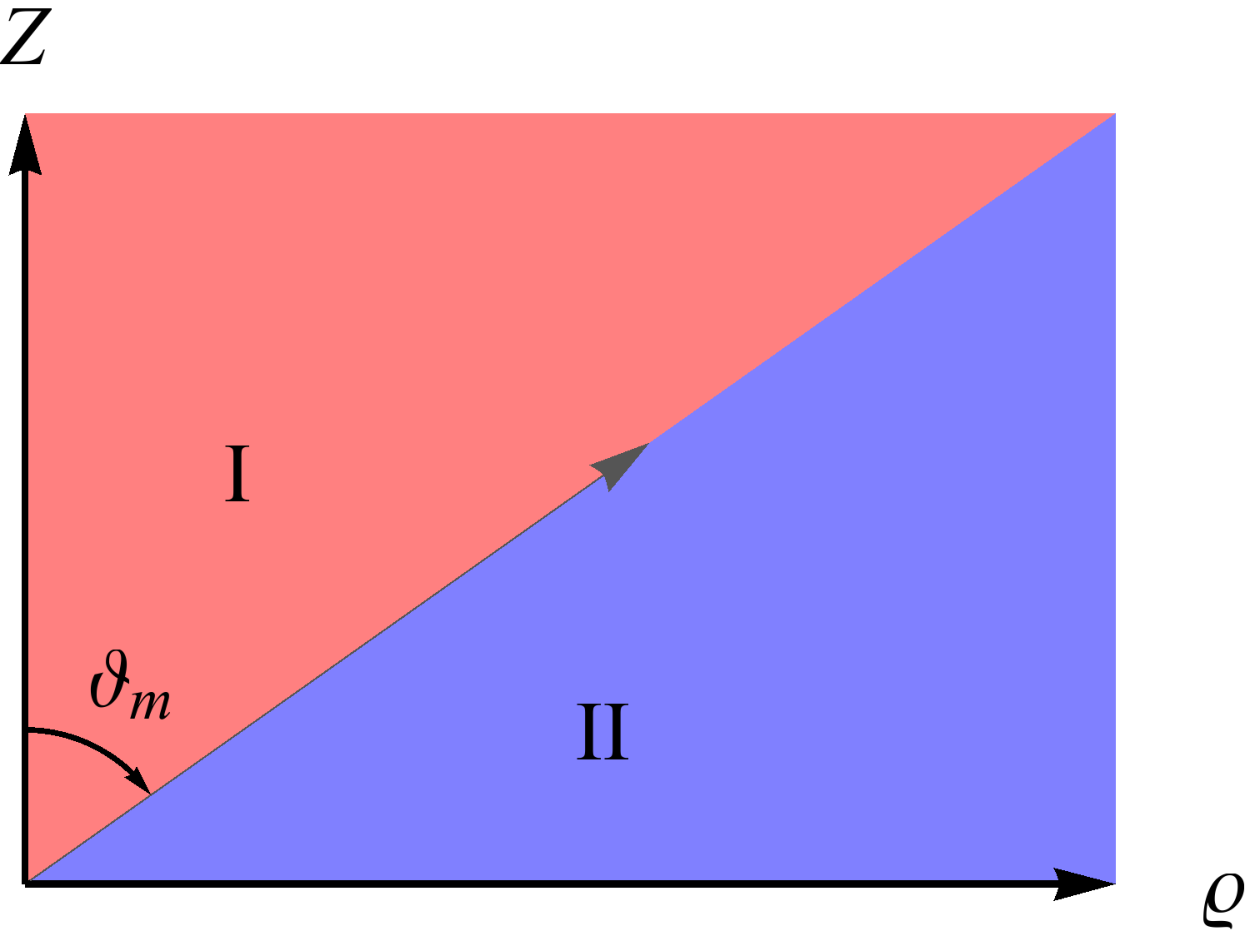}}
\end{tabular}
\caption{(Color online) (a) The magnetic field $\bfH$ precesses with an angle $\vartheta$ about the $Z$ axis. (b) The critical angle $\vartheta_m$, at which the magnetic modulus vanishes, divides the precession parameter space in two regions, the first one denoted by $I$, with $0 < \vartheta < \vartheta_m$ and $\calM >0$, and the second one denoted by $II$ with $\vartheta_m < \vartheta < \pi/2 $ and $\calM <0$.}\label{fig:FigPrecAng}
\end{figure}

\noindent
The magnetic filament is described by the curve $\Gamma: s \rightarrow \bfY(s) \in \mathbb{E}^3$, parametrized by arc length $s$ in three-dimensional Euclidean space and passing trough the center of the beads. Geometric quantities of the curve are expressed in terms of the Frenet-Serret (FS) frame adapted to the curve, denoted by $\{\bft, \bfn, \bfb\}$, see Fig. \ref{Fig:filament}.
\vskip0pc \noindent
The rotation of the FS frame along the curve is given by the FS formula
\begin{equation} \label{eq:FSformula}
\bft' = \bfD \times \bft\,, \quad \bfn' = \bfD \times \bfn\,, \quad \bfb'= \bfD 
\times \bfb\,,
\end{equation}
where $\bfD = \tau \bft + \kappa \bfb$ is the Darboux vector; $\kappa= \bft' \cdot \bfn$ and $\tau = \bfn' \cdot \bfb$ are the FS curvature and torsion, quantifying how the curve bends in the osculating and normal planes, respectively  \cite{KreyszigBook}.
\\
We consider paramagnetic filaments in the presence of a magnetic field $\bfH$ precessing at an angle $\vartheta$ about a direction we choose as the $Z$ axis, see Fig. \ref{fig:FigPrecAng}(a).
\vskip0pc  \noindent
The total energy density of the paramagnetic filament is the sum of the bending and magnetic energies $\mathcal{H}= \mathcal{H}_B + \mathcal{H}_M$,\footnote{We disregard the weight of the filament so we do not include gravitational effects; nor we include the magnetic dipole induced by the neighbors, whose influence would rescale the magnetic modulus \cite{Zhang1995, Cebers2003}} with $\mathcal{H}_B$ quadratic in the curvature \cite{Kratky1949, Kamien2002}, and $\mathcal{H}_M$ given by the time-averaged dipolar interactions between nearest neighbors induced by the magnetic field $\bfH$ (a derivation of $\calH_M$ is presented in Appendix \ref{sec:MagEn}) \cite{Goubalt2003, Cebers2003},
\begin{equation} \label{eq:lindensen}
\mathcal{H}_B = \frac{\calB}{2} \, \kappa^2\,, \quad \mathcal{H}_M = -\frac{\calM}{2}  \rmtz{}^2  \,,
\end{equation}
where $\rmtz = \bft \cdot \hat{\bf z}$ is the projection of the tangent vector onto the precession axis; $\calB$ is the bending modulus (with units of force times squared length); $\calM$ is 
the magnetic modulus (with units of force) defined by
\begin{equation} \label{def:magmodprec}
\calM(\vartheta) = \frac{\mu_0}{4 \pi} \left(\frac{3 \mu}{\Delta l^2}\right)^2 
\left(\cos^2 \vartheta - \frac{1}{3}  \right)\,,
\end{equation}
with $\mu_0$ the vacuum permeability, $\mu$ the magnitude of the magnetic dipoles, $\Delta l$ the separation between their centers and $\vartheta$ the precession angle.
\\
As shown below, configurations of the filaments depend sensitively on the sign of $\calM$, determined in turn by $\vartheta$. $\calM(\vartheta)$ vanishes at the so-called ``magic'' angle $\vartheta_m = \arccos\, (1/\sqrt{3}) \approx 3 \pi/10$, so in this case the leading order of the magnetic energy will be the quadrupolar term, which is of short range so filaments behave mostly as elastic curves \cite{Osterman2009}. In the regime $0 < \vartheta < \vartheta_m$, which will be termed as regime $I$, see Fig. \ref{fig:FigPrecAng}(b), the magnetic modulus is positive, $\calM(\vartheta)>0$ (the magnetic dipolar interactions are attractive), and from Eq. (\ref{eq:lindensen}) we see that in order to minimize $\calH_M$ the filaments will tend to align with the precession axis to maximize $\rmtz$ \cite{Martin2000}. By contrast in the regime $\vartheta_m < \vartheta <\pi/2$, termed as regime $II$, the magnetic modulus becomes negative, $\calM(\vartheta)<0$ (the magnetic dipolar interactions are repulsive), and $\calH_M$ is minimized when $\rmtz$ vanishes, so the filaments will tend to lie on the plane orthogonal to the precession axis \cite{Martin2000}.
\\
To reduce the material parameters space we rescale all quantities by the bending modulus $\calB$ and denote the rescaled quantity by an overbar. In particular, the rescaled quantity $\bar{\calM} = \calM/\calB$ possesses units of inverse squared length, so the inverse of its square root, $\ell = 1/\sqrt{|\bar{\calM}|}$, provides the characteristic length scale at which buckling occurs. The dimensionless parameter $\gamma = \bar{\calM} L^2 = \mathrm{sign}(\calM) (L/\ell)^2$ is known as the magnetoelastic parameter. This parameter quantifies the ratio of magnetic to bending energies: bending and magnetic energy scale as $H_B \sim \calB /L$ and $H_M \sim \calM L$, so their ratio scale as $H_M/H_B \sim \gamma$. Below, we use $\gamma$ as a parameter to characterize the conformations of the filaments. Typical experimental values of these paramagnetic filaments\footnote{For beads with diameter $2 a = \Delta l \approx 1 \upmu \mathrm{m}$, magnetic susceptibility $\chi = 0.67$ in a magnetic field $\mathrm{H}\approx 1000 \mathrm{A}/\mathrm{m}$, the magnitude of the induced dipole is $\mu = 4/3 \pi a^3 \chi \mathrm{H} \approx 1/3 \times 10^{-15} \mathrm{A} \mathrm{m}^2 $.} are $L \approx 10 - 100 \upmu \textrm{m}$,  $\calB \approx 10^{-25} -10^{-21} \mathrm{N} \mathrm{m^2}$, $\calM \approx 10^{-13} \mathrm{N}$, so $\ell \approx 1-100 \upmu \mathrm{m}$, and $|\gamma| \approx 10^{-2}- 10^{4}$, \cite{Biswal2003, Goubalt2003, Biswal2004}.
\\
The total bending and magnetic energies of the filament are given by the line integrals of the corresponding energy densities
\begin{equation}
H_B = \int_\Gamma \rmd s \, \mathcal{H}_B \,, \quad H_M = \int_\Gamma \rmd s \, \mathcal{H}_M\,.
\end{equation}
Thus the total energy is $H[\bfY] = H_B + H_M$, but since the filament is inextensible, we consider the effective energy
\begin{equation} \label{def:toten}
H_E[\bfY] = H[\bfY] + \Lambda (L-L_0) \,,
\end{equation}
where $\Lambda$ is a Lagrange multiplier fixing total length which acts as an intrinsic line tension.
\\
The change of the energy $H_E$ under a deformation of the curve $\bfY \rightarrow \bfY + \delta \bfY$ is given by 
\cite{CapoChryssGuv2002, GuvVaz2012, GuvValVaz2014}
\begin{equation} \label{eq:1stvarH}
\delta H_E = \int \rmd s \, \mathbf{F}' \cdot \delta \bfY + \int \rmd s \, \delta Q'\,.
\end{equation}
In the first term, which represents the response of the energy to a deformation in the bulk, $\bfF$ is the force vector, given by the sum of the bending and magnetic forces, $\bfF = \bfF_B + \bfF_M - \Lambda \bft $, defined by \cite{Langer1996, CapoChryssGuv2002, GuvVaz2012, GuvValVaz2014, Dempster2017}
\begin{subequations} \label{def:FBFM}
\begin{eqnarray} 
\bfF_B &=&  \calB \left( \frac{\kappa^2}{2}\, \bft + \kappa' \bfn + \kappa \tau \bfb \right)\,,\\ 
\bfF_M &=&   \calM \, \rmtz  \left(\hat{\bf z} - \frac{\rmtz}{2} \bft \right) 
= \calM \,  \rmtz \left(\frac{\rmtz}{2} \, \bft + \rmnz \, \bfn + \rmbz \, \bfb \right)\,,
\end{eqnarray}
\end{subequations}
where $\rmnz = \bfn \cdot \hat{\bf z}$ and $\rmbz = \bfb \cdot \hat{\bf z}$. $\bfF$ is the force exerted by the line element at $s$ on the neighboring line element at $s+ \rmd s$, so $F>0$ ($F<0$) represents compression (tension). From the force balance at the boundaries follows that $-\bfF$ is the external force on the filament \cite{Langer1996}.
We see in Eq. (\ref{def:FBFM}) that the magnitudes of the bending and magnetic forces scale as $F_B \sim \calB/L^2$ and $F_M \sim \calM$. Thus, the magnetoelastic parameter also quantifies the ratio of magnetic to bending forces, $F_M /F_B \sim \gamma$, \cite{Cebers2003}. If the filaments are immersed in a medium of viscosity $\eta$, we have from the balance of bending and viscous forces that the characteristic bending relaxation time is $T_B \sim \eta L^4 / \calB$,  \cite{Powers2010}, whereas the characteristic magnetic relaxation time is $T_M \sim \eta L^2/\calM$ \cite{Dempster2017}. The ratio of bending to magnetic relaxation times is the magnetoelastic parameter $T_B/ T_M \sim \gamma$. For a filament of length $L =10- 100 \upmu \mathrm{m}$, bending and magnetic moduli $\calB \approx 10^{-25} - 10^{-21} \mathrm{N} \mathrm{m}^2$ and $\calM \approx 10^{-13} \mathrm{N}$, in water ($\eta \approx 10^{-3} \, \mathrm{N s}/\mathrm{m}^2$) we have $T_B \approx 10^{-2}-10^6 \, \mathrm{s}$ and $T_M \approx 1-10^2 \, \mathrm{s}$. Therefore, in order to be legitimate, the use of the time-averaged magnetic energy density given in Eq. (\ref{eq:lindensen}) is justified if the precessing period $T$ is less than a millisecond, $T < 10^{-3} \, \mathrm{s}$ (frequency $\nu=1/T > 1 \, \mathrm{kHz}$ \cite{Dempster2017}), so that the characteristic relaxation times are much larger, $T_B, T_M \gg T$.
\\
The second term in Eq. (\ref{eq:1stvarH}) contains quantities arising after integration by parts and is given by the total derivative of 
\begin{equation} \label{def:deltaQ}
\delta Q = -\bfF \cdot \delta \bfY + \calB \kappa \bfn \cdot \delta \bft\,,
\end{equation}
so it represents the change of the energy due to boundary deformations.
\\
Stationarity of the energy implies that in equilibrium the force vector is conserved along the filament, $\bfF' = \mathbf{0}$, a consequence of the translational invariance of the total energy. By contrast, the torque vector, $\bfM = \bfY \times \bfF + \bfS$, with  $\bfS =- \calB \kappa \bfb$, is not conserved: $\bfM' = \bfY \times \bfF' + \bft \times \bfF_M$,\footnote{In this expression we have used the identity $\bfS' + \bft \times \bfF =0$, \cite{CapoChryssGuv2002}.} while the first term vanishes in equilibrium, the second term, representing a torque per unit length due to the magnetic field, $\bft \times \bfF_M = - \calM \, \rmtz (\rmbz \bfn - \rmnz \bfb)$,  does not vanish in general. However, the component of the torque along the precession axis, $M^z = \bfM \cdot \hbfz$, is conserved on account of the rotational symmetry of the energy about such direction: $M^z{}' = \bfM' \cdot \hbfz = \hbfz \cdot \bfY \times \bfF'$, which vanishes in equilibrium. 
\\
Spanning the derivative of $\bfF$ in terms of the two normals as $\mathbf{F}' = \varepsilon_\bfn \bfn + \varepsilon_\bfb \bfb$,\footnote{The projection onto the tangent vanishes identically due to the reparametrization invariance of the energy \cite{CapoChryssGuv2002}.} so the normal projections of the conservation law provide the Euler-Lagrange (EL) equations satisfied by equilibrium configurations, which read \cite{Dempster2017}
\begin{subequations} \label{eq:ELNB}
\begin{eqnarray}
\bar{\varepsilon}_\bfn &=&  \kappa'' + \kappa \, 
\left( \frac{\kappa^2}{2} - \tau^2 - \bar{\calM} \left(\frac{\rmtz{}^2}{2} - 
\rmnz{}^2\right) - \bar{\Lambda} \right)  = 0\,, \label{eq:ELN}\\
\bar{\varepsilon}_\bfb &=&  \kappa \, \tau' + 2 
\kappa' \tau + \, \bar{\calM} \, \kappa \, \rmnz \, \rmbz=0\,. \label{eq:ELB}
\end{eqnarray}
\end{subequations}
In solving these equations, the Lagrange multiplier $\Lambda$ is determined from boundary or periodicity conditions.
\\
The squared magnitude of the force vector
\begin{equation} \label{eq:F2}
\bar{F}^2 = \bar{\bfF} \cdot \bar{\bfF} = \left(\frac{\kappa^2}{2} + 
\frac{\bar{\calM}}{2} \,  \rmtz{}^2  - \bar{\Lambda} \right)^2 + \left(\kappa' +  
\bar{\calM} \, \rmtz \rmnz  \right)^2 + \left(\kappa \, \tau + \bar{\calM}\, 
\rmtz \rmbz 
\right)^2 \,,
\end{equation}
is constant on account of the conservation law of $\bfF$. This constant corresponds to the first Casimir of the Euclidean group and provides a first integral of the EL equations.\footnote{EL Eq. (\ref{eq:ELB}) can be written as $\kappa \, \bar{\varepsilon}_\bfb = -(\bar{\bfF} \cdot \bar{\bfS})' - \bar{\calM} \rmtz (\kappa \rmbz)'$. Thus, the scalar quantity $\bfF \cdot \bfS$, corresponding to the second Casimir in the case of Euler Elastica, is not conserved in equilibrium because the magnetic field breaks the rotational invariance of the energy and introduces a source of stresses.}
Below we analyze solutions of two families of curves satisfying these equations with their ends held along the precession axis: curves lying on a plane passing through the precession axis and helices whose axis is parallel to the precession axis.

\section{Planar cuves} \label{sec:PlanarCuves}
\begin{figure}[htbp]
\begin{center}
\includegraphics[height=7cm]{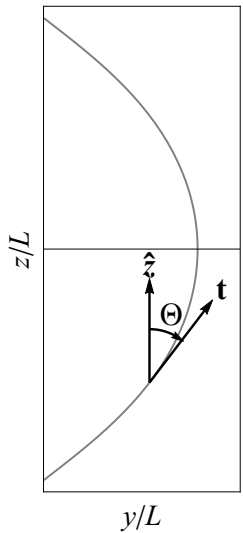}
\end{center}
\caption{The planar curve can described by the angle $\Theta$, formed by $\bft$ and $\hbfz$.}\label{Fig:defTheta}
\end{figure}

\noindent
Let us consider curves on a plane, say $Y$-$Z$, so the embedding functions are $\bfY = y \hat{\bf y} + z \hat{\bf z}$ and the tangent vector is $\bft = y' \hat{\bf y} + z' \hat{\bf z}$. Since the curve lies on a plane, it has vanishing torsion, $\tau = 0$, and the EL equation associated with deformations along $\bfn$ reduces to
\begin{equation} \label{eq:ELPlanar}
\bar{\varepsilon}_\bfn = \kappa'' + \kappa \left( \frac{\kappa^2}{2} - \bar{\calM}\left(\frac{\rmtz{}^2}{2} - \rmnz{}^2\right) - \bar{\Lambda} \right) = 0\,,
\end{equation}
whereas the EL corresponding to deformations along $\bfb$ is satisfied identically, because $\varepsilon_\bfb$ vanishes on account of the orthogonality of the binormal vector to the plane of the curve, i.e. $\rmbz=0$. The force vector, defined in Eq. (\ref{def:FBFM}), lies on the osculating plane of the curve
\begin{equation} \label{eq:Fplanar}
\bar{\bfF} := \bar{F}^y \hat{\bf y} + \bar{F}^z \hat{\bf z}= \left(\frac{1}{2} (\kappa^2 + \bar{\calM} \, \rmtz{}^2) - \bar{\Lambda} \right) \bft 
+ \left( \kappa' + \bar{\calM} \,  \rmtz \rmnz \right) \bfn \,,
\end{equation}
Projecting $\bfF$ onto the FS basis $\{\bft, \bfn \}$ we obtain
\begin{subequations} \label{eq:FprojsTN}
\begin{eqnarray} 
\frac{\kappa^2}{2} +  \frac{\bar{\calM}}{2} \,  \rmtz{}^2 - \bar{\Lambda} &=& \bar{\bfF} \cdot \bft \,, \label{eq:FprojsT} \\
 \kappa' + \bar{\calM} \,  \rmtz \rmnz &=& \bar{\bfF} \cdot \bfn \,, 
\label{eq:FprojsN}
\end{eqnarray}
\end{subequations}
Differentiating Eq. (\ref{eq:FprojsT}) with respect to $s$ and using the FS formula we obtain Eq. (\ref{eq:FprojsN}), whereas differentiation of Eq. (\ref{eq:FprojsN}) reproduces the EL Eq. (\ref{eq:ELPlanar}). Therefore Eq. (\ref{eq:FprojsT}) provides a second integral of the EL Eq. (\ref{eq:ELPlanar}), which permit us to express the difference between the bending and magnetic energy densities as the sum of the tangential component of the force and the constant $\Lambda$. Moreover, this relation can be used to eliminate the curvature in favor of the projections of the tangent vector, for instance, the total energy density can be recast as
\begin{equation} \label{def:endensimp}
\mathcal{H} = \mathcal{H}_B + \mathcal{H}_M = \bfF \cdot \bft - \calM \,  \rmtz{}^2 + \Lambda \,.
\end{equation}
In order to solve the second integral (\ref{eq:FprojsT}), it is convenient to parametrize the planar curve by the angle $\Theta$ that the tangent makes with the precession axis ($\Theta>0$ in the clockwise sense), see Fig. \ref{Fig:defTheta}.
\vskip0pc \noindent
In terms of $\Theta$ the tangent and normal vectors are $\bft = \sin \Theta \hbfy + \cos \Theta \hbfz$ 
and $\bfn = -\cos \Theta \hbfy + \sin \Theta \hbfz$, whereas the FS curvature is 
$\kappa=-\Theta'$. Expressing Eq. (\ref{eq:FprojsT}) in terms of $\Theta$, it reduces to a quadrature for $\Theta$:
\begin{equation} \label{eq:quadpsi}
\frac{1}{2} (\Theta')^2 + V(\Theta) = \bar{\Lambda}\,,
\end{equation}
where we have defined
\begin{equation} \label{eq:Vvert}
 V(\Theta) = - \bar{F}^y \sin \Theta - \bar{F}^z \cos \Theta + \frac{\bar{\calM}}{2} \cos^2 \Theta\,.
\end{equation}
Regarding $\Theta$ as the position of a unit mass particle and $s$ as time, Eq. (\ref{eq:quadpsi}) represents the total energy $\bar{\Lambda}$ given by the sum of the kinetic energy $(\Theta')^2/2$ and the potential energy $V(\Theta)$. Likewise, Eq. (\ref{eq:FprojsN}) reads 
\begin{equation} \label{eq:motion}
 \Theta''- \bar{F}^y \cos \Theta +\bar{F}^z \sin \Theta - \bar{\calM} \sin \Theta \cos \theta =0 \,,
\end{equation}
which in the mechanical analogy would represent the corresponding equation of motion, see Appendix \ref{sec:Hamform}.
\\
We consider filaments with their boundaries fixed, but not the tangents. Thus the variation vanishes at the boundaries (which we set at $s=\pm s_b$), i.e. $\delta \bfY(\pm s_b) = \mathbf{0}$, and from Eq. (\ref{def:deltaQ}) we have that the stationarity of the energy at the boundaries, $\delta Q(\pm s_b)=0$, imply the vanishing of the curvature at those points. Therefore the appropriate boundary conditions (BC) for equilibrium configurations is
\begin{equation} \label{def:BCfixedends}
\kappa (\pm s_b) = - \Theta'(\pm s_b)=0 \,.
\end{equation}
In consequence, the intrinsic torque $\bar{\bfS} = -\kappa \bfb$ vanishes at the ends and only the torque coupling position and force contributes. Furthermore, the quadrature implies that the maximum value of the angle, say $\Theta_M$, occurs at the boundaries, i.e. $\Theta (\pm s_b) = \Theta_M$. Thus at the turning points the ``kinetic'' energy vanishes and the ``potential'' energy is equal to the ``total energy'' \cite{AudolyBook}, which determines the Lagrange multiplier $\Lambda$ in terms of the angle $\Theta_M$:
\begin{equation}
V(\Theta_M) =  \bar{\Lambda} \,.
\end{equation}
Once $\Theta$ has been determined as a function of $s$, the coordinates are obtained by integrating the tangential components 
\begin{equation} \label{eq:ypzpvert}
t^y= y' = \sin \Theta\,, \quad t^z = z' = \cos \Theta\,. 
\end{equation} 
In the next section we apply these results to the case of filaments aligned with the precession axis.

\section{Vertical filaments} \label{sec:vetfils}
\noindent
Here we consider a curve resulting from a deformation of a straight filament lying along the precession axis, chosen as the $Z$ axis, such that the end points remain along this axis (see Fig. \ref{Fig:defTheta}). In consequence the force is also along the precession axis: $F^y=0$ and $F^z =F$. Thus, the potential reduces to $ V(\Theta) = - \bar{F} \cos \Theta + \frac{\bar{\calM}}{2} \cos^2 \Theta$, which has period $2 \pi$ and left-right symmetry $\Theta \rightarrow -\Theta$.  
\\
We set the mid point of the curve at $Z=0$ from where arc length is measured, being positive (negative) above (below) the $Y$ axis, i.e. $s \in [-s_b,s_b]$ with $s_b = L/2$. We denote the height of the boundary by $\pm z_b = z(\pm s_b)$ so that the height difference is $\Delta z = 2 z_b \leq L$. We characterize the curves by the height difference rescaled with the total length $L$
\begin{equation} \label{def:xi}
 \xi = \frac{\Delta z}{L} \leq 1 \,.
\end{equation}
To gain some insight about how the magnetic field modifies the behavior of the filaments, we first solve the quadrature (\ref{eq:quadpsi}) in the regime of small deviations from a vertical straight line.

\subsection{Perturbative regime} \label{sec:PertAnvert}
\begin{figure}[htbp]
\centering
\includegraphics[width=0.43\textwidth]{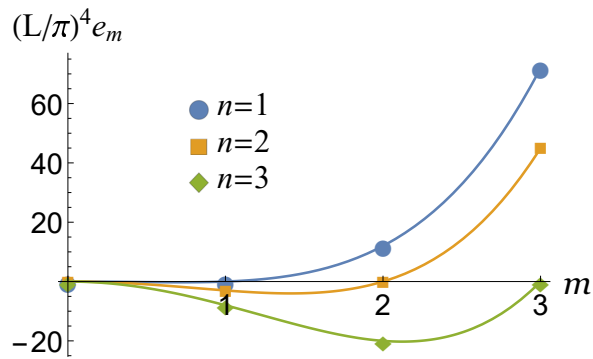}
\caption{(Color online) Eigenvalues corresponding to the first deformation modes. Only the 
state $n=1$ is 
stable in the perturbative regime.} \label{fig:Eigenvalueshor}
\end{figure}

\noindent
We consider a small perturbation $ \Theta_{(1)} \leq \Theta_{M(1)} \ll 1$  of a straight line with $\Theta = 0$ and we expand the constants perturbatively as $F = F_{(0)} + F_{(2)}$, $\calM = \calM_{(0)} + \calM_{(2)}$.\footnote{$F$ and $\calM$ are constants, however, we are interested in determining the corrections as functions of a small parameter, determined below, required by deviations from a straight line.} At quadratic order, the quadrature describes a harmonic motion
\begin{equation} \label{eq:linquadvert}
\Theta'_{(1)}{}^2 + (\bar{F}_{0} - \bar{\calM}_{0}) \Theta_{(1)}^2 =  \left(\bar{F}_{(0)}-\bar{\calM}_{(0)}\right) \Theta_{M(1)}^2\,.
\end{equation}
Only for $\bar{F}_{0}  > \bar{\calM}_{0}$\footnote{If $\bar{F}_{0} = \bar{\calM}_{0}$, then $\kappa_{(1)} = -\Theta'_{(1)}=0$ and the curve is a straight line.} the quadratic potential is positive and bounded solutions are possible, given by
\begin{equation} \label{eq:pertvertsols}
\Theta_{(1)} = \Theta_{M(1)} \sin q (s - s_0)\,, \quad q^2 = \bar{F}_{0} - \bar{\calM}_{0} > 0\,.
\end{equation} 
If the filament develops $n$ half periods,\footnote{For instance the filament shown in Fig. \ref{Fig:defTheta} completes only one half-period ($n=1$).} the wave number $q$ is given by
\begin{equation} \label{def:qpert}
q = \frac{n \pi}{L}, \quad n \in \mathbb{N}. 
\end{equation}
The boundary condition (\ref{def:BCfixedends}), requiring the vanishing FS curvature, $\kappa_{(1)} = -\Theta_{(1)}' = - q \, \Theta_{M(1)} \cos q (s-s_0)$ at the ends, determines $s_0 = - \mathrm{mod}(n-1,2)L/(2 n)$, where $\mathrm{mod}(a,b)$ stands for $a$ $\mathrm{module}$ $b$.
\\
Combining expressions (\ref{eq:pertvertsols}) and (\ref{def:qpert}) for $q$, we determine the magnitude of the force
\begin{equation} \label{eq:pertwnvert}
\bar{F}_{(0)} = \left(\frac{n \pi}{L}\right)^2 + \bar{\calM}_{(0)}\,, 
\quad \mbox{or} 
\quad L^2 \bar{F}_{(0)}  = (n \pi)^2 + \gamma_{(0)}\,.
\end{equation}
We see that the magnetic field modifies the minimum force required to trigger an Euler buckling instability. Furthermore, unlike the purely elastic case where the force on the filaments is always compressive, the magnetic contribution enables the force to be either tensile or compressive depending on the value of the magnetoelastic parameter relative to the squared number of half-periods: for $\gamma_{(0)} > - (n \pi)^2$ (precession regime $I$ or $II$) the force is positive, $F_{(0)}>0$, so the filament is under compression, whereas for $\gamma_{(0)}< -(n \pi)^2<0$ (precession regime $II$) the force becomes negative, $F_{(0)}<0$, and the filament is under tension. In the particular case with $\gamma_{(0)}= -(n \pi)^2$, free filaments with $F_{(0)}=0$ are possible \cite{Cebers2005a}.
\vskip0pc \noindent
The coordinates can be obtained by integrating Eq. (\ref{eq:ypzpvert}), obtaining
\begin{subequations} \label{eq:pertyzvert}
\begin{eqnarray}
y &=& - \frac{\Theta_{M(1)}}{q} \cos q (s-s_0) \,, \\
z &=& s + \frac{\Theta_{M(1)}^2}{8 q} \left(\sin 2 q (s-s_0) - 2 q s \right)\,.
\end{eqnarray}
\end{subequations}
Evaluating the second expression at the boundaries we determine the amplitude $\Theta_{M(1)}$ in terms of the scaled height difference $\xi$ defined in (\ref{def:xi}):
\begin{equation} \label{eq:betaM}
\Theta_{M(1)} = 2 \sqrt{1 - \xi}\,.
\end{equation}
The total energy of the filament is $\bar{H} = \bar{H}_{(0)} + \bar{H}_{(2)}$, where $L \bar{H}_0 = -\gamma_{(0)}/2$ is the scaled energy of the original straight vertical state and the second order correction is
\begin{equation} \label{def:2ndordchangeEn}
L \bar{H}_{(2)} =\left(\frac{\Theta_{M(1)}}{2} \right)^2 \left((n\pi)^2 + \gamma_{(0)}\right) = (1-\xi) L^2 \bar{F}_{(0)} \,.
\end{equation}
Since $H_{(2)}$ increases linearly with the magnitude of the force it can be either positive or negative.
\\
Let us now look at the stability of these states. To lowest order, the differential operator of the second variation of the energy, (derived in Appendix \ref{Sect:app2ndvar}, Eq. (\ref{eq:defLdiff})), reads
\begin{equation} \label{eq:L0thhor}
\mathcal{L} = \frac{\partial^2}{ \partial s^2} \left(\frac{\partial^2}{ \partial 
s^2} + q^2 \right)\,.
\end{equation}
The two trivial zero modes (with vanishing eigenvalues), $\delta \phi=0,a$, with $a$ constant, are associated to the translational invariance of the energy: at lowest order they correspond to infinitesimal vertical and horizontal translations 
respectively. To analyze the eigenmodes of $\mathcal{L}$ we use the basis $\delta \phi_k = \sin k s, \cos k s$, which in order to preserve the periodicity of the original curves should have wave numbers $k = m \pi/L$, $m\in \mathbb{N}$, so the corresponding eigenvalues for each case are
\begin{equation} \label{def:evLhorpert}
e_{m} = k^2 \left(k^2 - q^2 \right) = \left(m \pi^2/L^2\right)^2 \left( m^2 -n^2 
\right)\,.
\end{equation}
The two non-trivial zero modes with $m =n$ correspond to infinitesimal rotations in the $Y-Z$ plane, but for finite rotations such eigenmodes will not be zero modes, because the energy is only invariant under rotations about the precession axis (in the $X$-$Y$ plane). The eigenvalues corresponding to states with $n=1,2,3$ are shown in Fig. \ref{fig:Eigenvalueshor}. We see that, like in the purely elastic case, only the eigenvalues of the ground state $n=1$ are all positive, so it is the only stable state in the perturbative regime. Therefore, any excited state $n>1$ would decay recursively to the next intermediate state with the more negative eigenvalue until the ground state is reached, \cite{Guven2012}. As we will see below, comparison of the total energy of successive states leads suggests that the state $n=1$ is still the ground state in the non-linear regime.

\subsection{Non-linear regime} 
\begin{figure}[htbp]
\centering
\begin{tabular}{cccc}
\includegraphics[scale=0.375]{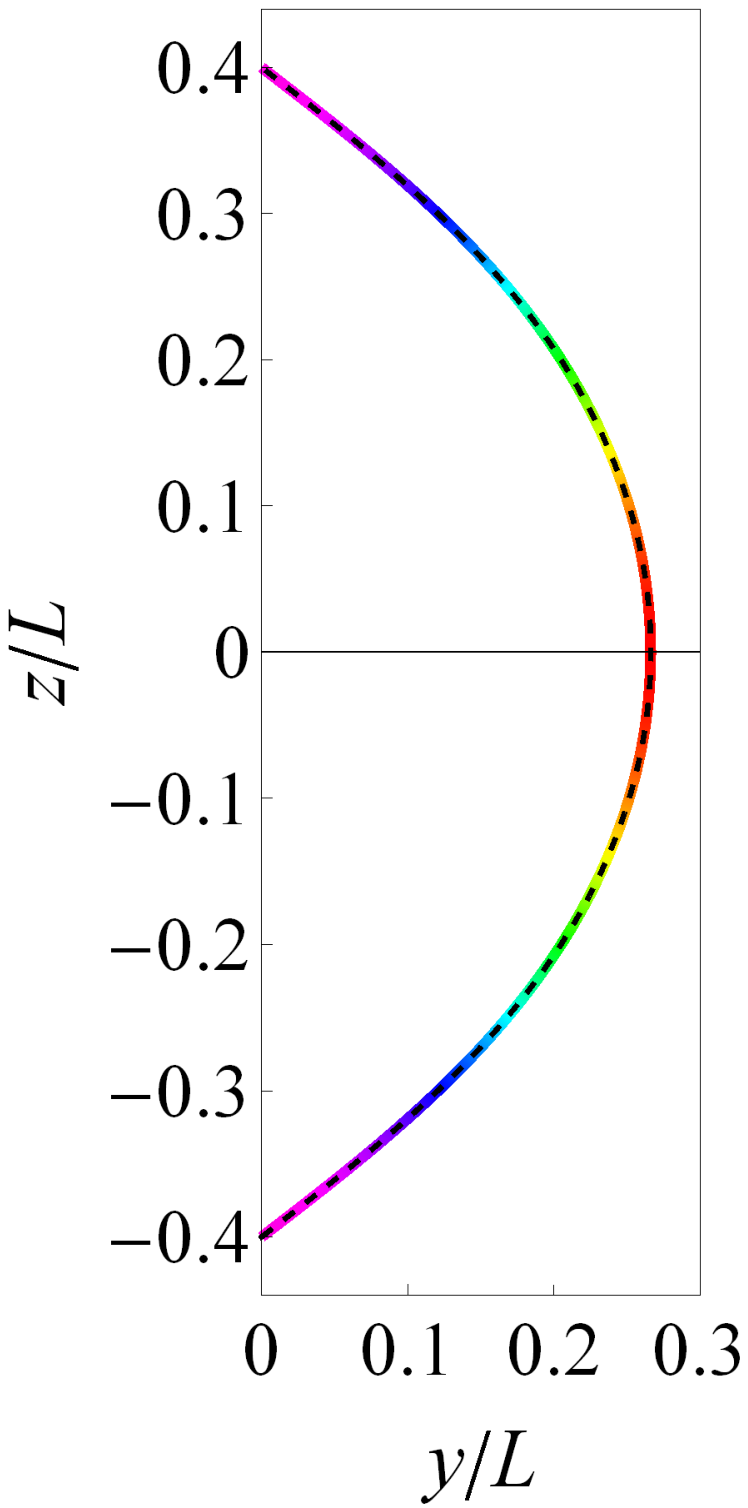} &
\includegraphics[scale=0.375]{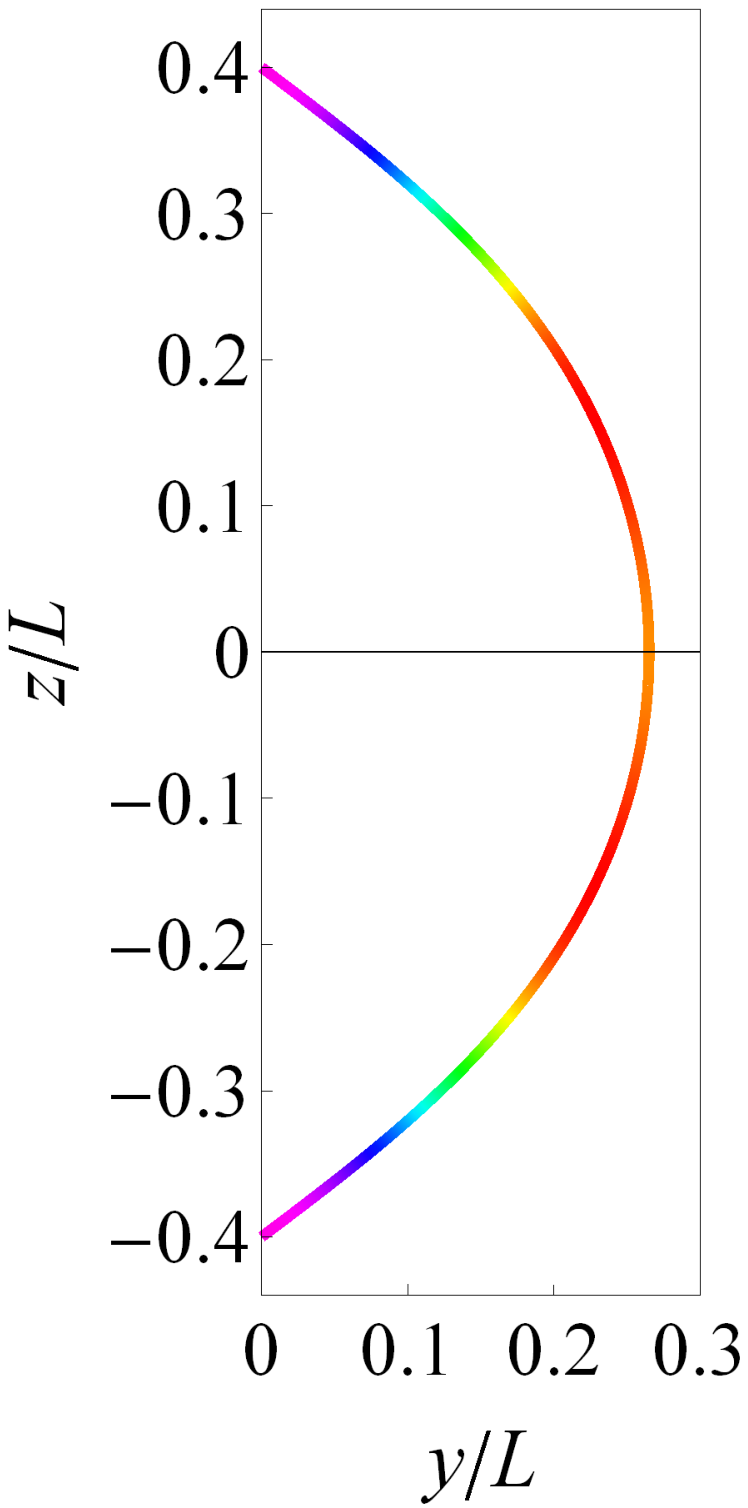} &
\includegraphics[scale=0.375]{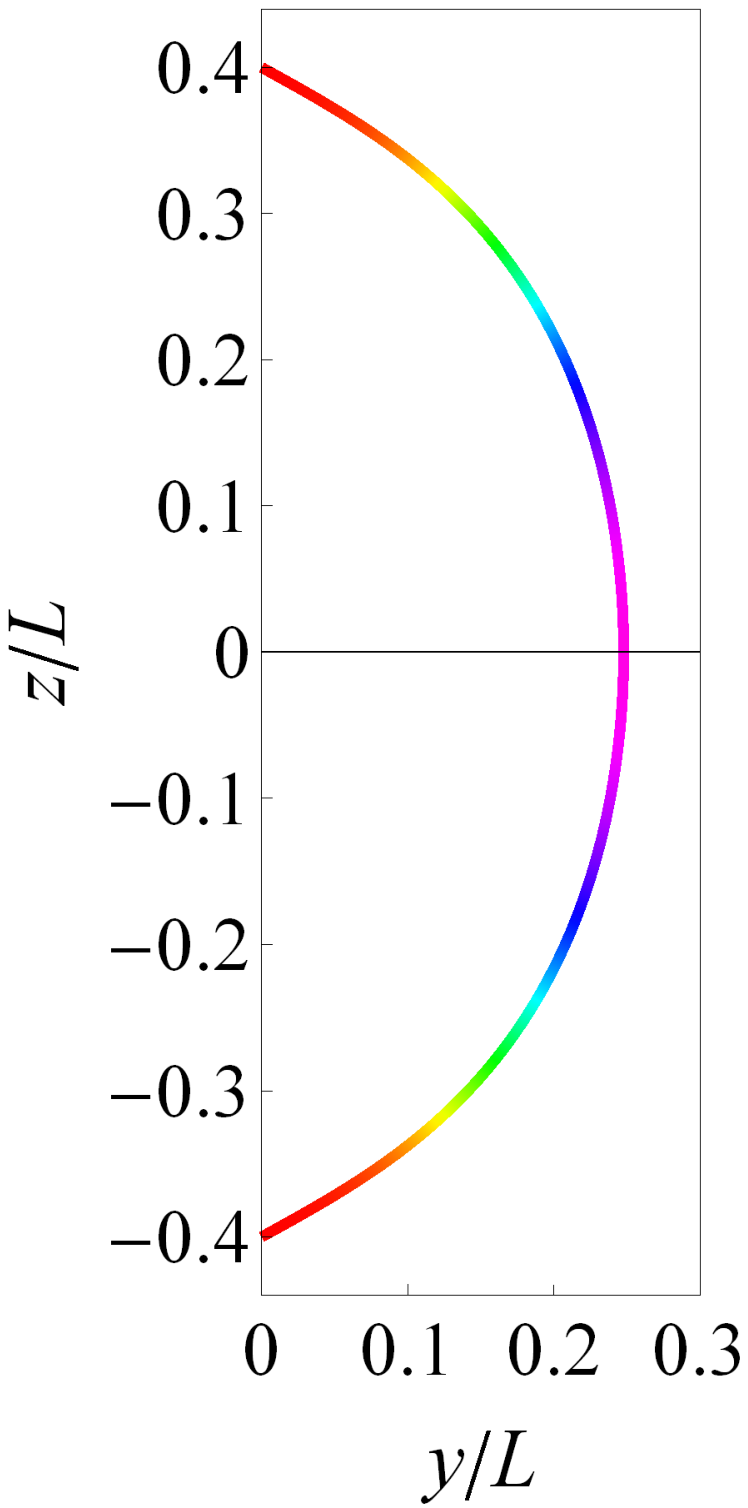} &
\includegraphics[scale=0.375]{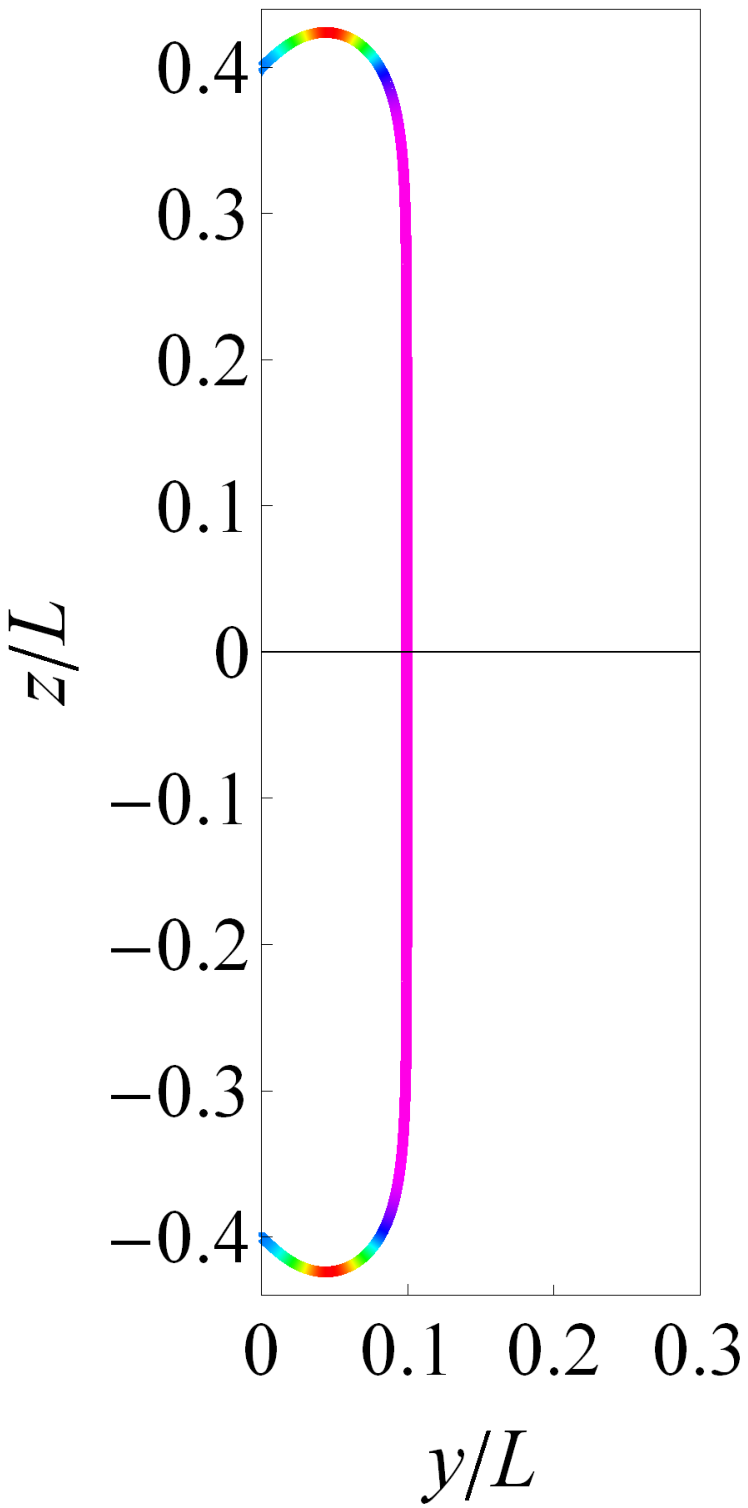} \\
\includegraphics[scale=0.35]{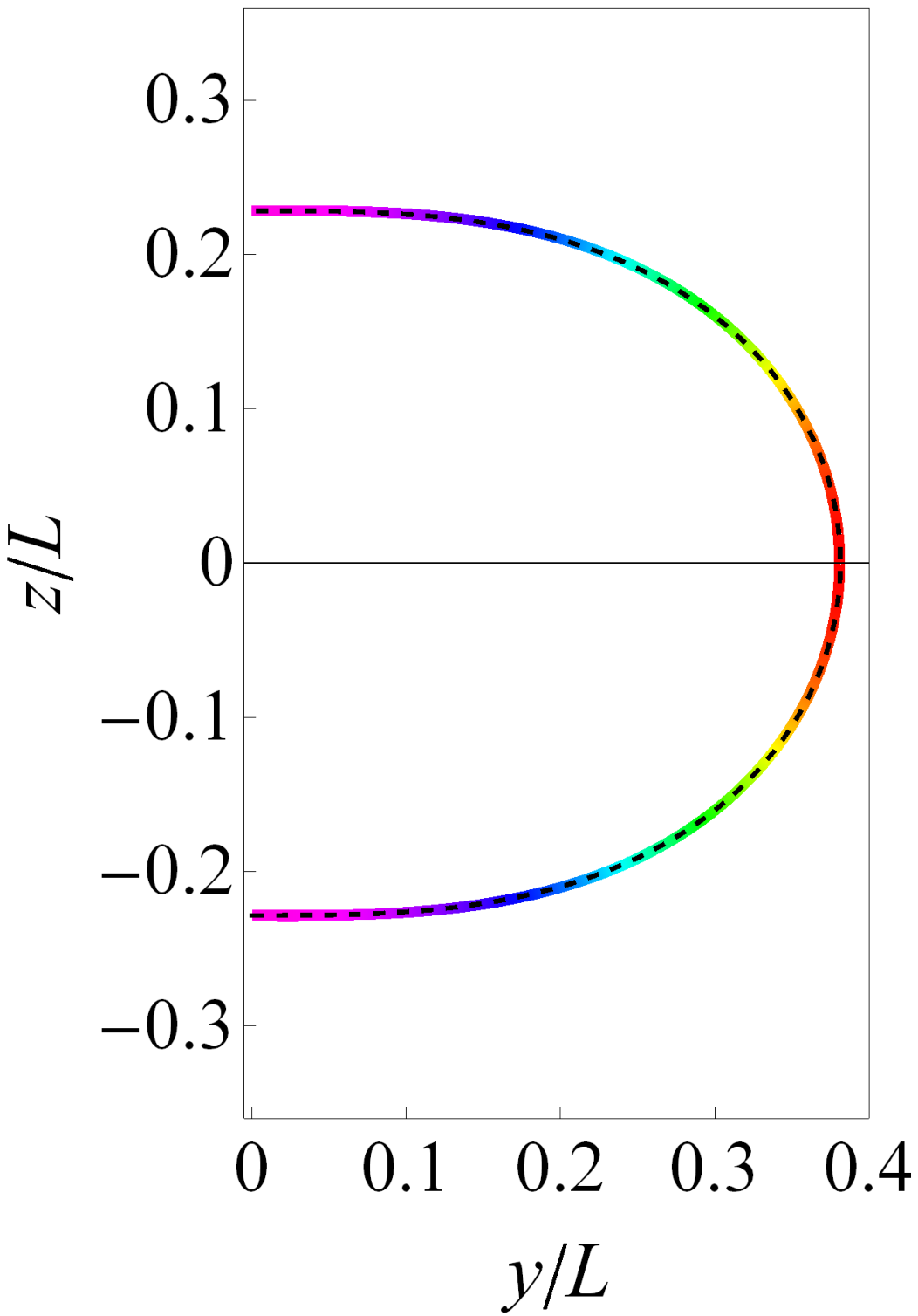}  &
\includegraphics[scale=0.35]{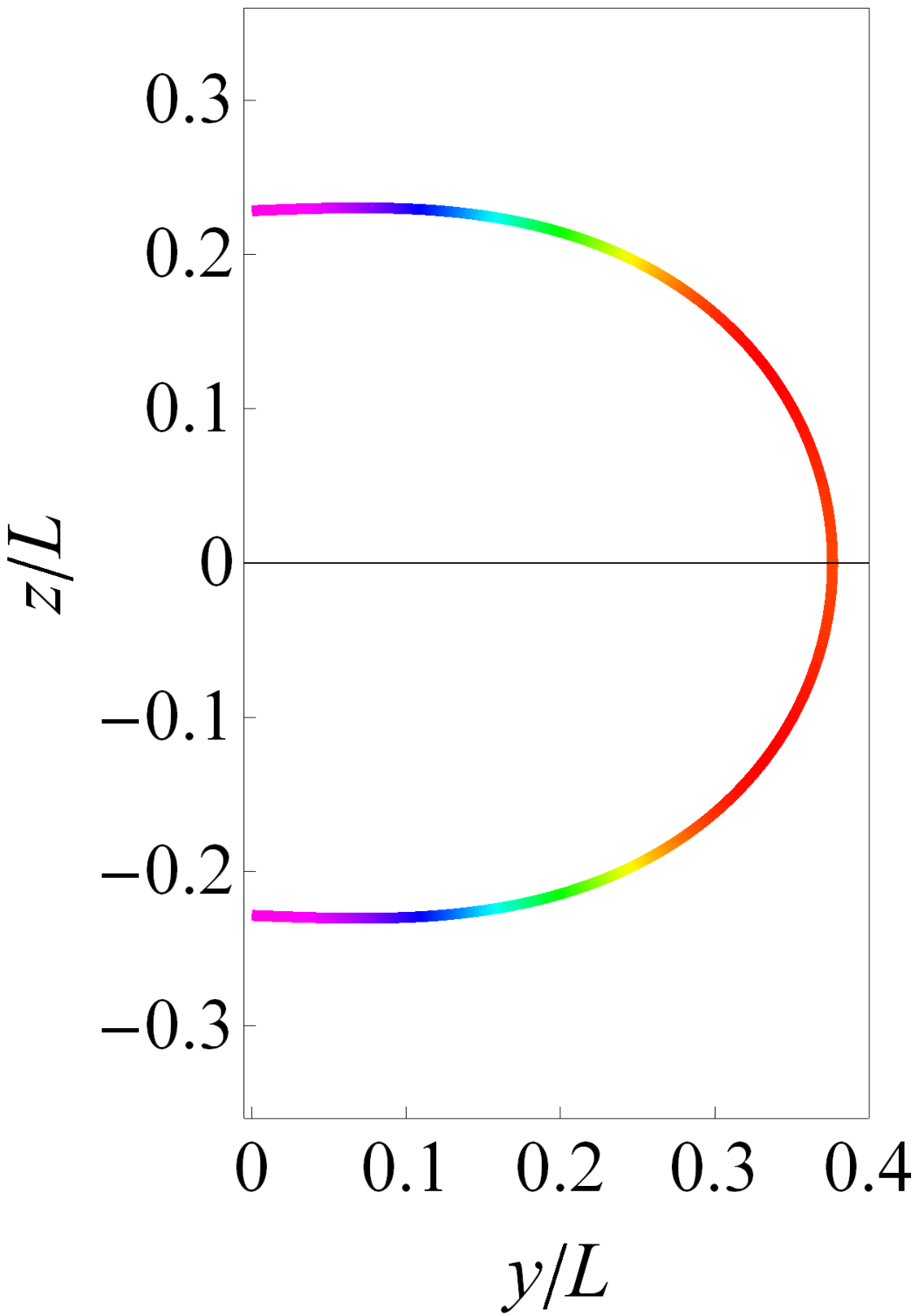}  &
\includegraphics[scale=0.35]{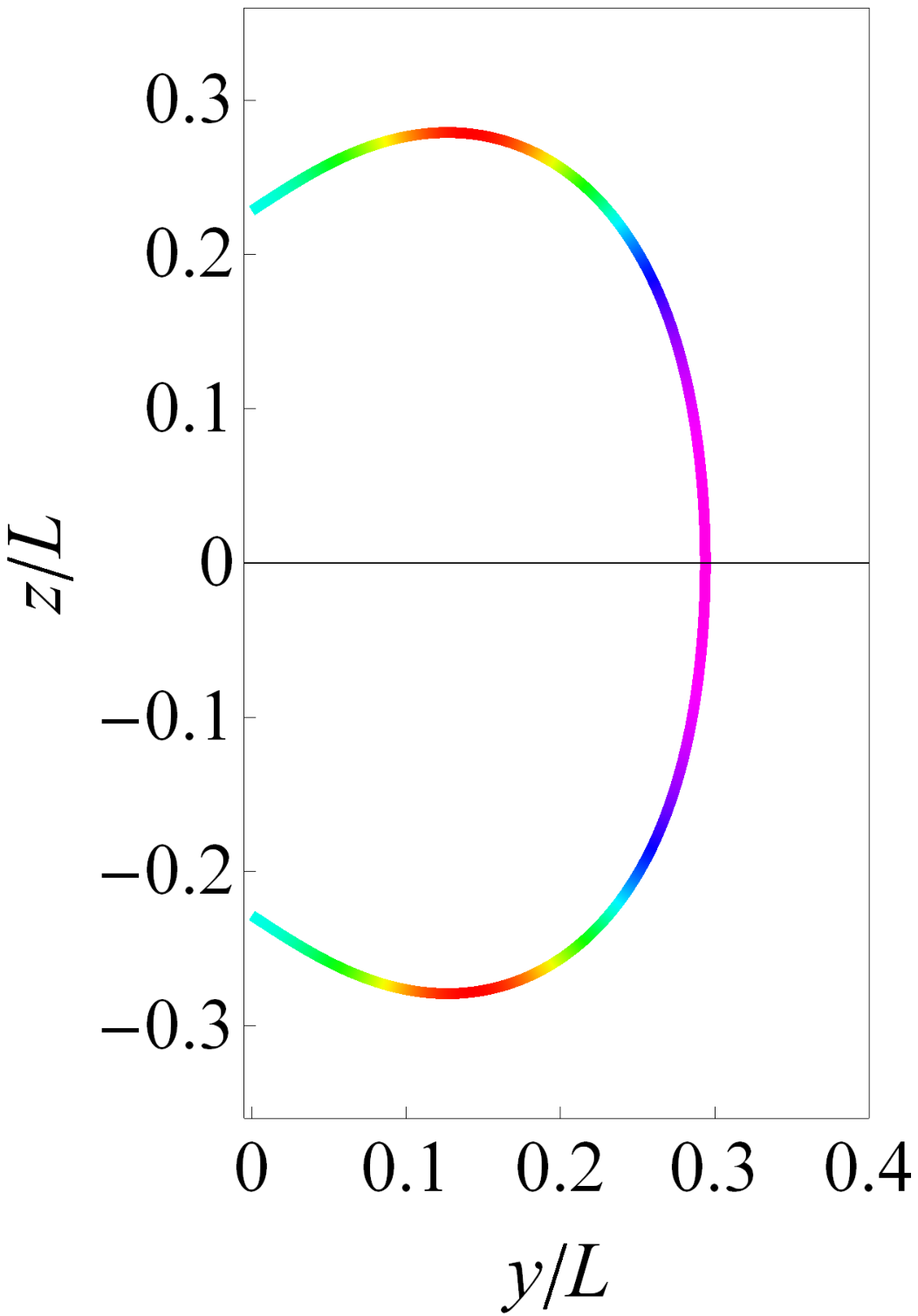}  &
\includegraphics[scale=0.35]{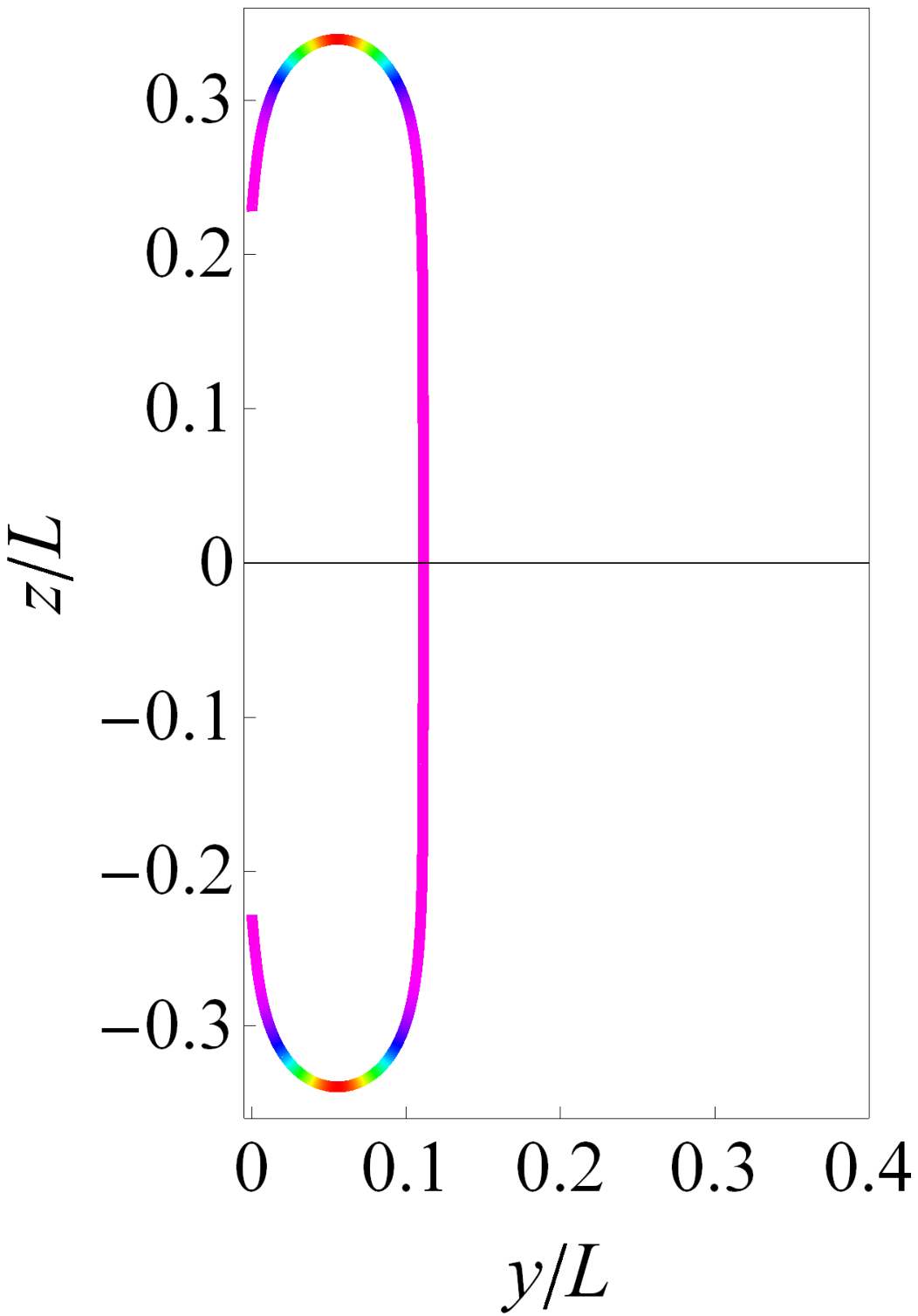}  \\
\includegraphics[scale=0.325]{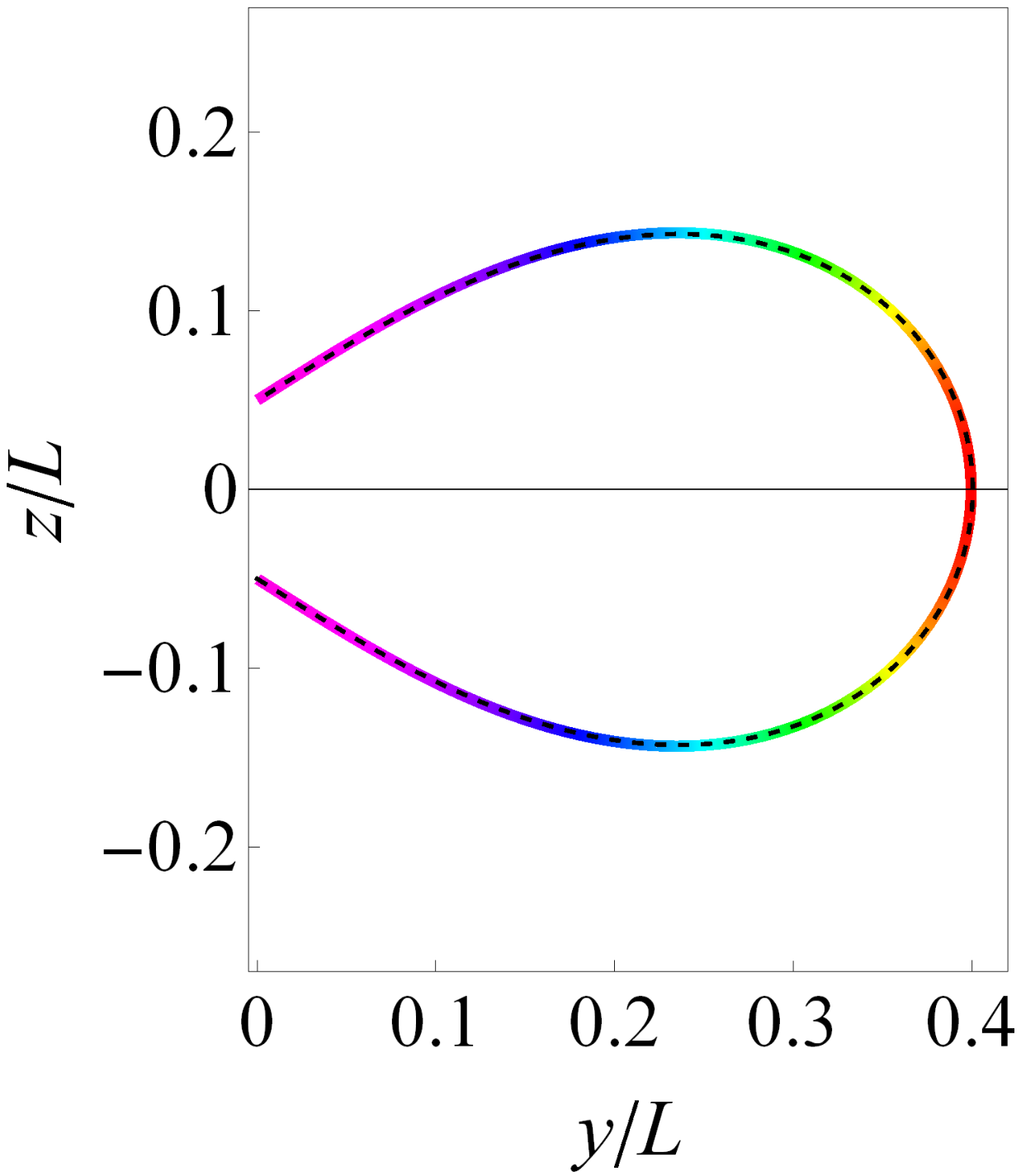} &
\includegraphics[scale=0.325]{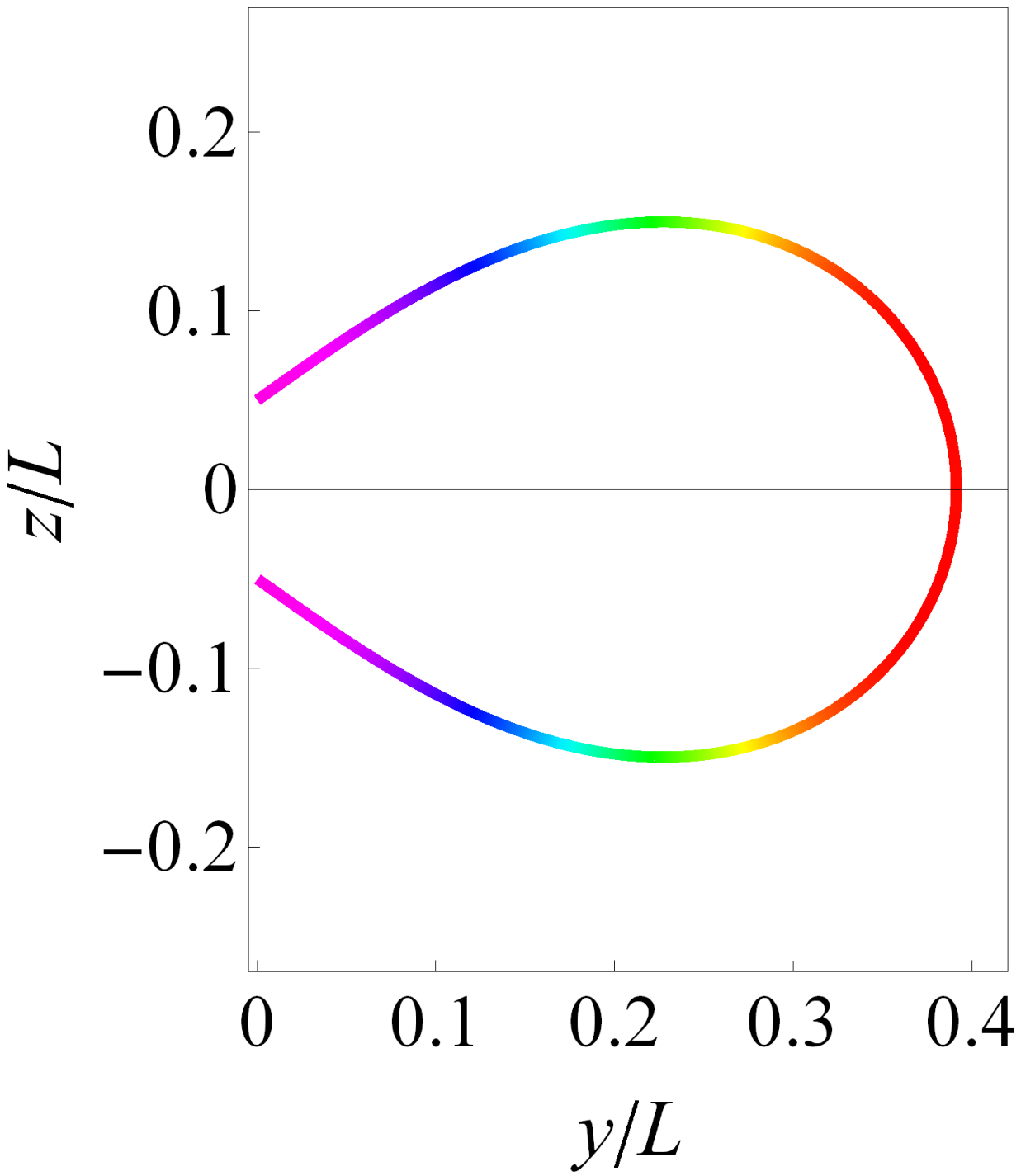} &
\includegraphics[scale=0.325]{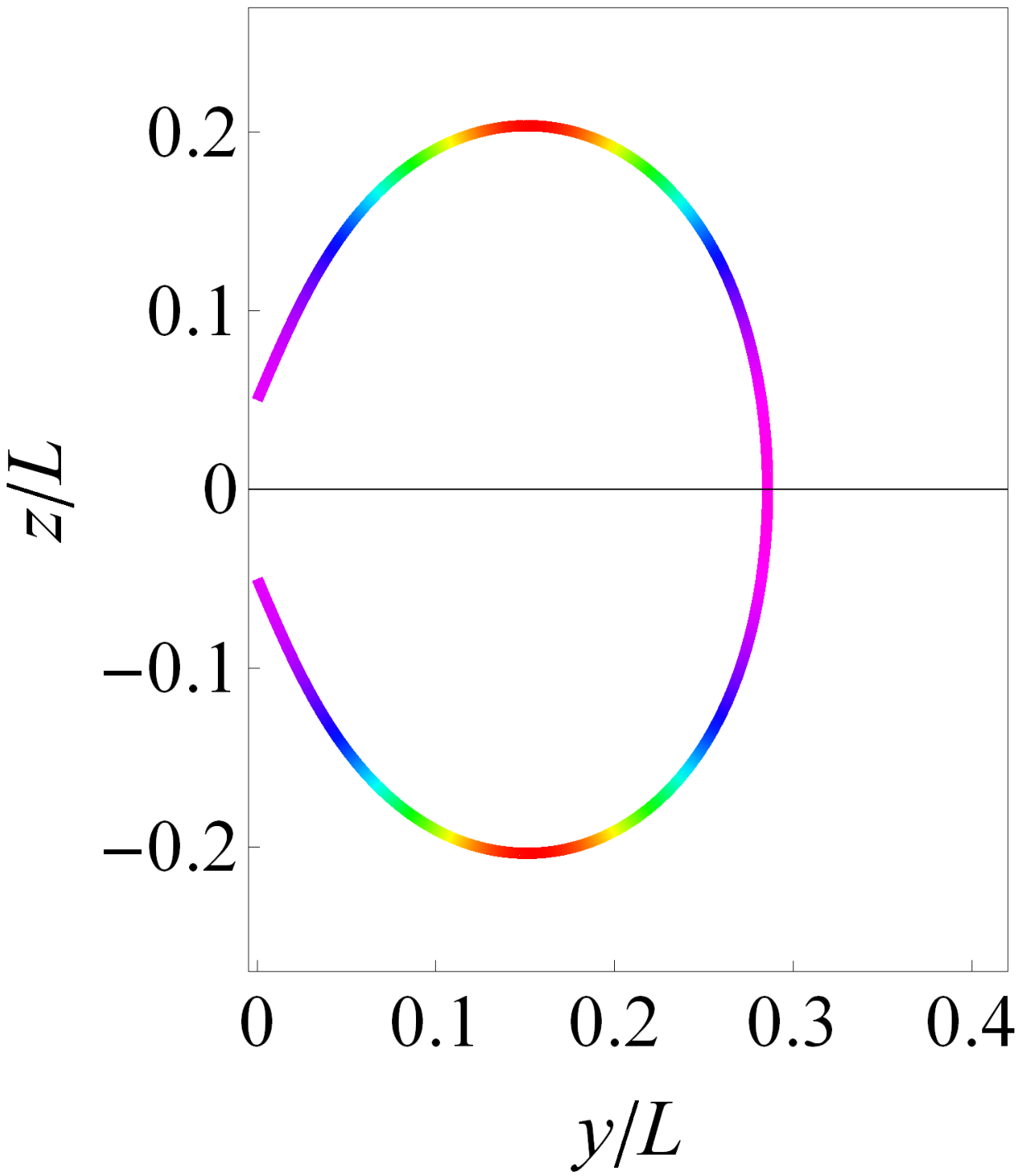} &
\includegraphics[scale=0.325]{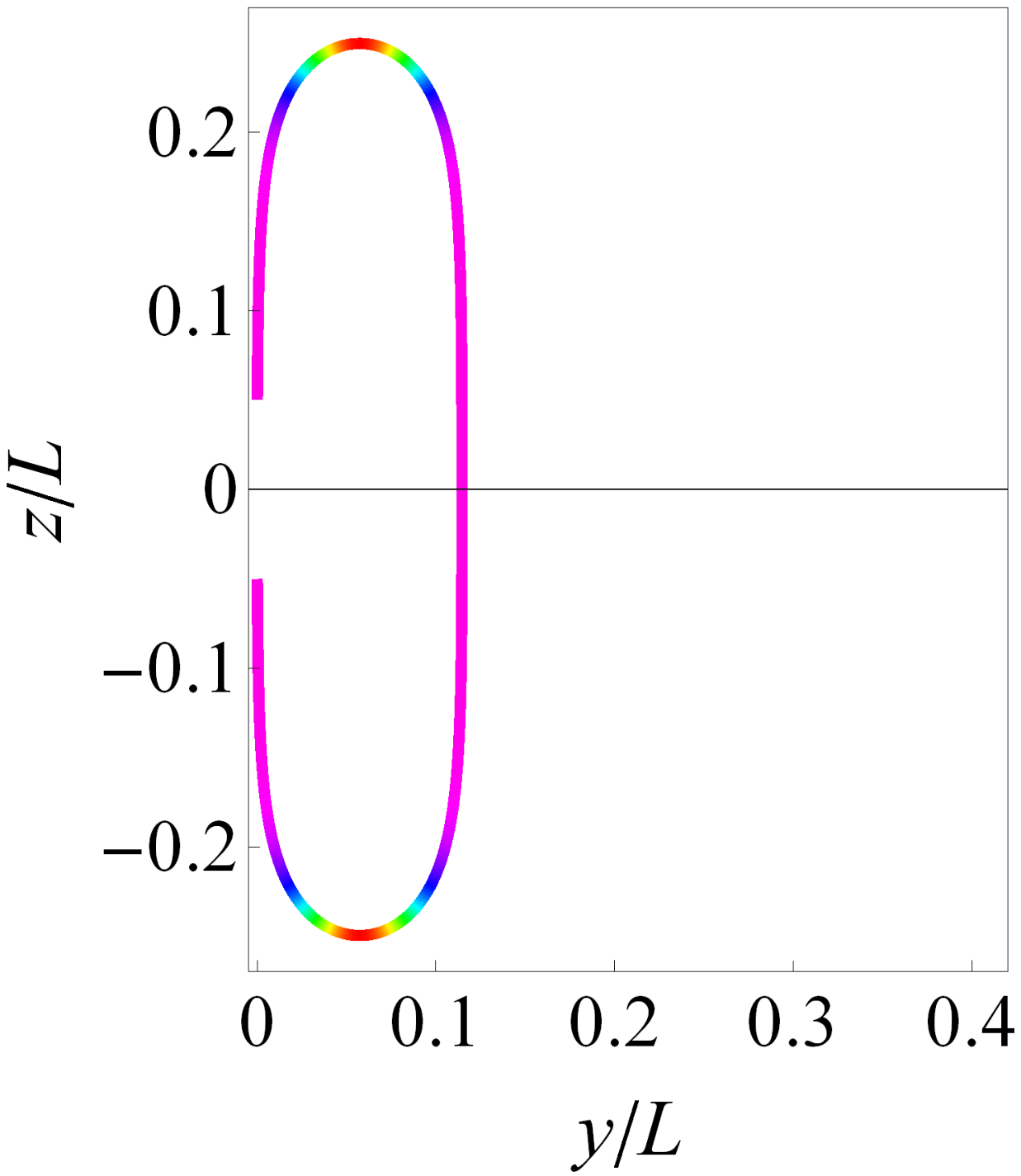} \\
{$\gamma = 1$} & {$\gamma = 10$} &  {$\gamma = 100$} & {$\gamma=1000$}
\end{tabular}
\includegraphics[scale=0.15]{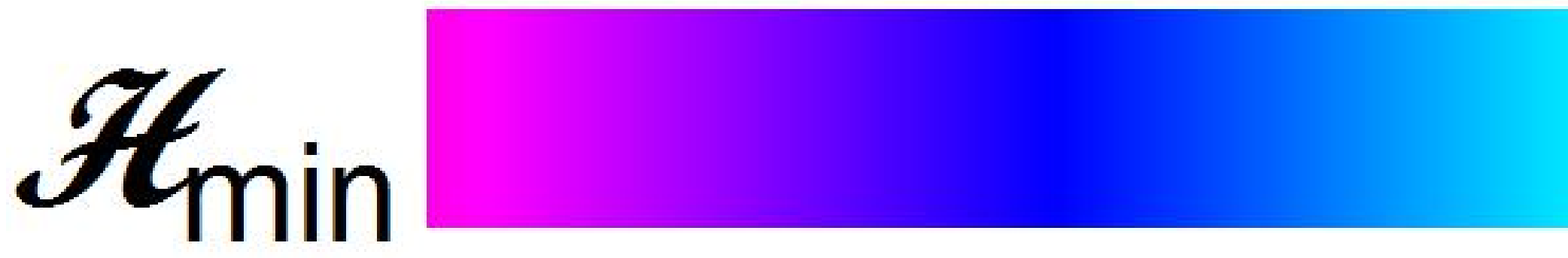}
 \caption{(Color online) Planar configurations with $n=1$ in regime $I$ ($\calM>0$) for different values of the separation between boundaries: $\xi = 0.8$ (top row), $\xi=0.457$ (middle row), and $\xi=0.1$ (bottom row). The Euler elastica for which $\gamma=0$ is shown with a dashed line in the fist column. The local energy density is color-coded in these figures.} \label{fig:VertFilsn1Mpos}
\end{figure}

\begin{figure}[htbp]
\centering
\begin{tabular}{cccc}
\includegraphics[scale=0.375]{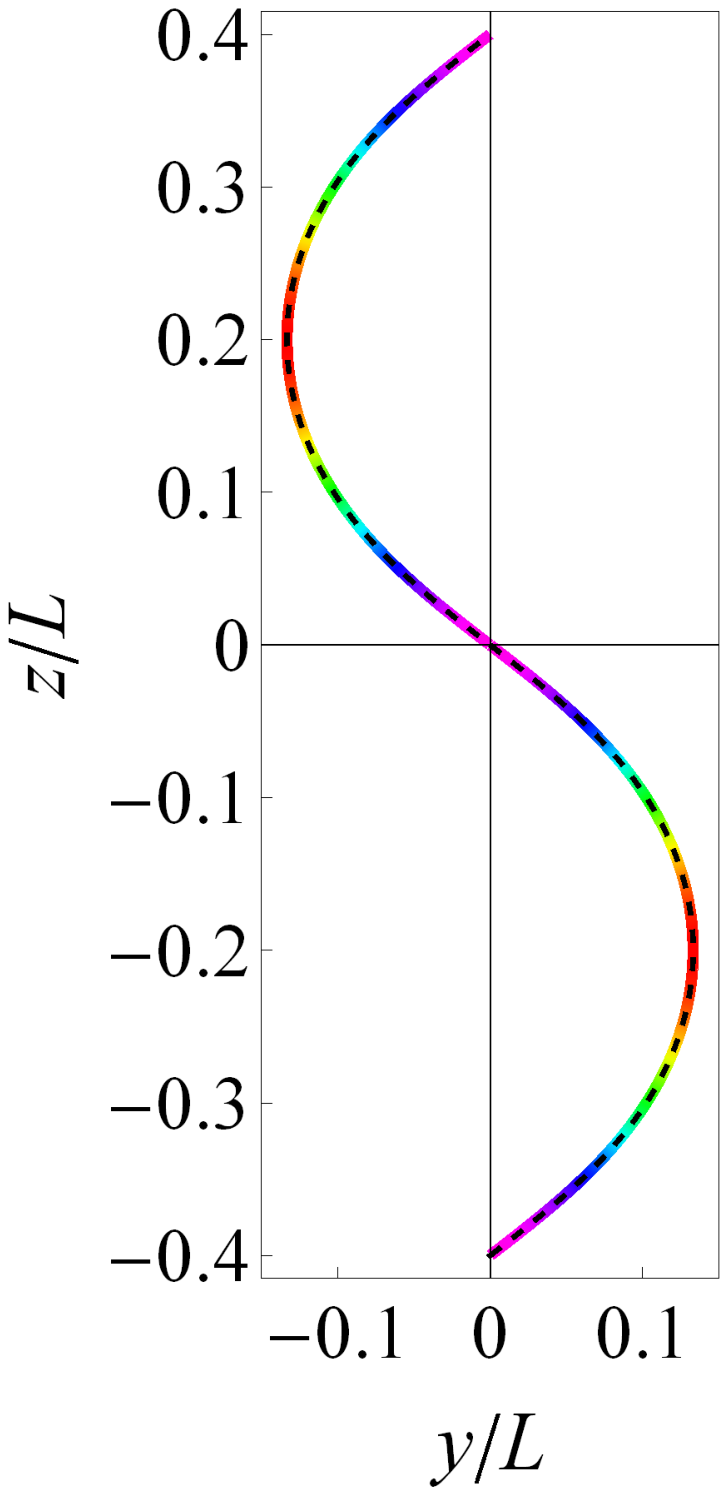} &
\includegraphics[scale=0.375]{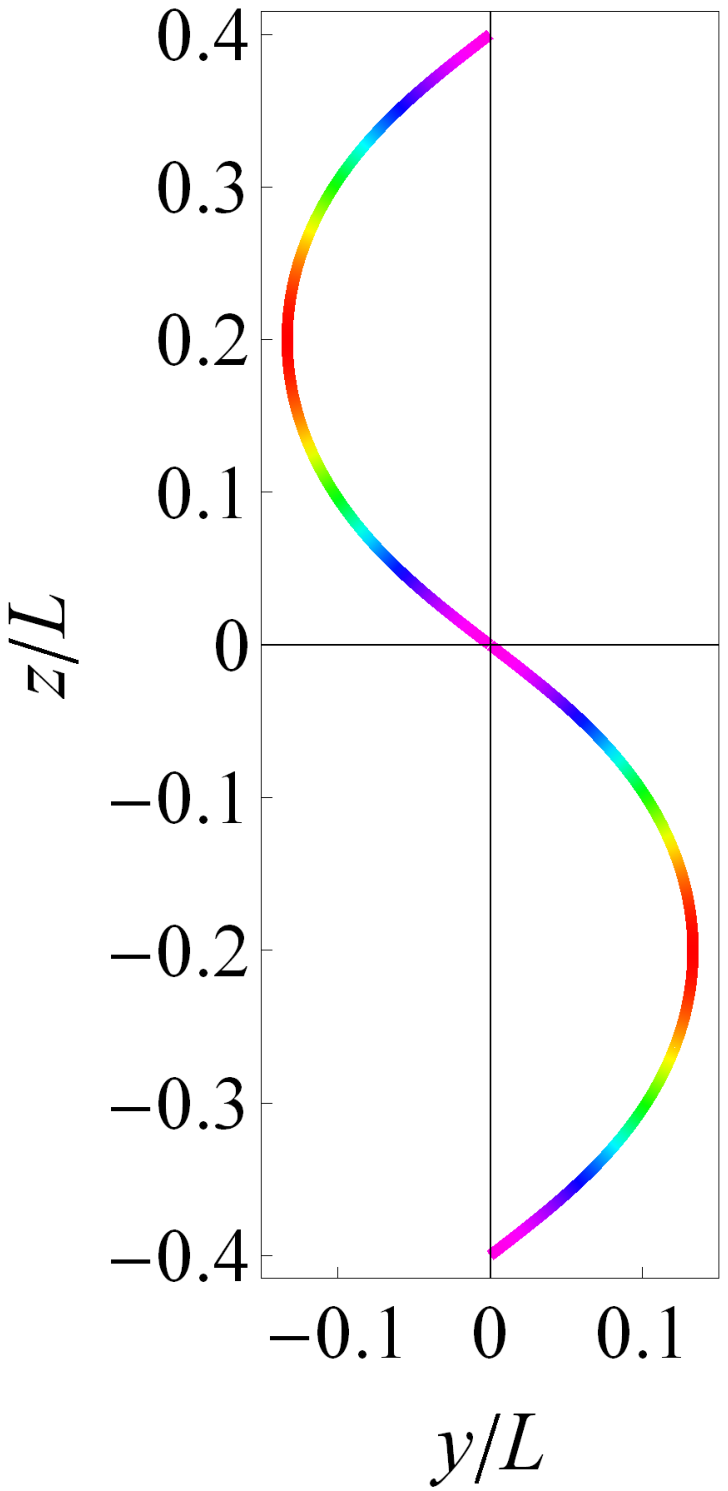} &
\includegraphics[scale=0.375]{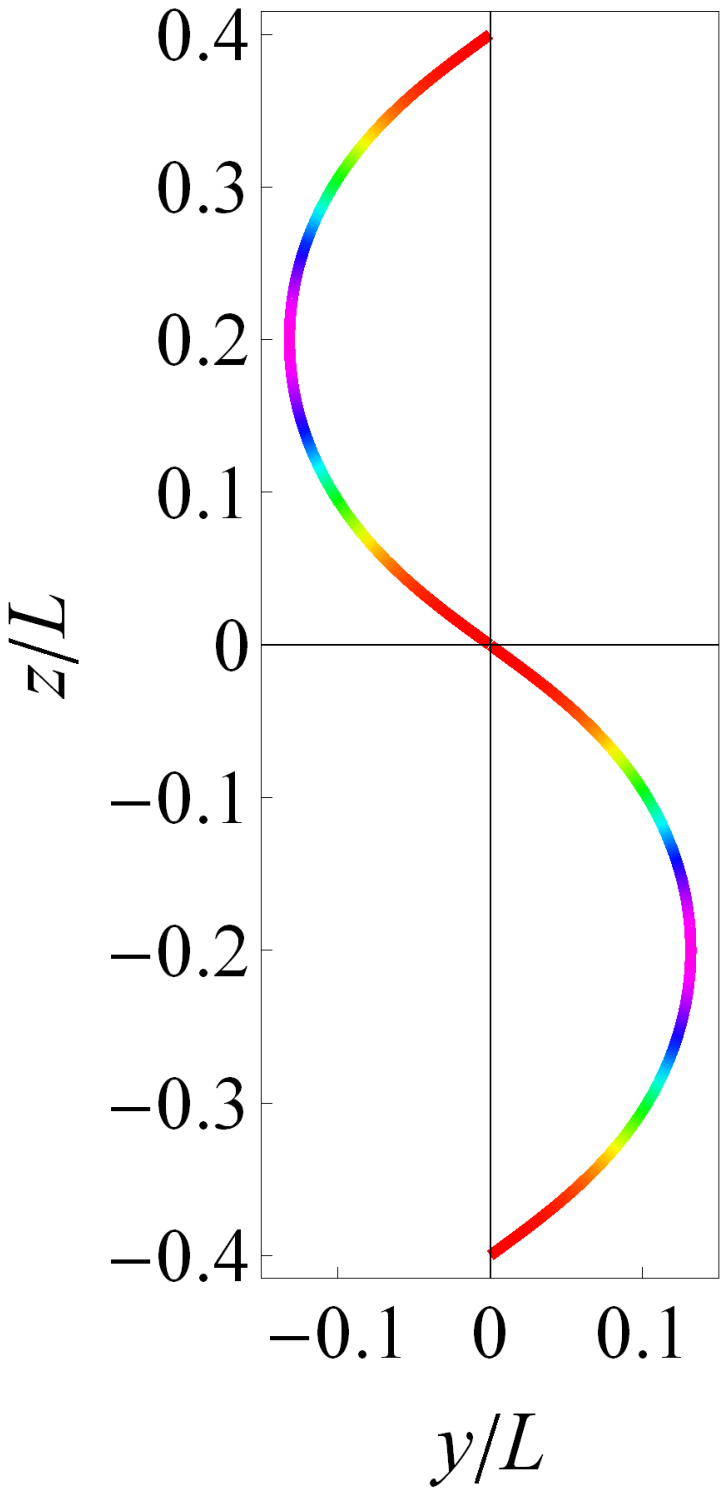} &
\includegraphics[scale=0.375]{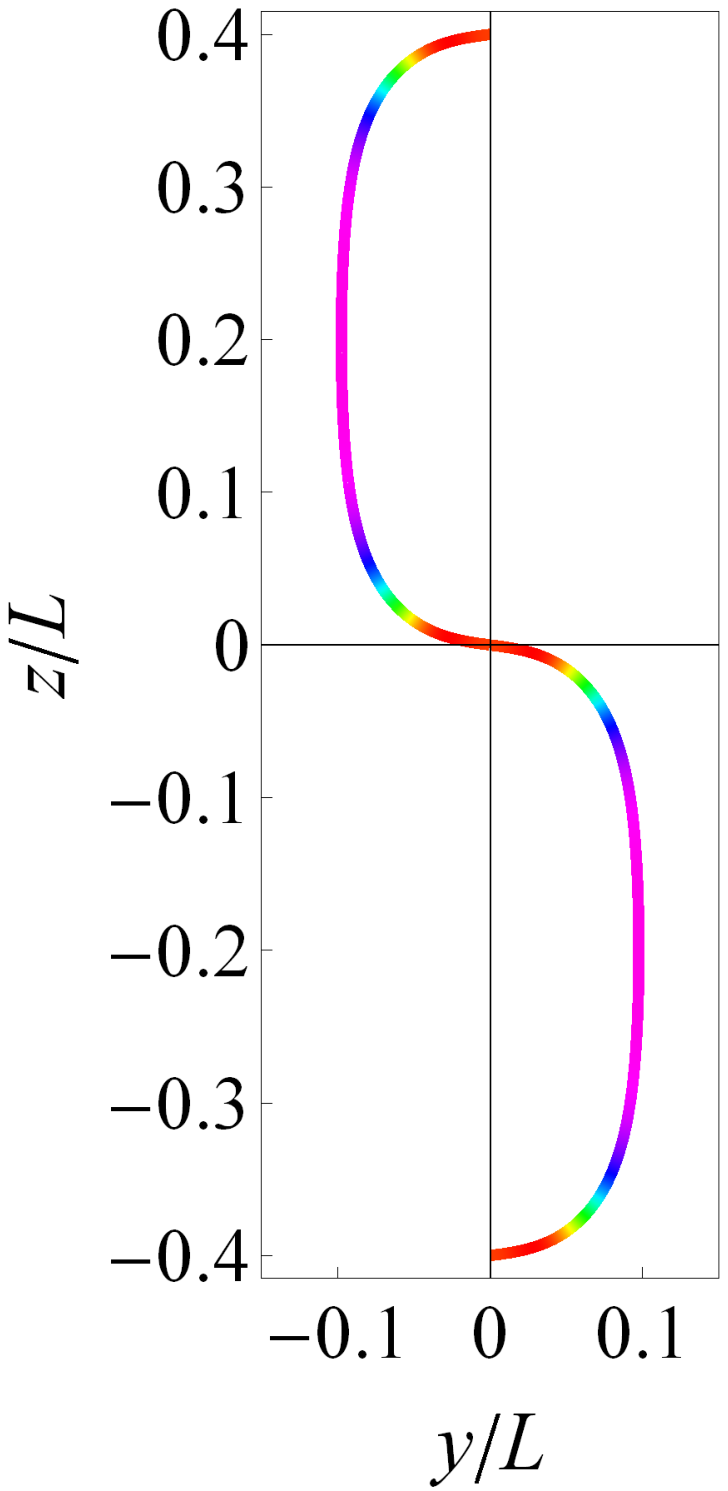} \\
\includegraphics[scale=0.35]{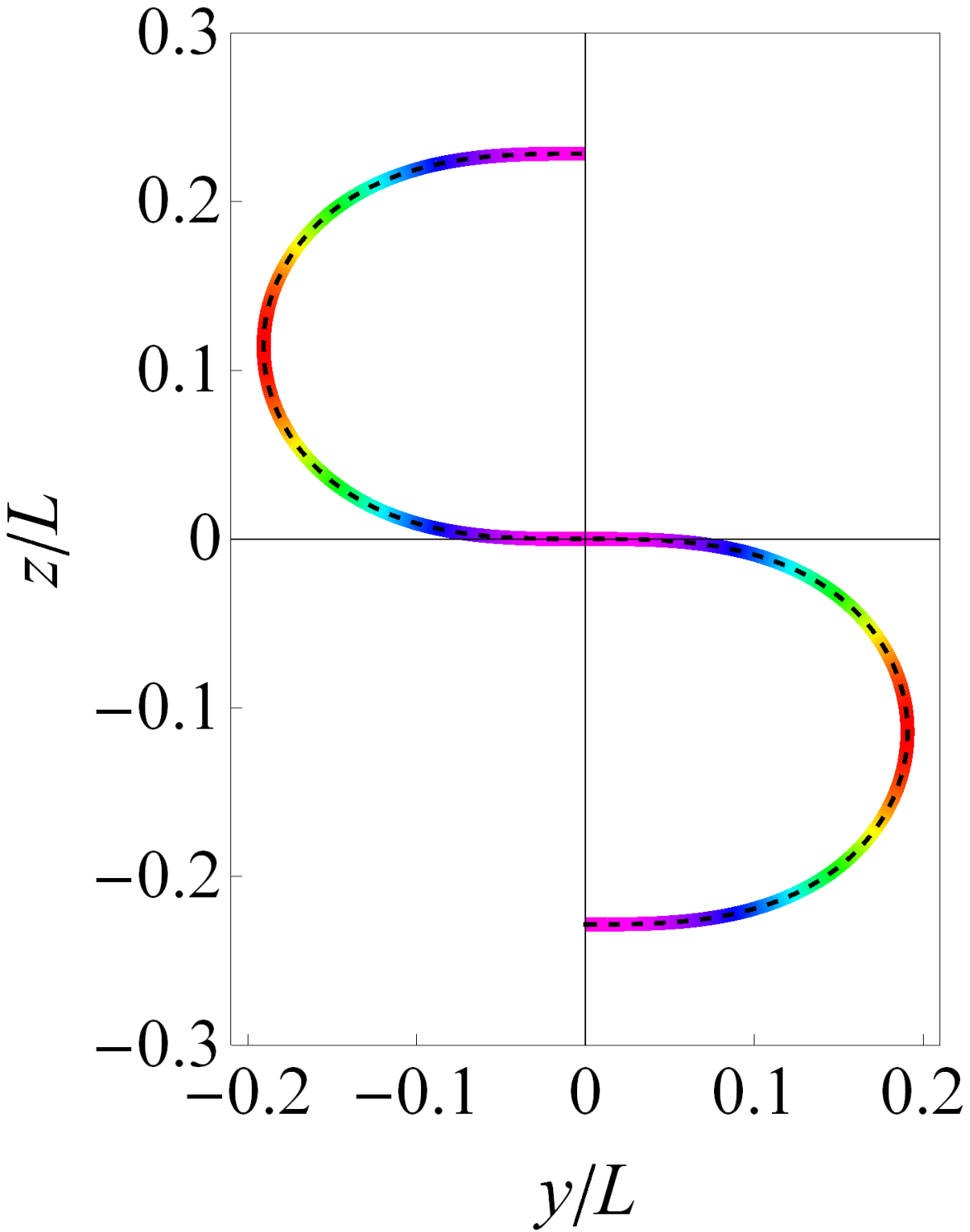} &
\includegraphics[scale=0.35]{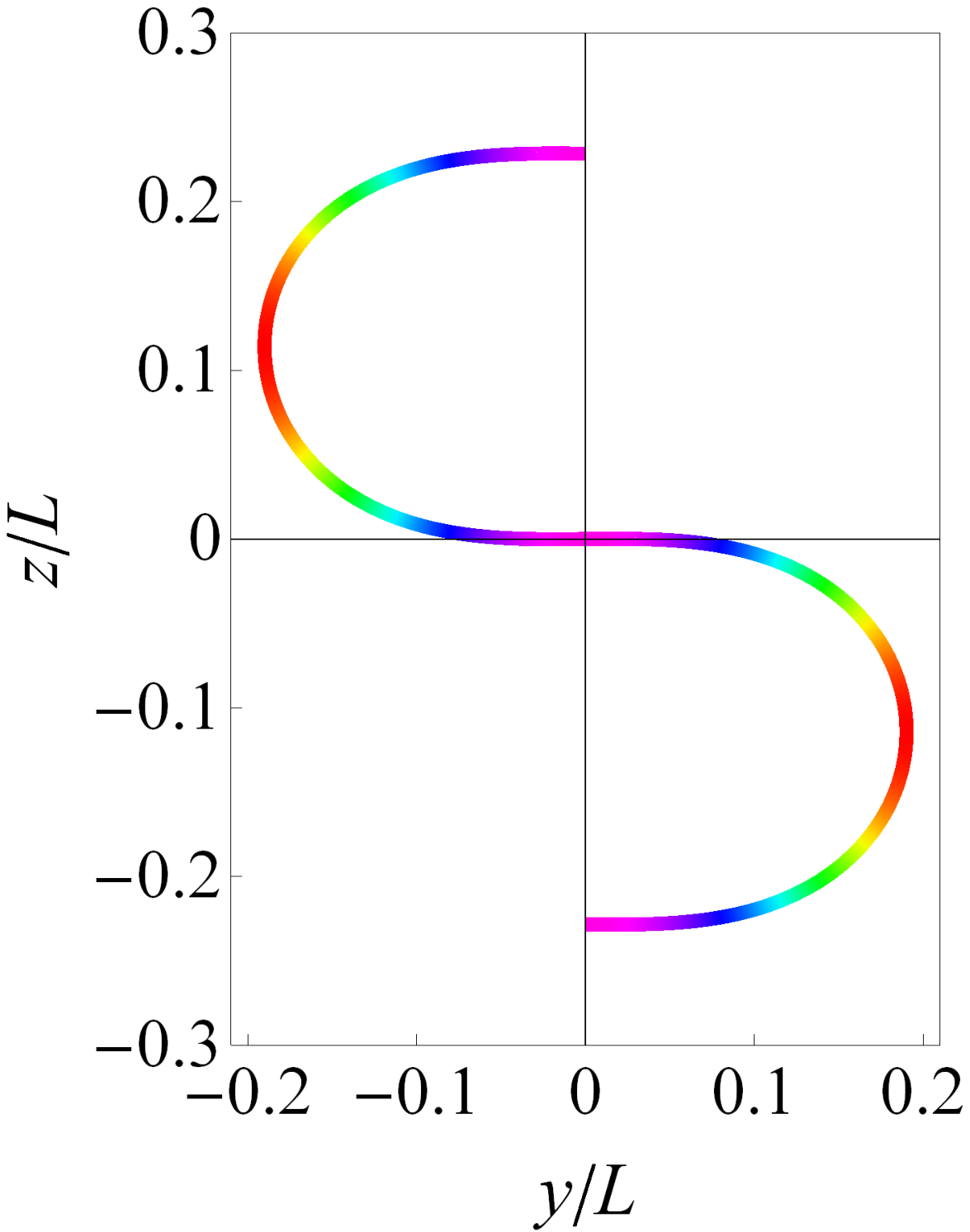} &
\includegraphics[scale=0.35]{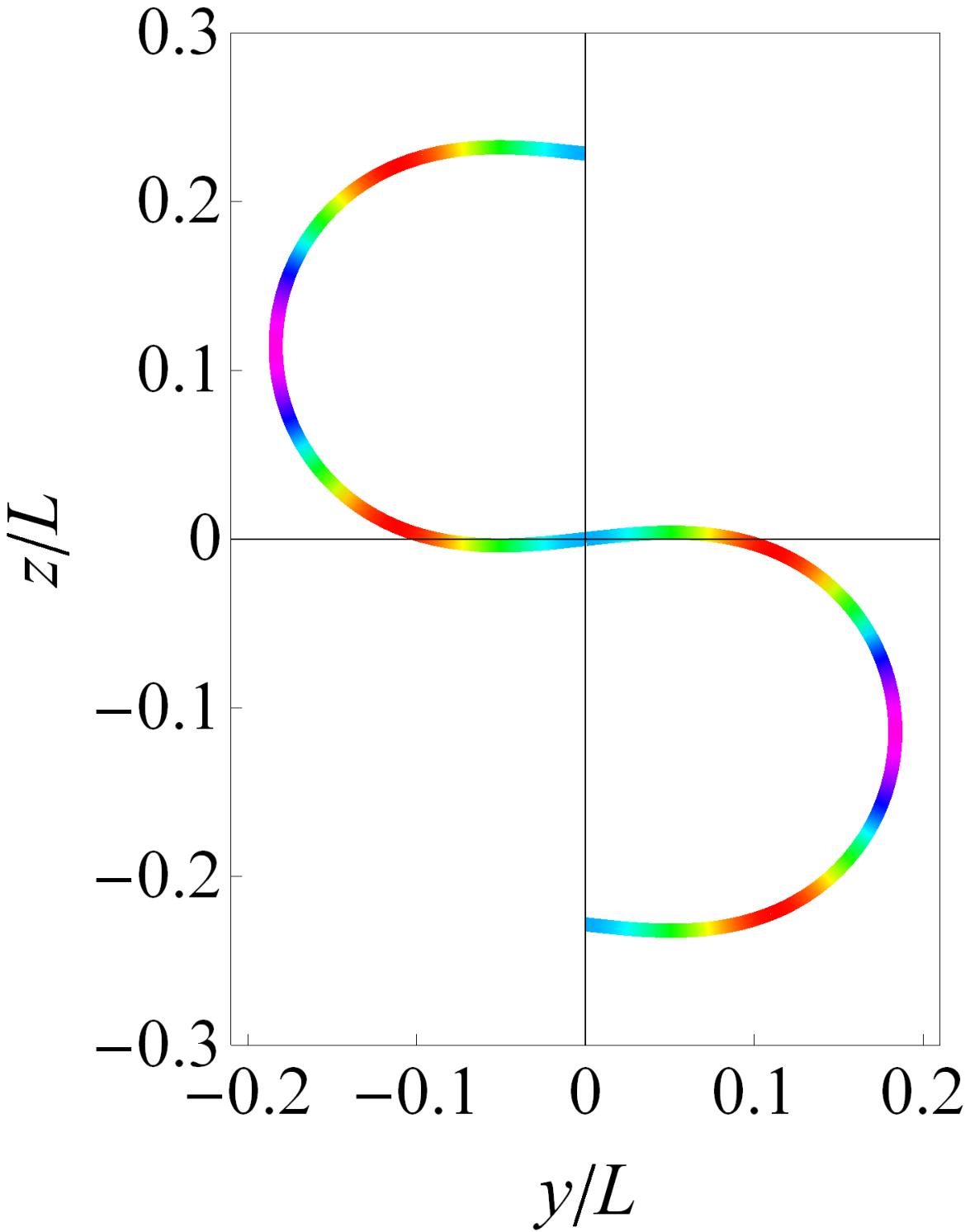} &
\includegraphics[scale=0.35]{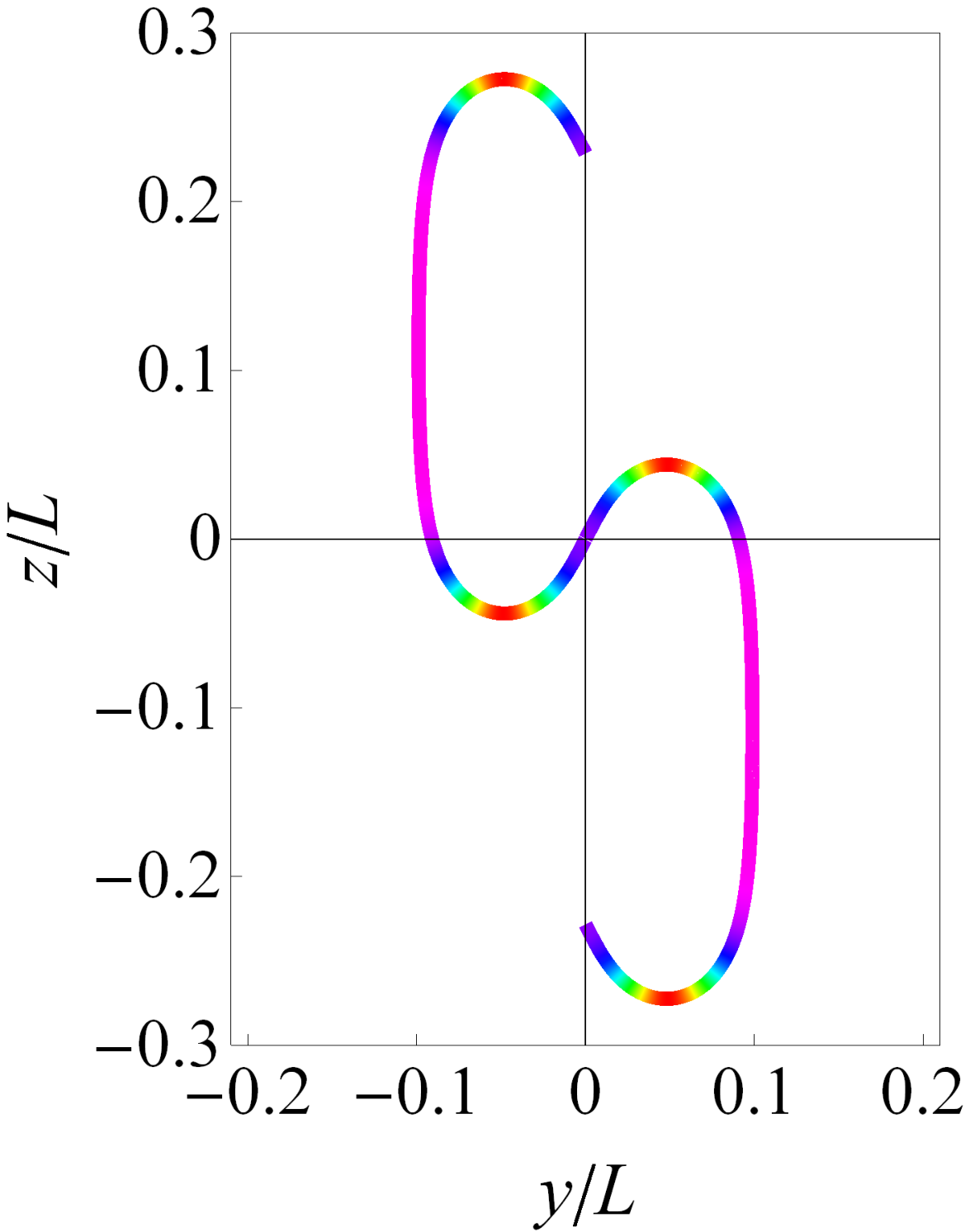} \\
\includegraphics[scale=0.325]{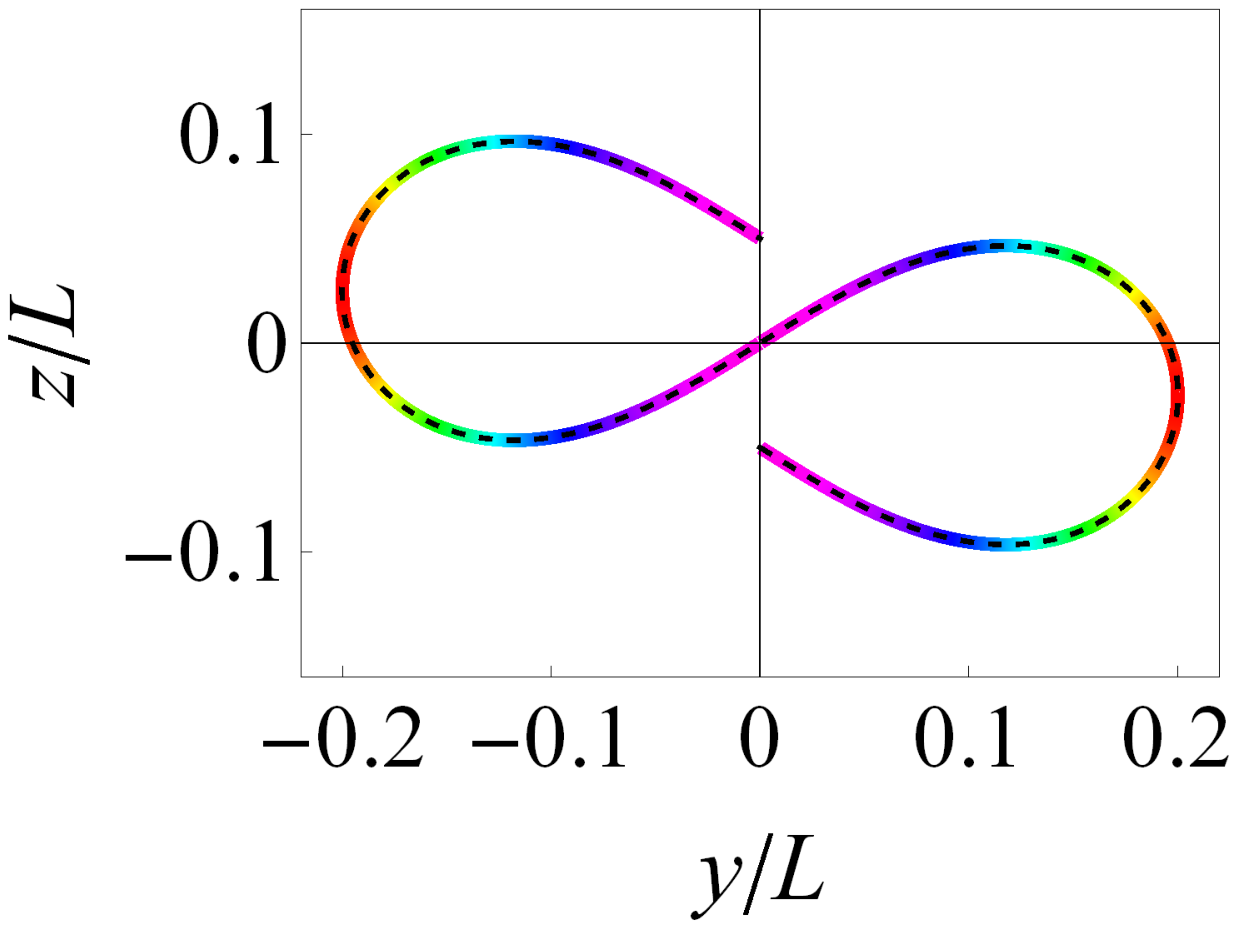} &
\includegraphics[scale=0.325]{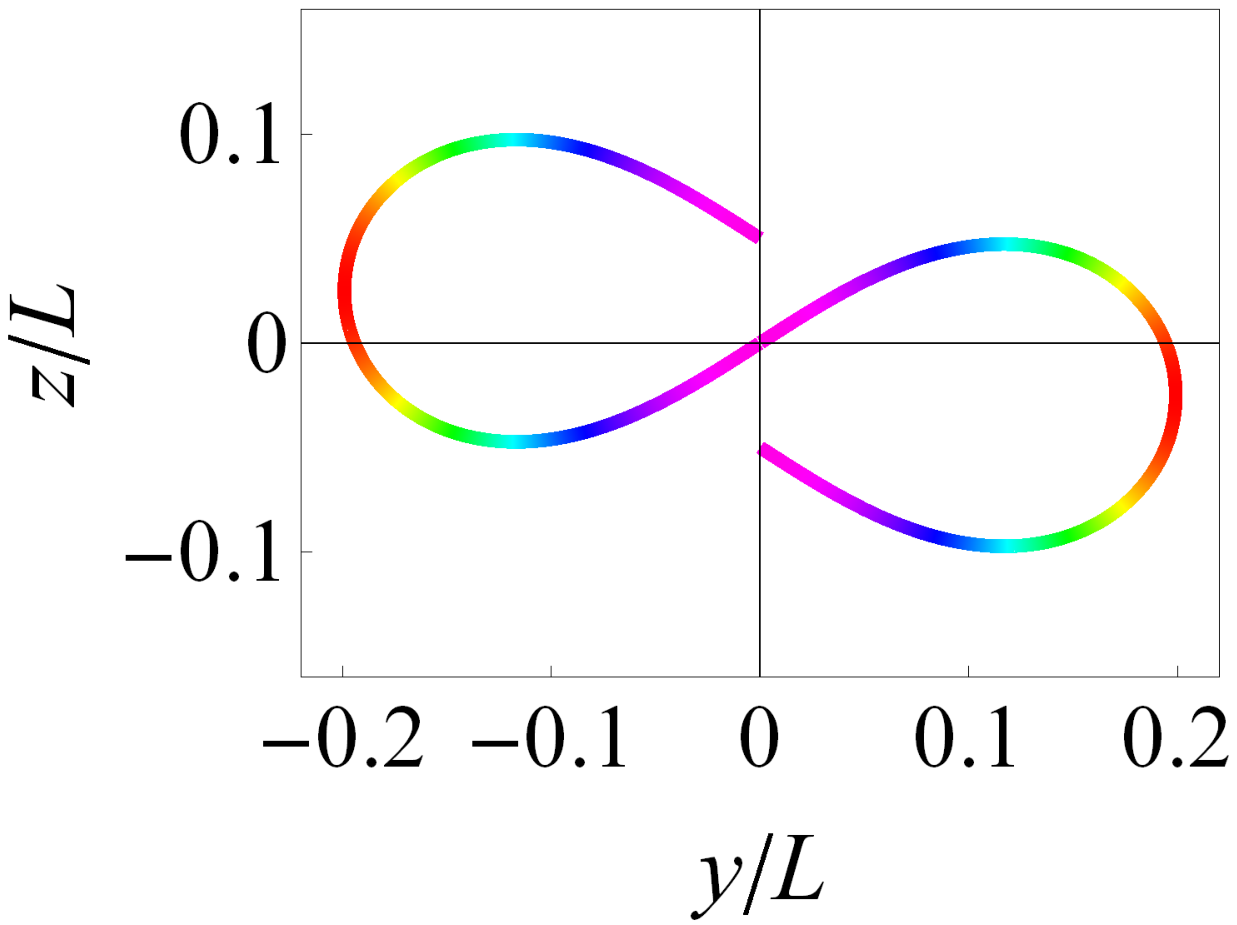} &
\includegraphics[scale=0.325]{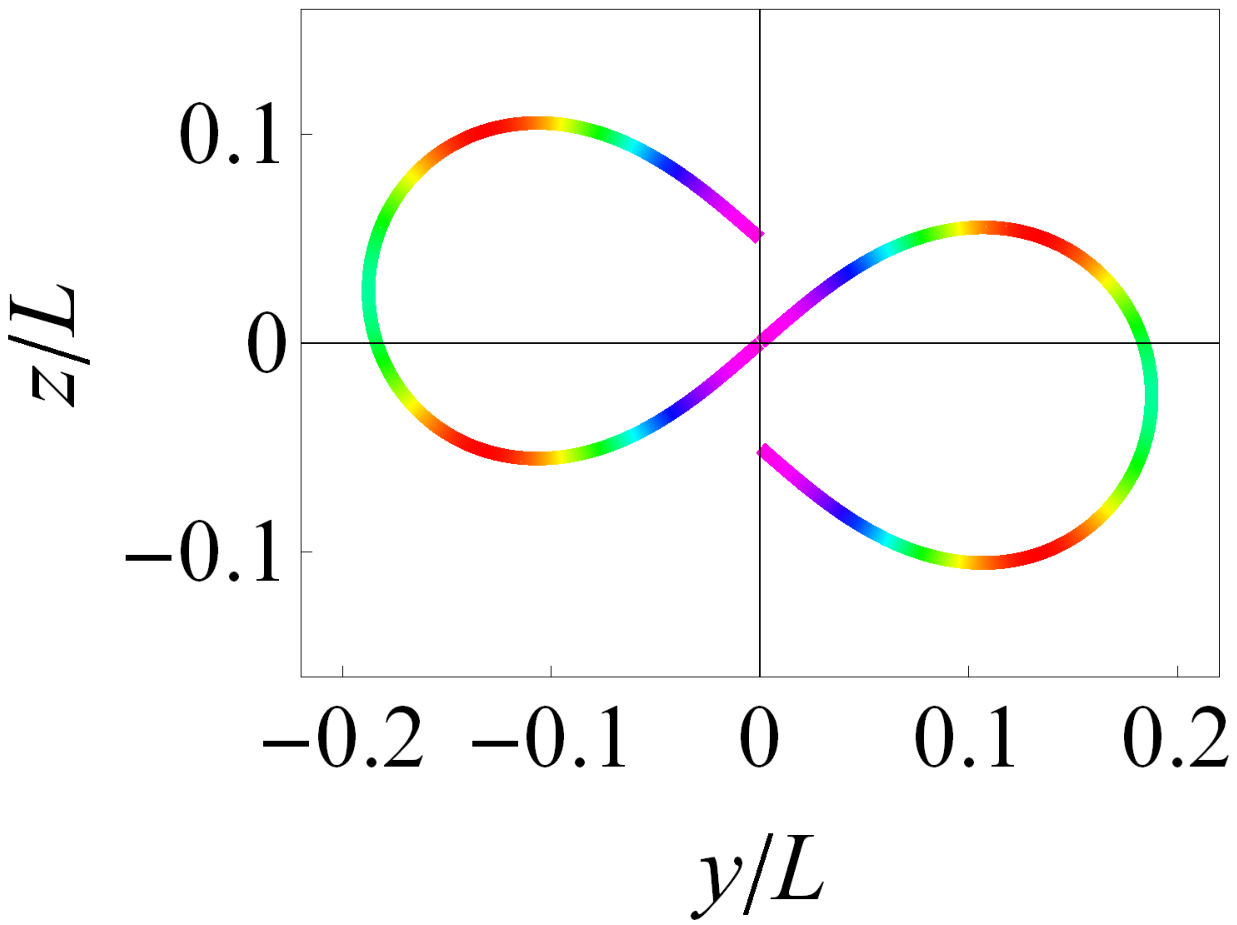} &
\includegraphics[scale=0.325]{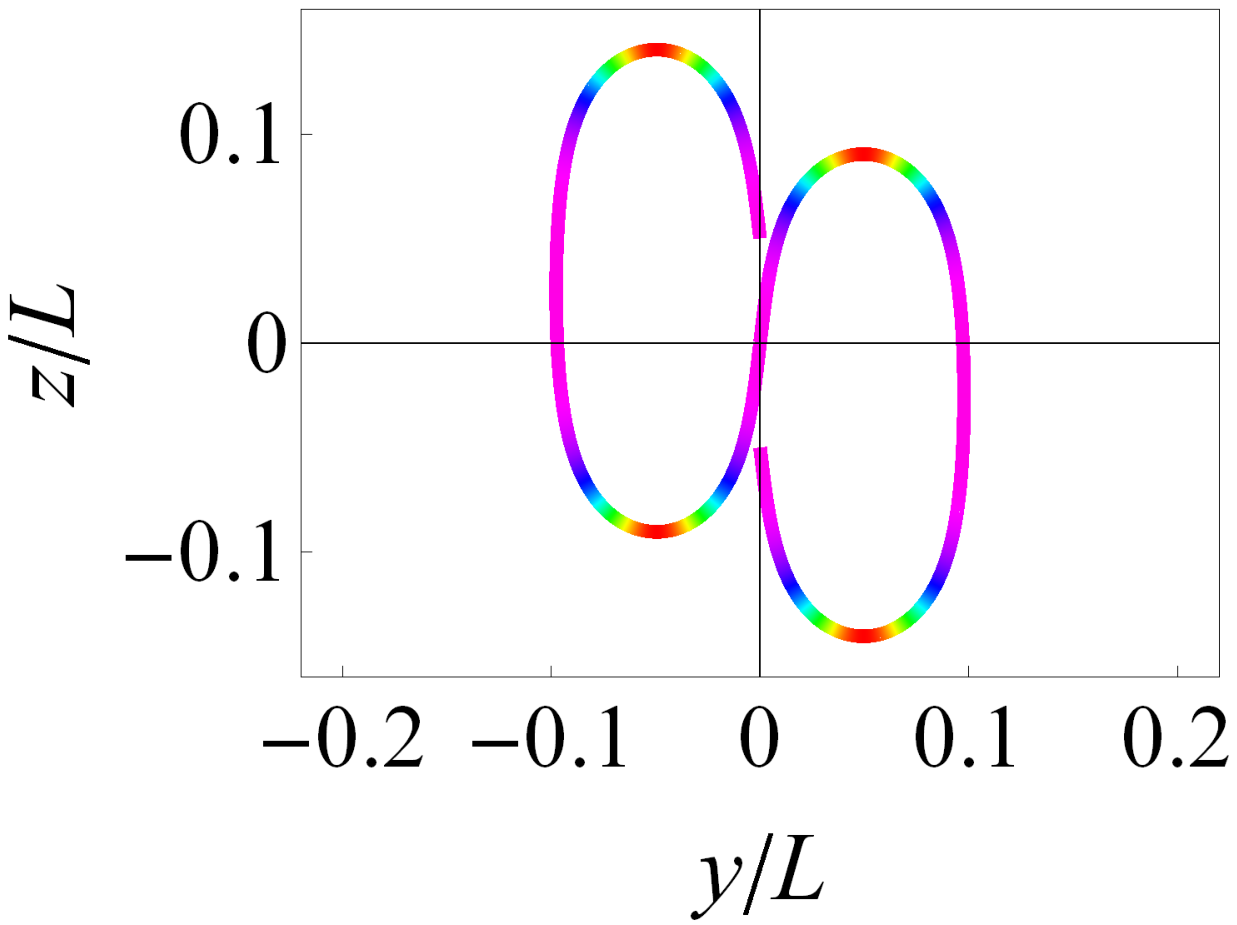} \\
{$\gamma = 1$} & {$\gamma = 10$} & {$\gamma = 100$} & {$\gamma=1000$}
\end{tabular}
\includegraphics[scale=0.15]{Fig5m}
 \caption{(Color online) Planar configurations with $n=2$ in regime $I$ ($\calM>0$) for different values of the separation between boundaries: $\xi = 0.8$ (top row), $\xi=0.457$ (middle row), and $\xi=0.1$ (bottom row). The local energy density is color-coded in these figures.} \label{fig:VertFilsn2Mpos}
\end{figure}

\begin{figure}[htbp]
\centering
\begin{tabular}{cccc}
\includegraphics[scale=0.375]{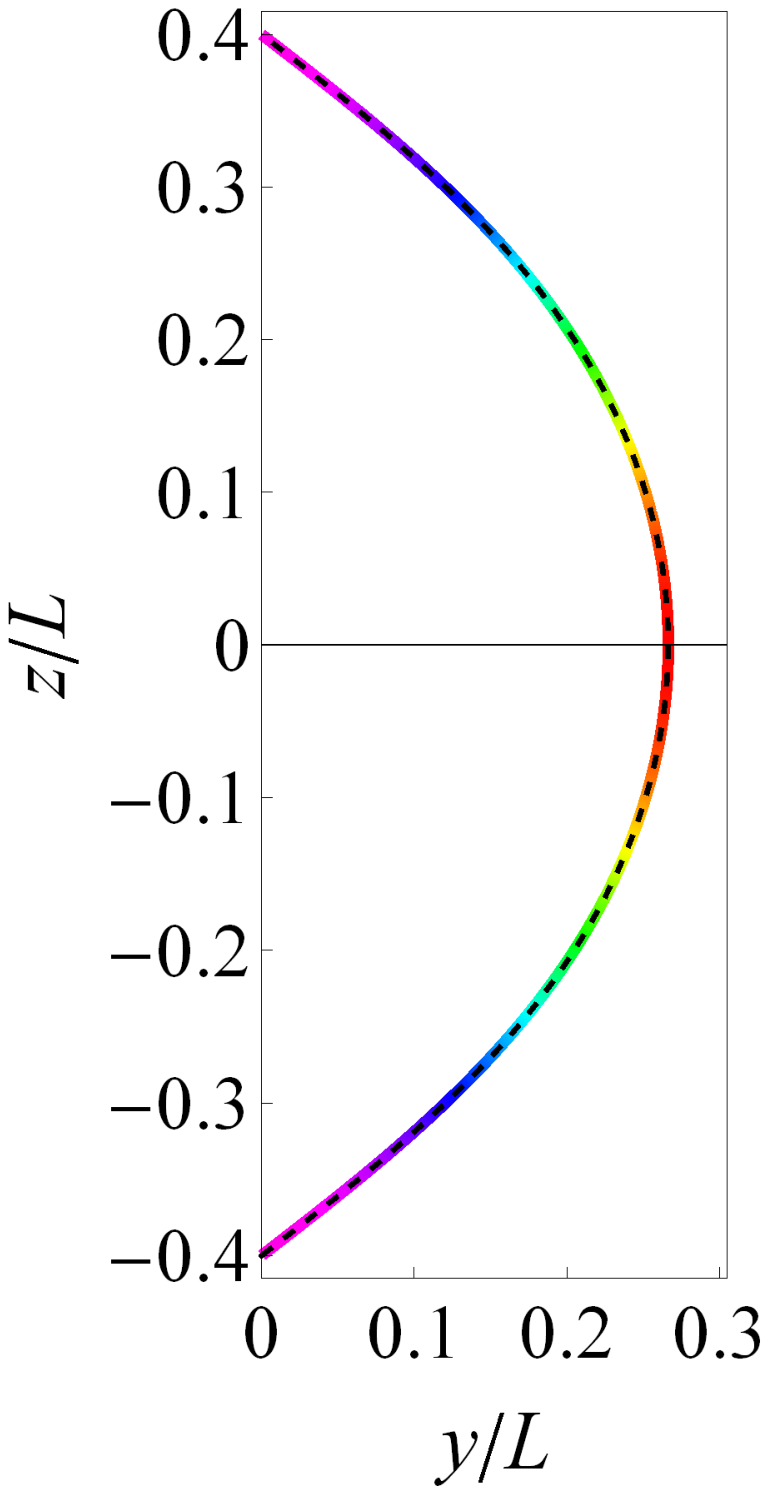} &
\includegraphics[scale=0.375]{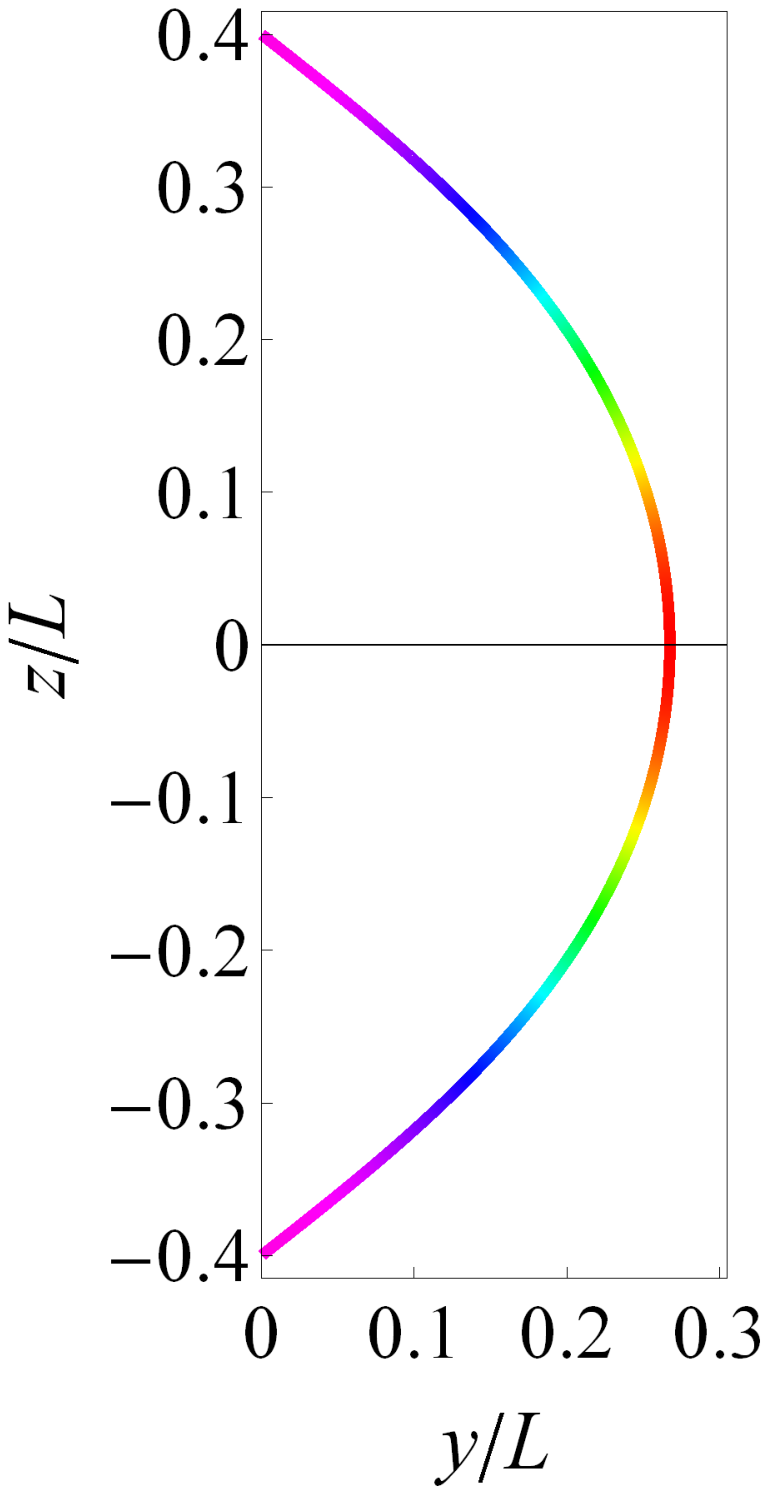} &
\includegraphics[scale=0.375]{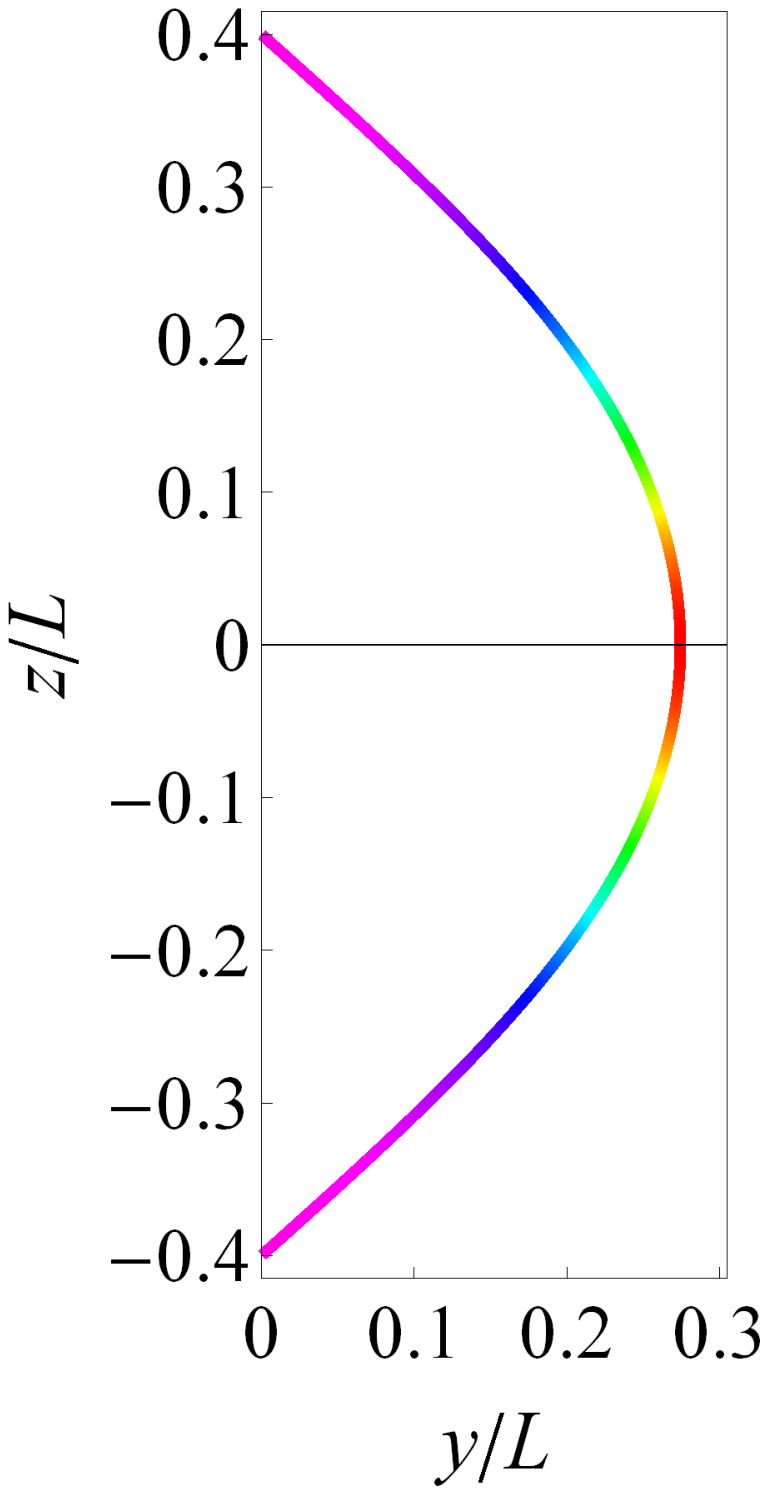} &
\includegraphics[scale=0.375]{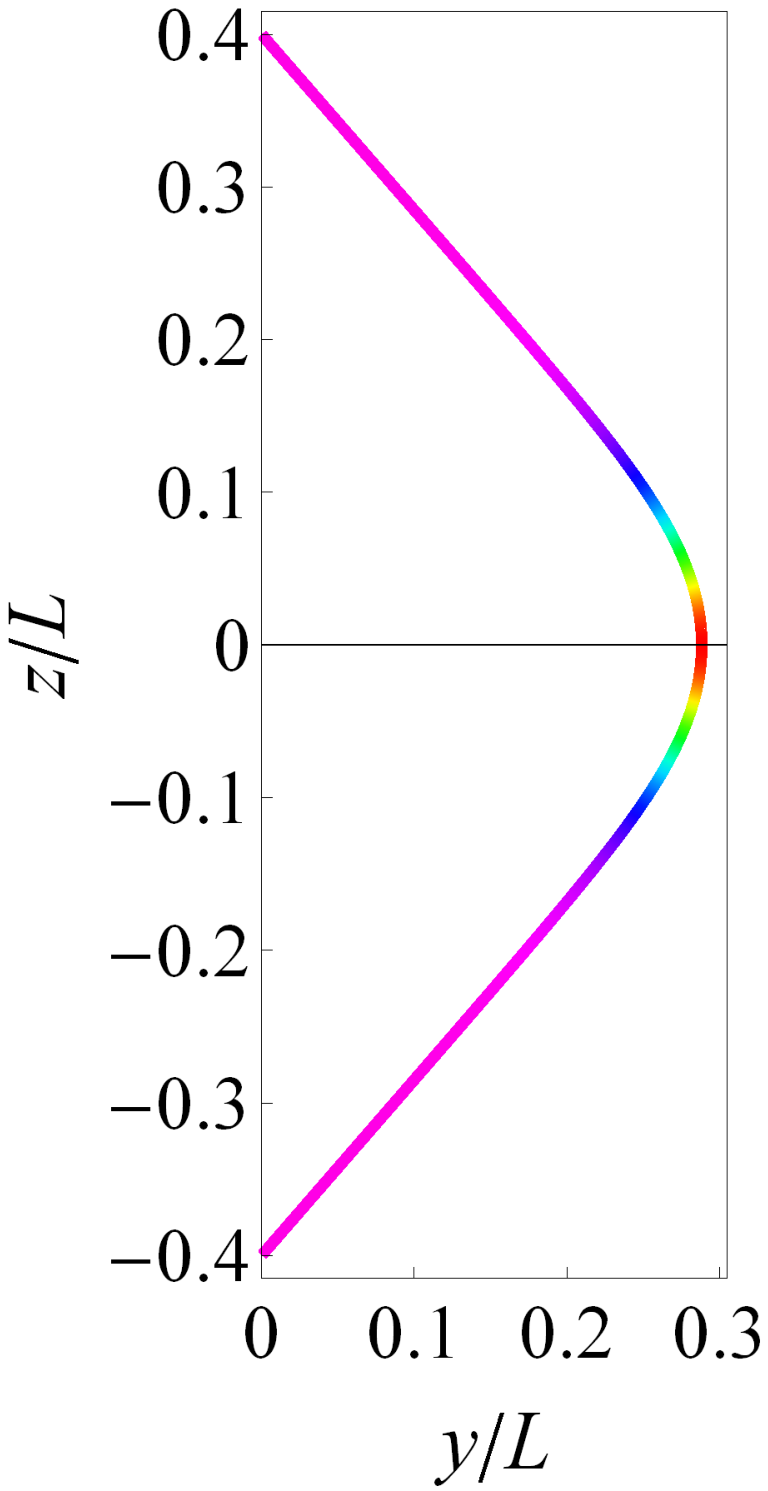} \\
\includegraphics[scale=0.325]{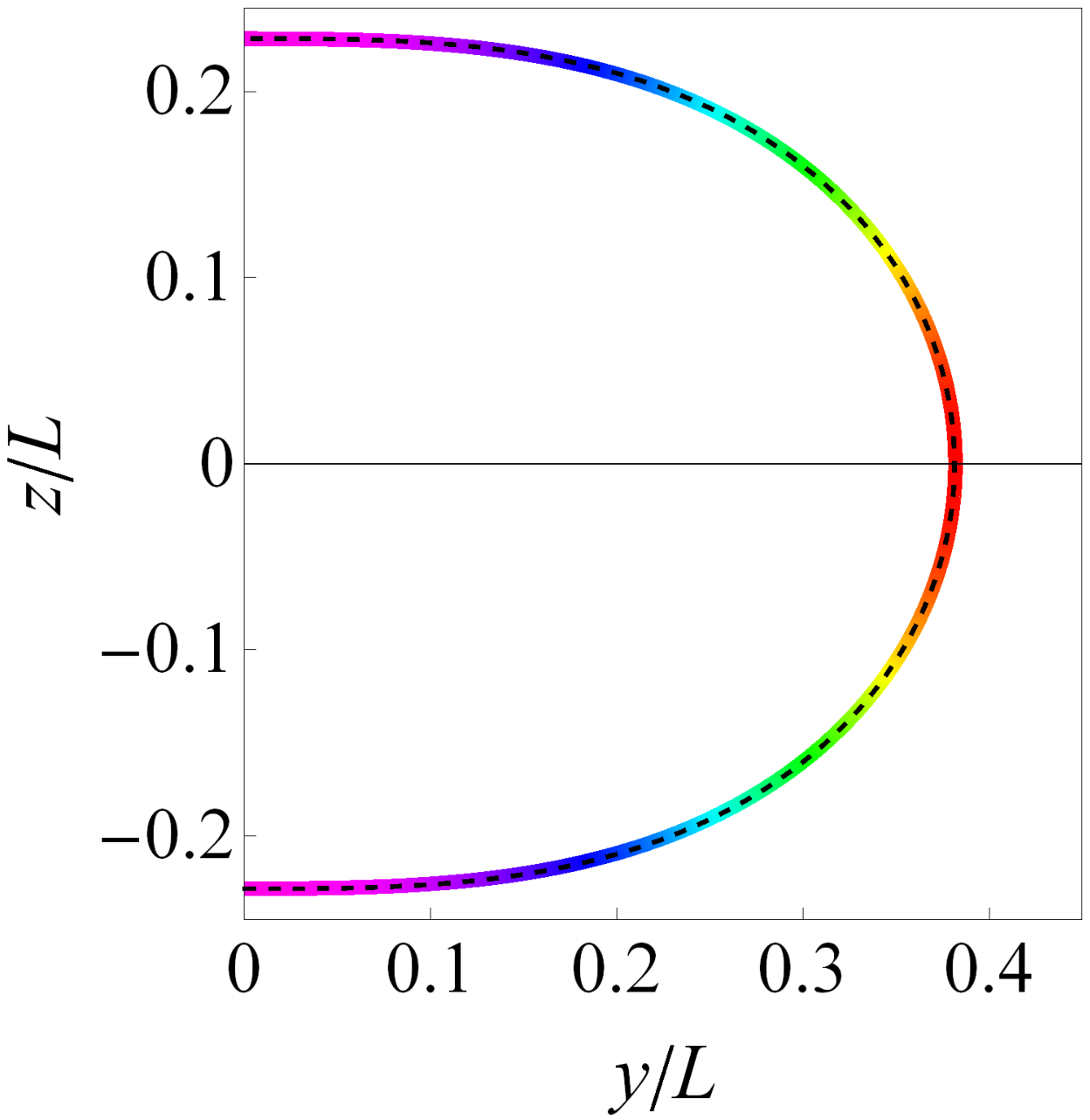} &
\includegraphics[scale=0.325]{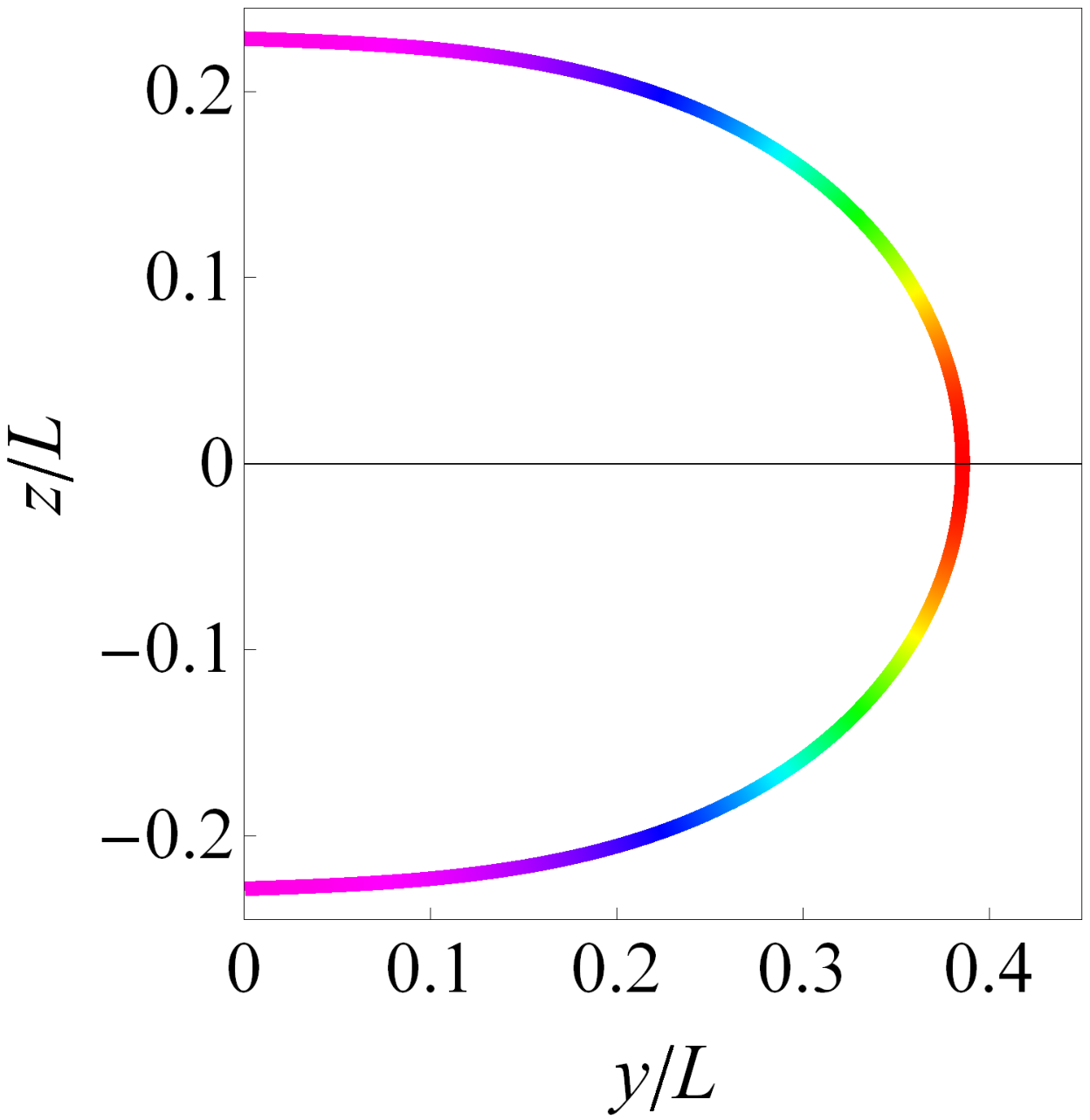} &
\includegraphics[scale=0.325]{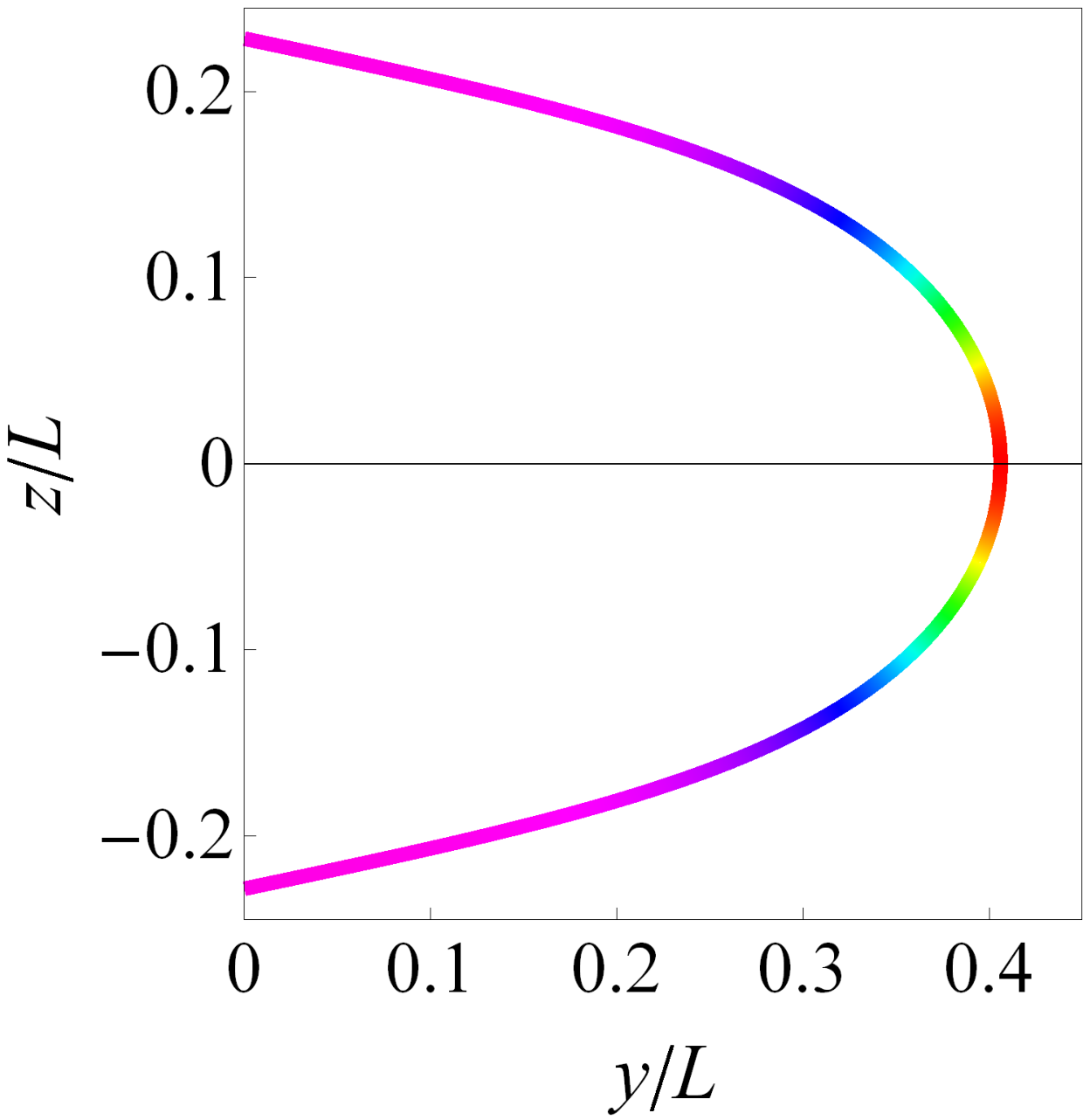} &
\includegraphics[scale=0.325]{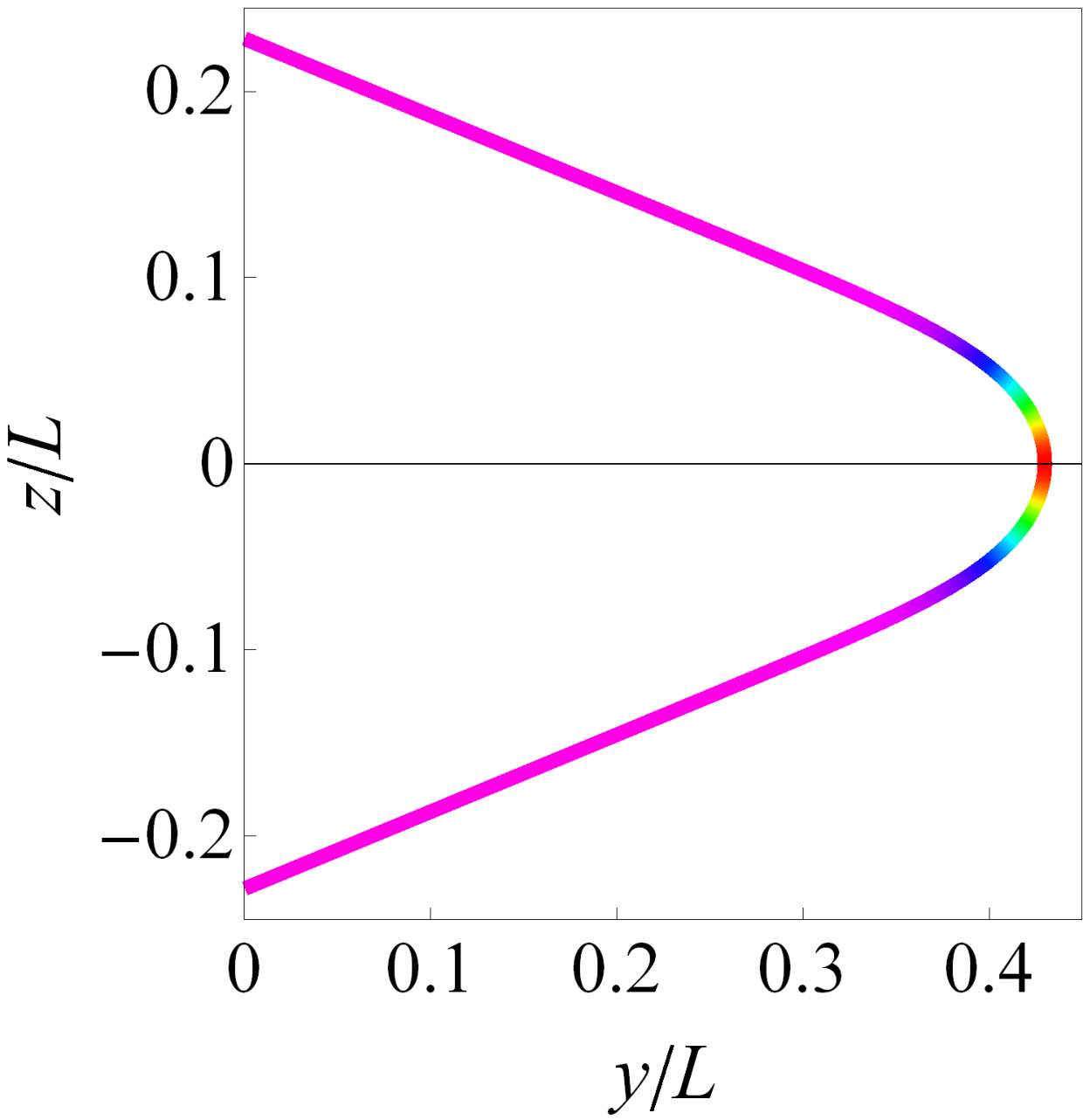} \\
\includegraphics[scale=0.33]{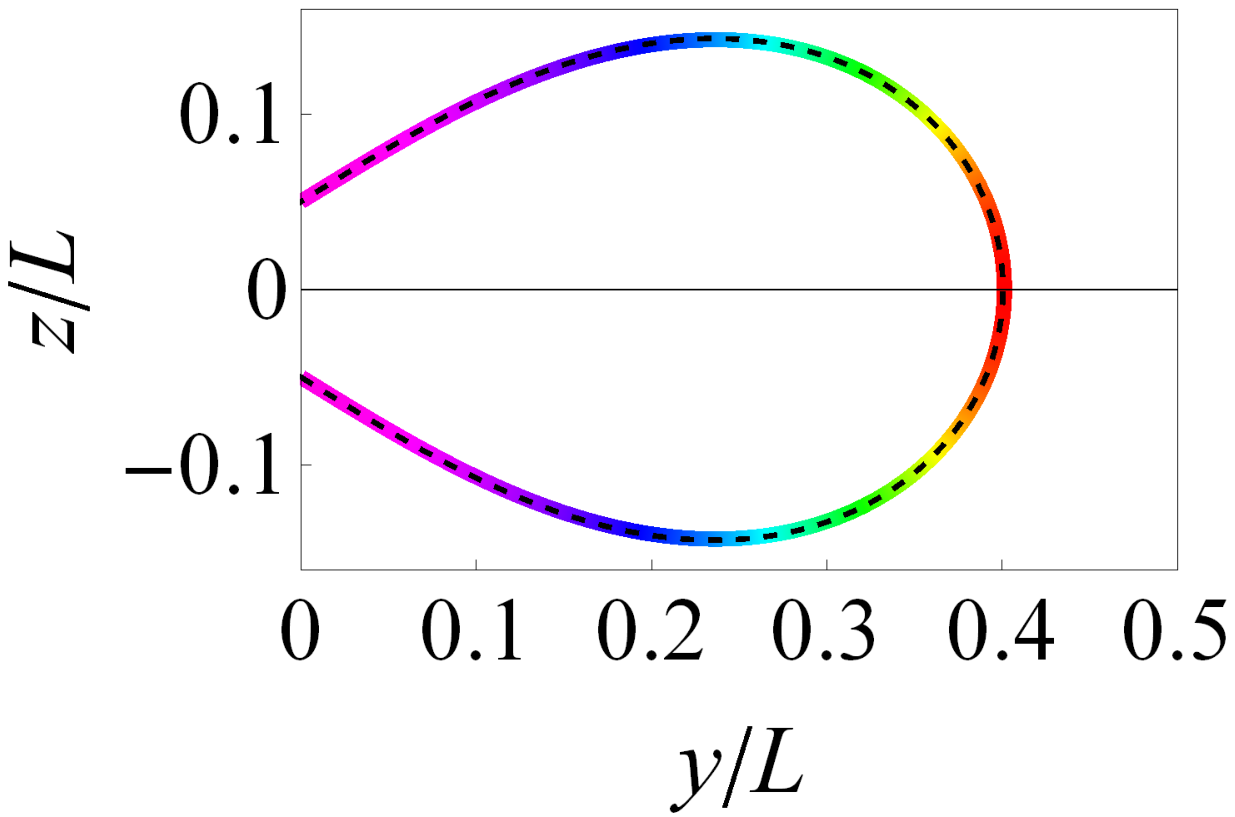} &
\includegraphics[scale=0.33]{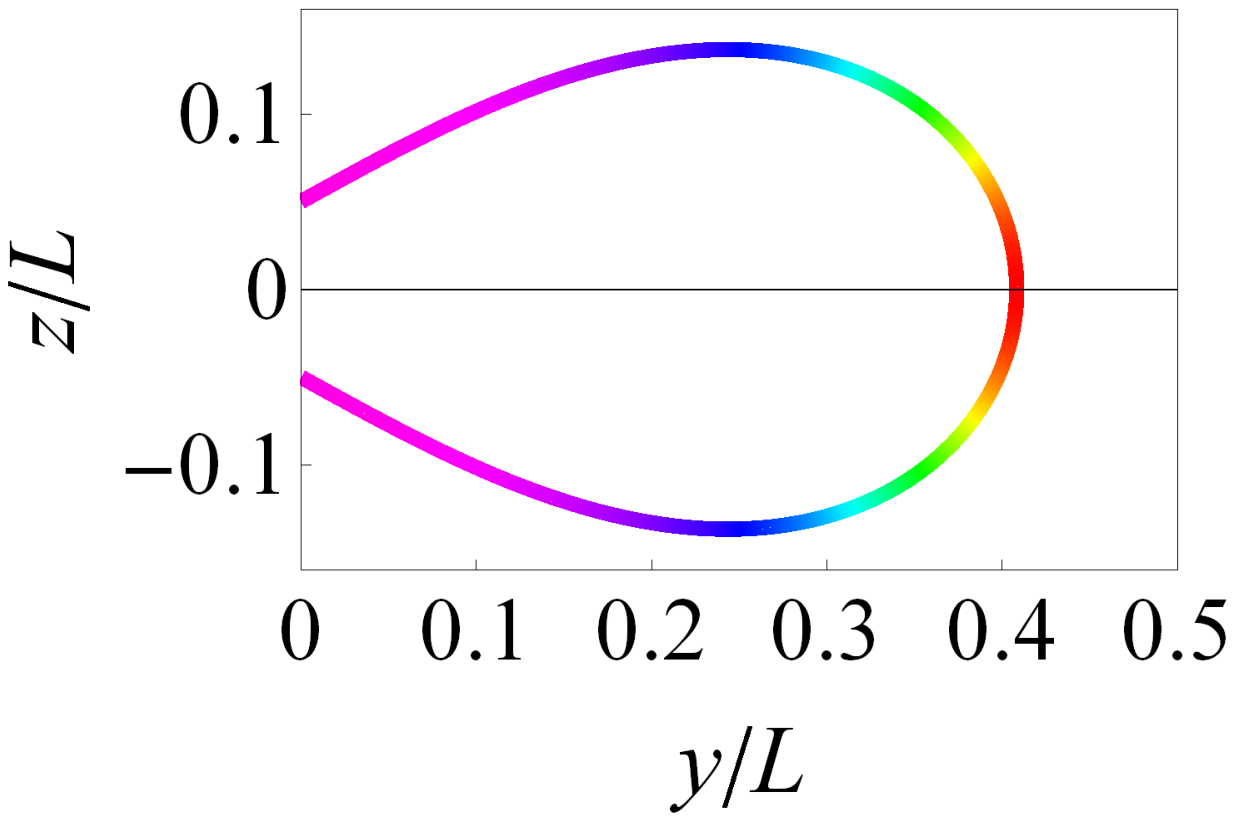} &
\includegraphics[scale=0.33]{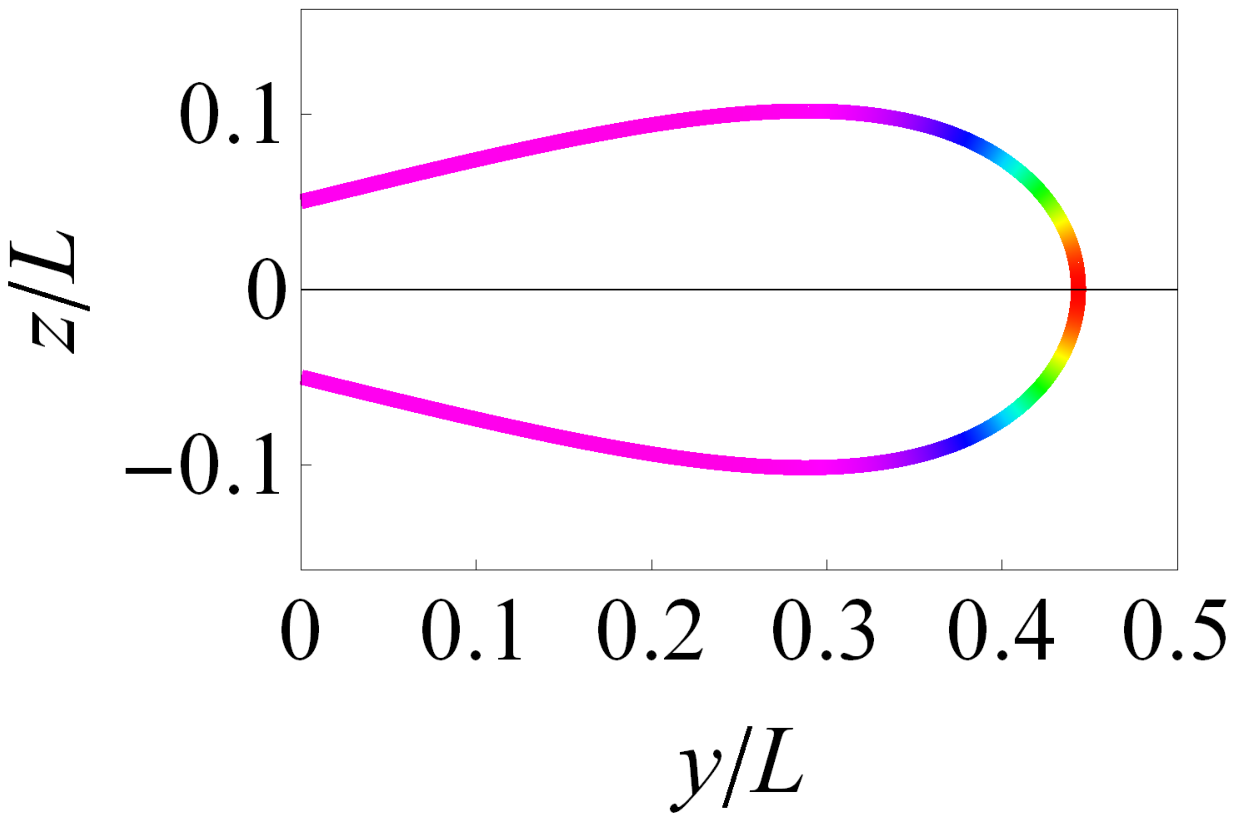} &
\includegraphics[scale=0.33]{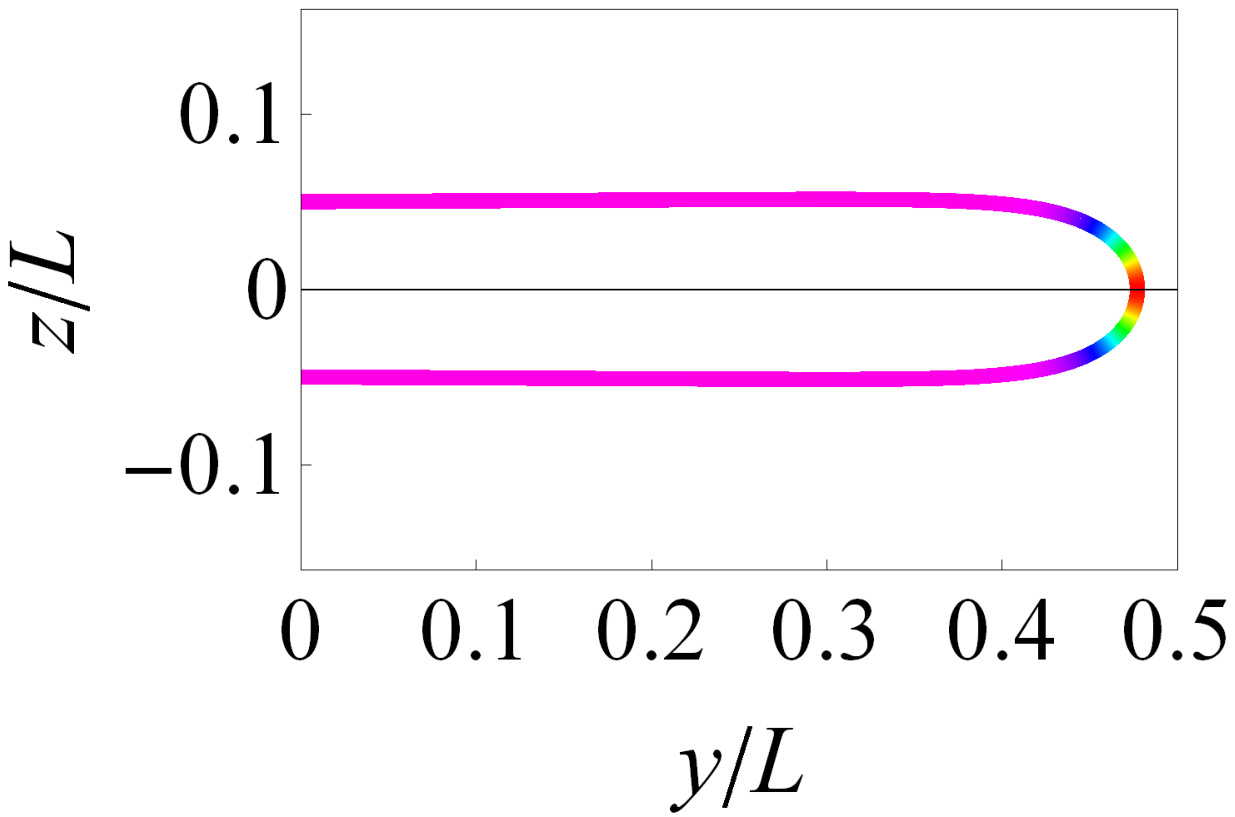} \\
{$\gamma = -1$} & {$\gamma = -10$} &  {$\gamma = -100$} & {$\gamma=-1000$}
\end{tabular}
\includegraphics[scale=0.15]{Fig5m}
\caption{(Color online) Planar configurations with $n=1$ in regime $II$ ($\calM < 0$) for different values of the separation between boundaries: $\xi = 0.8$ (top row), $\xi=0.457$ (middle row), and $\xi=0.1$ (bottom row). The local energy density is color-coded in these figures.}\label{fig:VertFilsn1Mneg}
\end{figure}

\begin{figure}[htbp]
\centering
\begin{tabular}{cccc}
\includegraphics[scale=0.375]{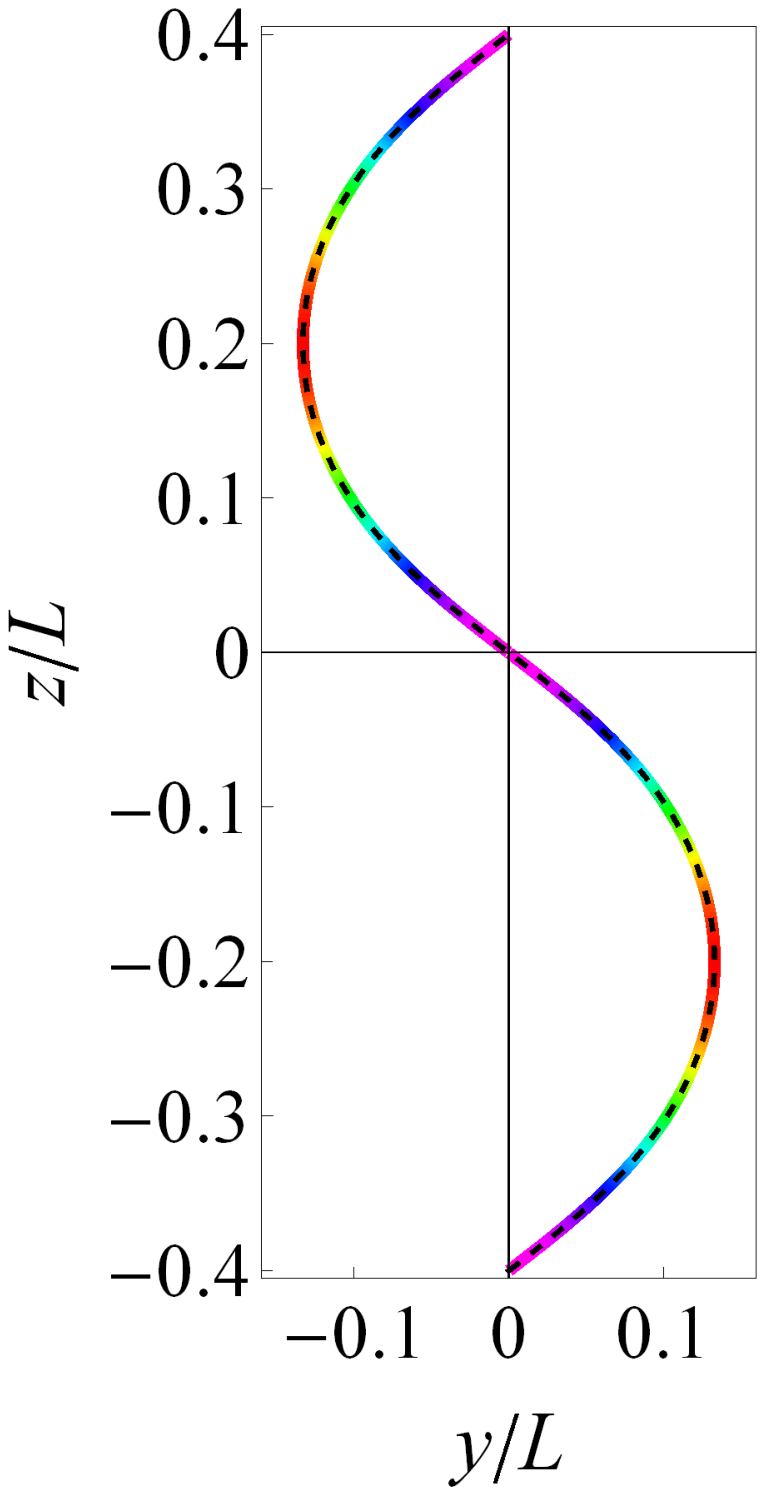} &
\includegraphics[scale=0.375]{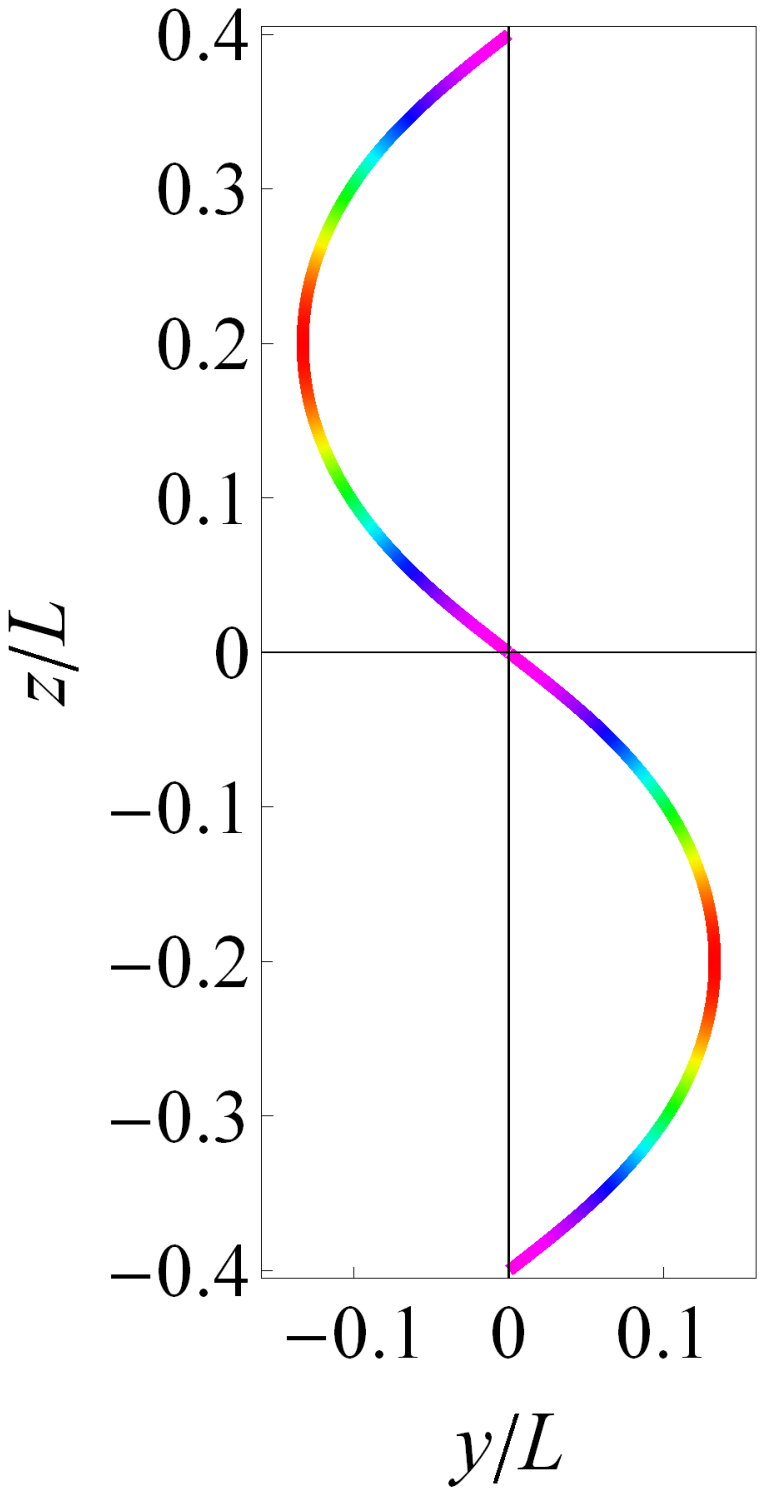} &
\includegraphics[scale=0.375]{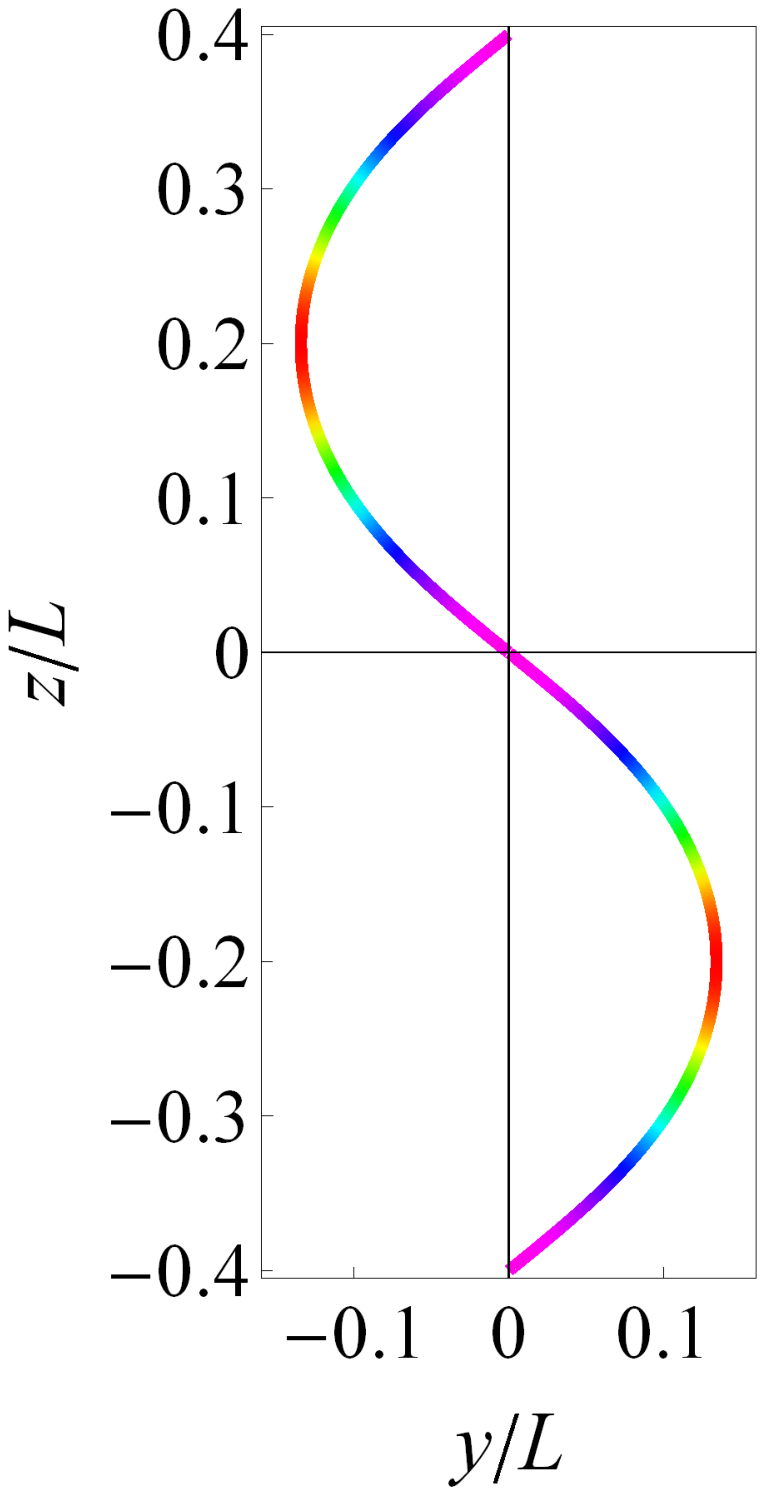} &
\includegraphics[scale=0.375]{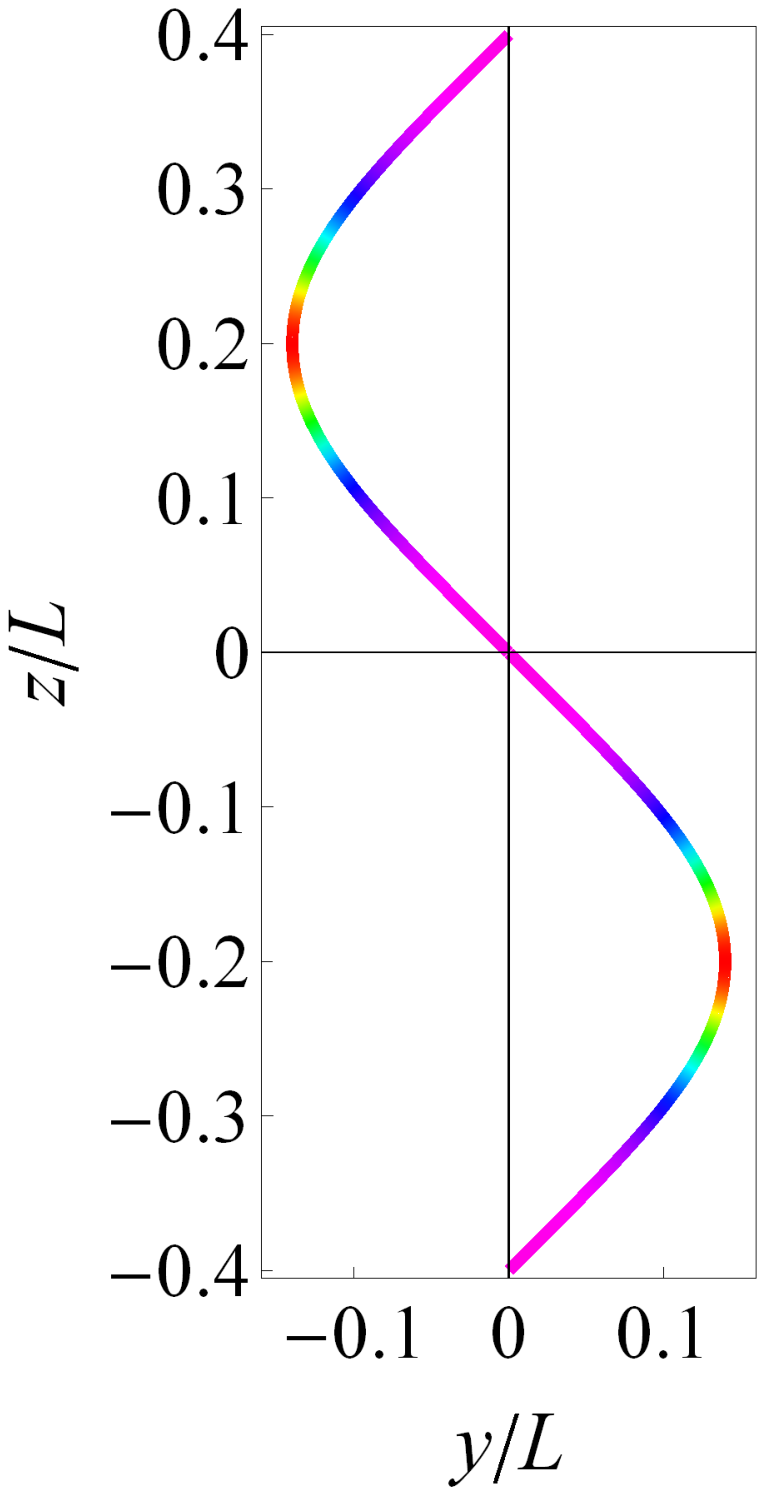} \\
\includegraphics[scale=0.34]{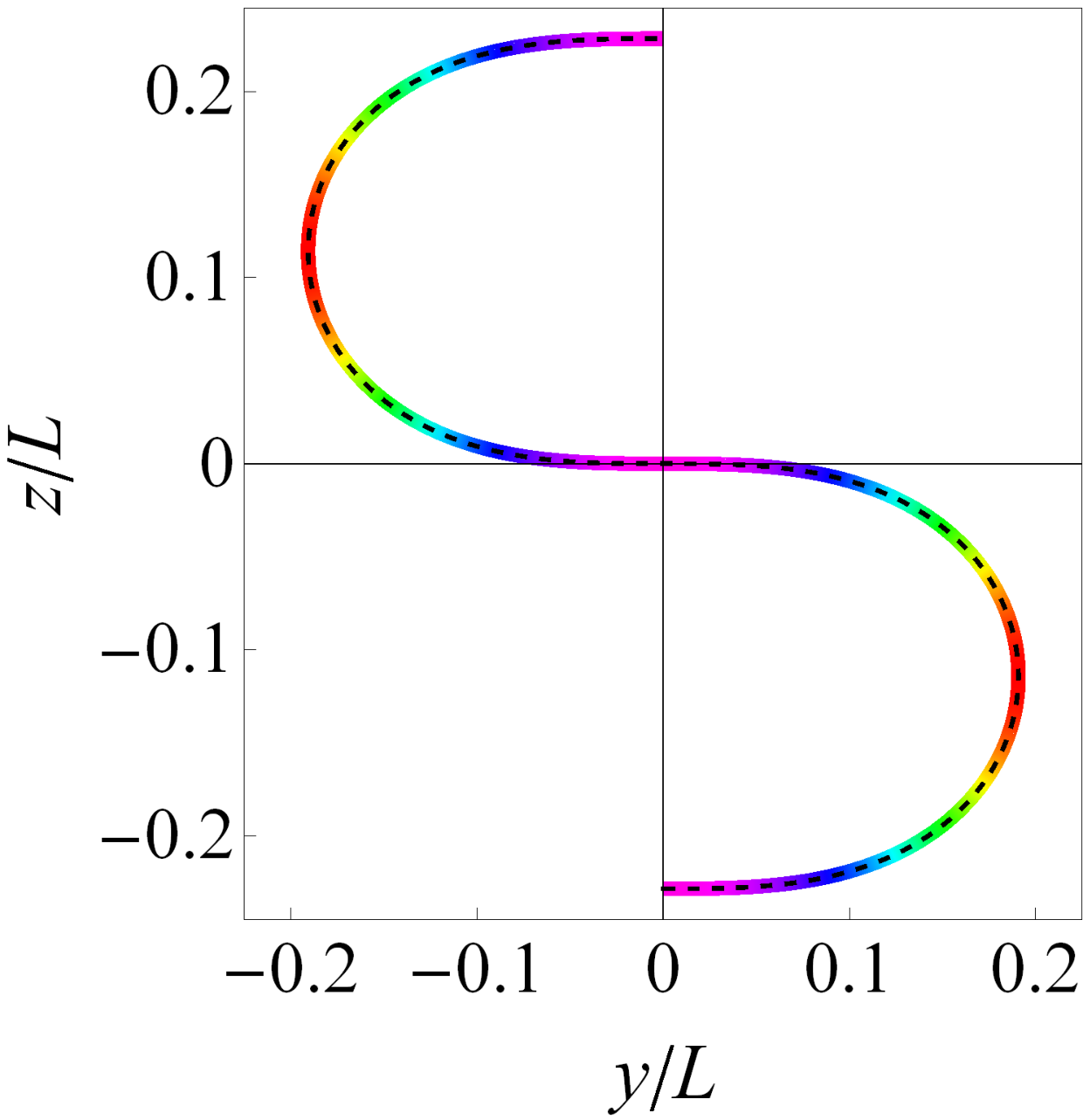} &
\includegraphics[scale=0.34]{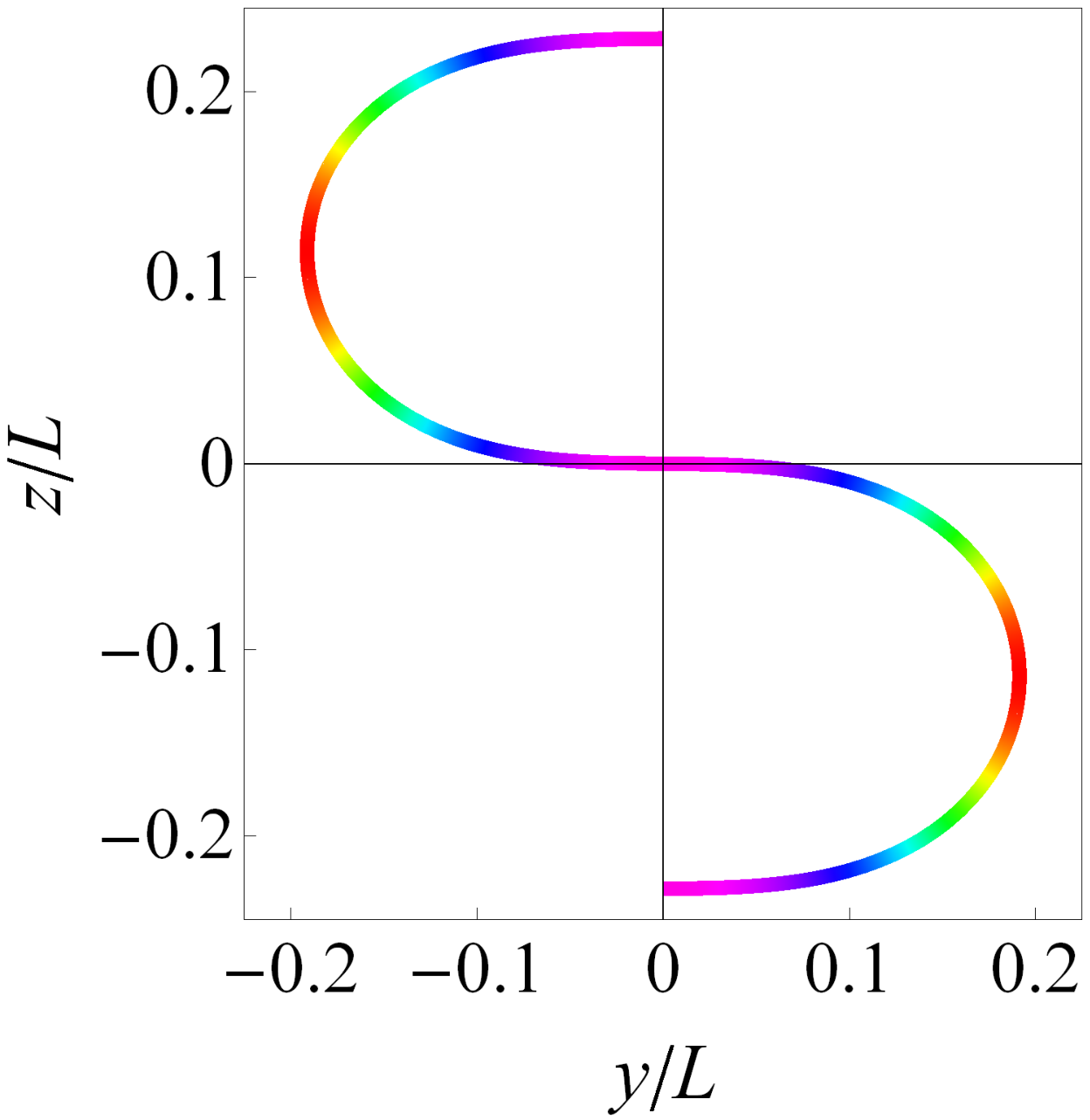} &
\includegraphics[scale=0.34]{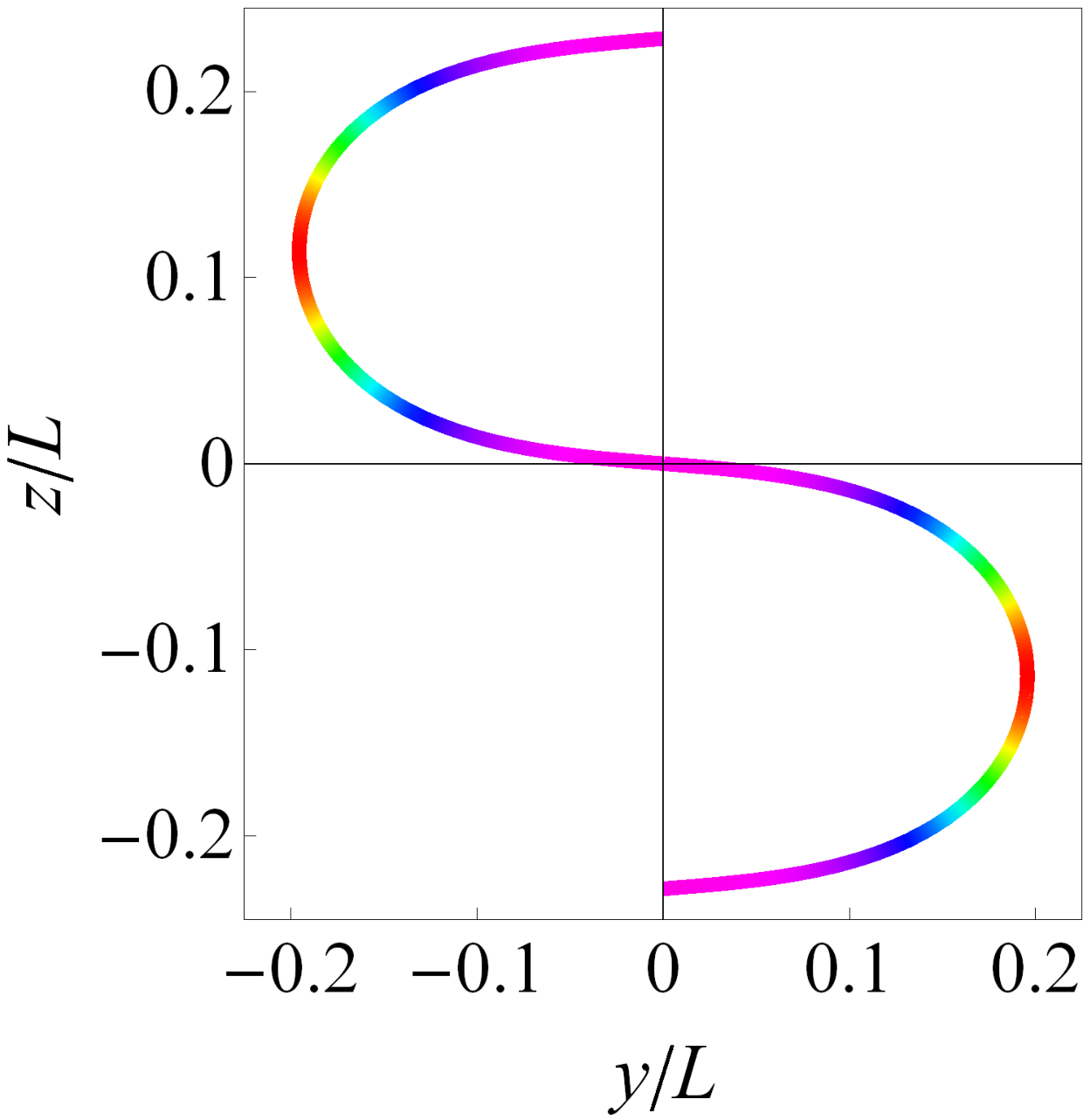} &
\includegraphics[scale=0.34]{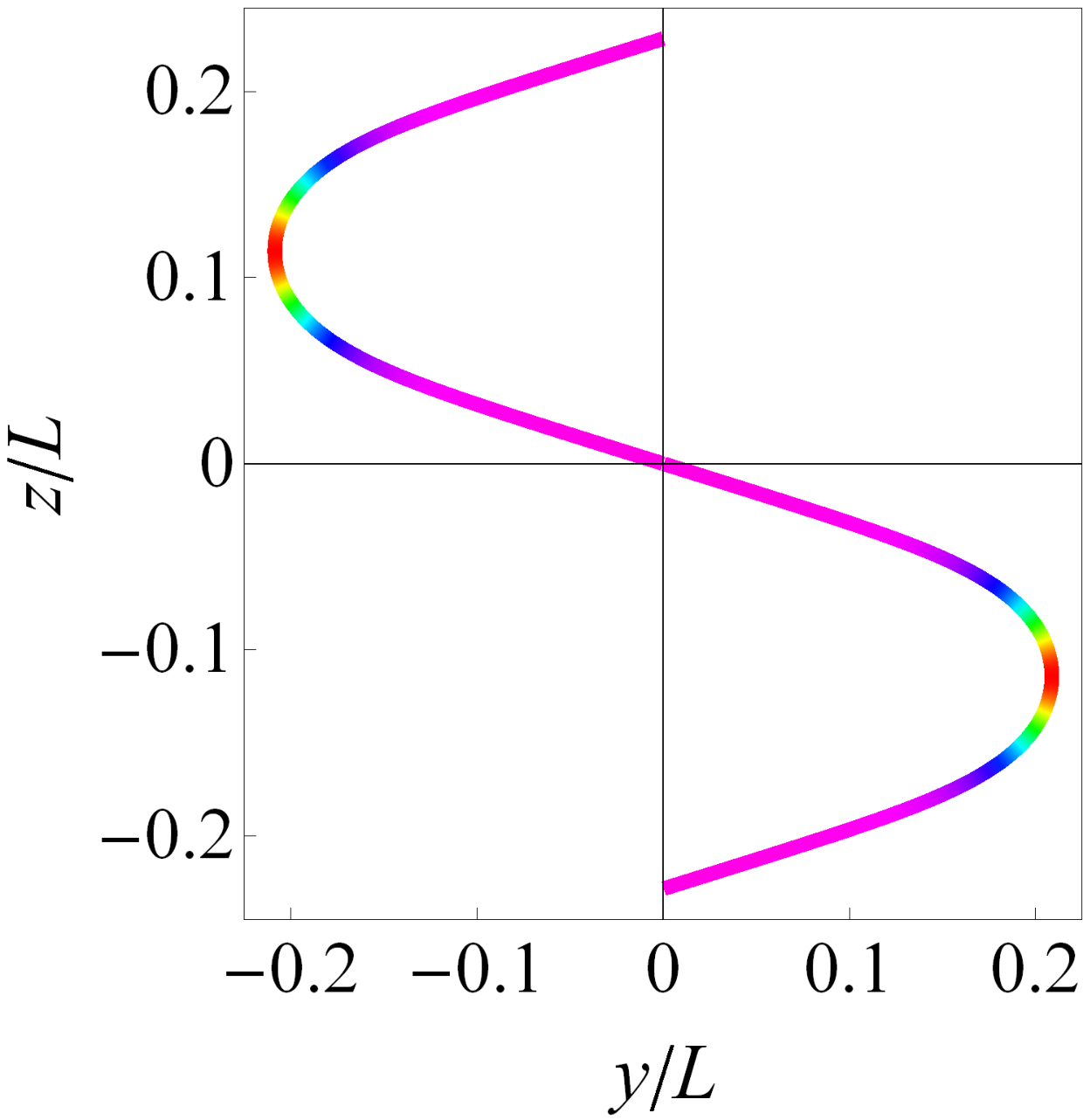} \\
\includegraphics[scale=0.33]{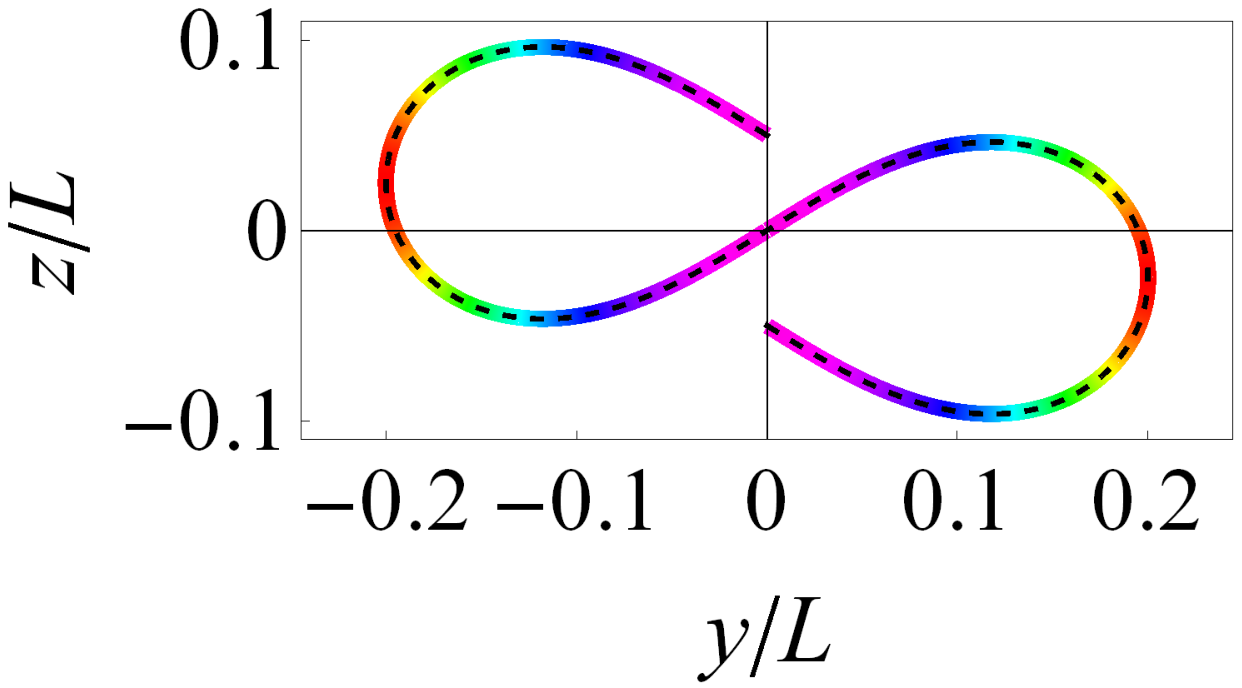} &
\includegraphics[scale=0.33]{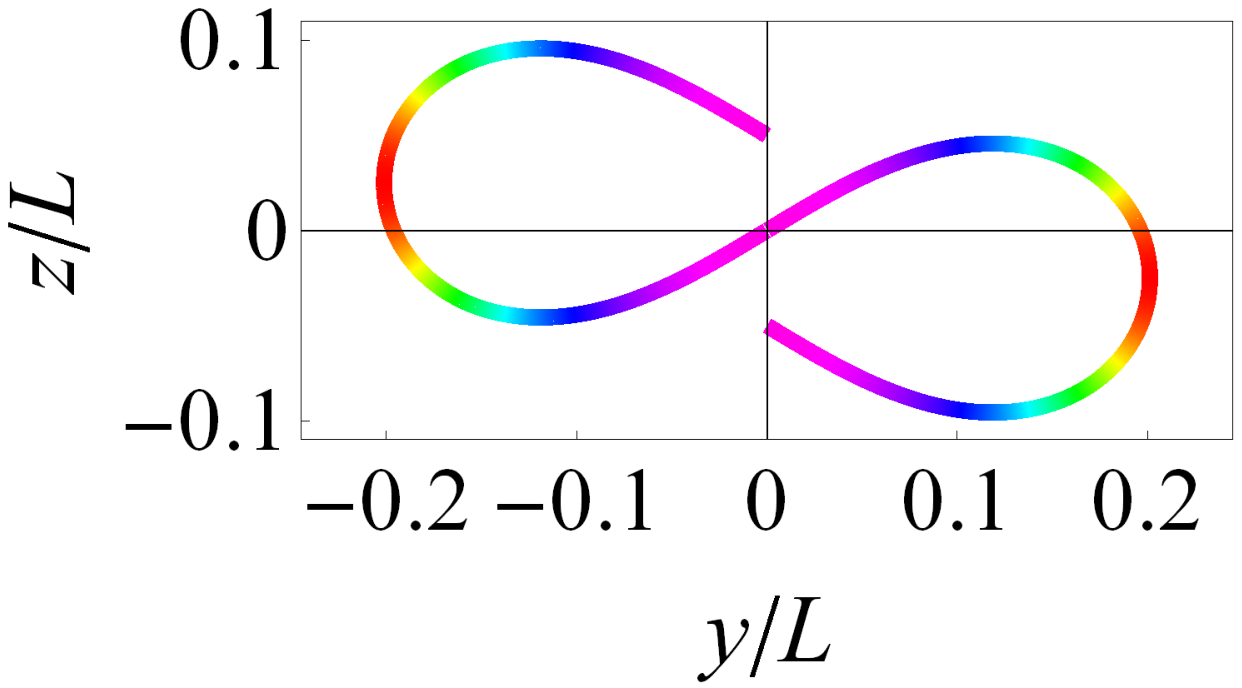} &
\includegraphics[scale=0.33]{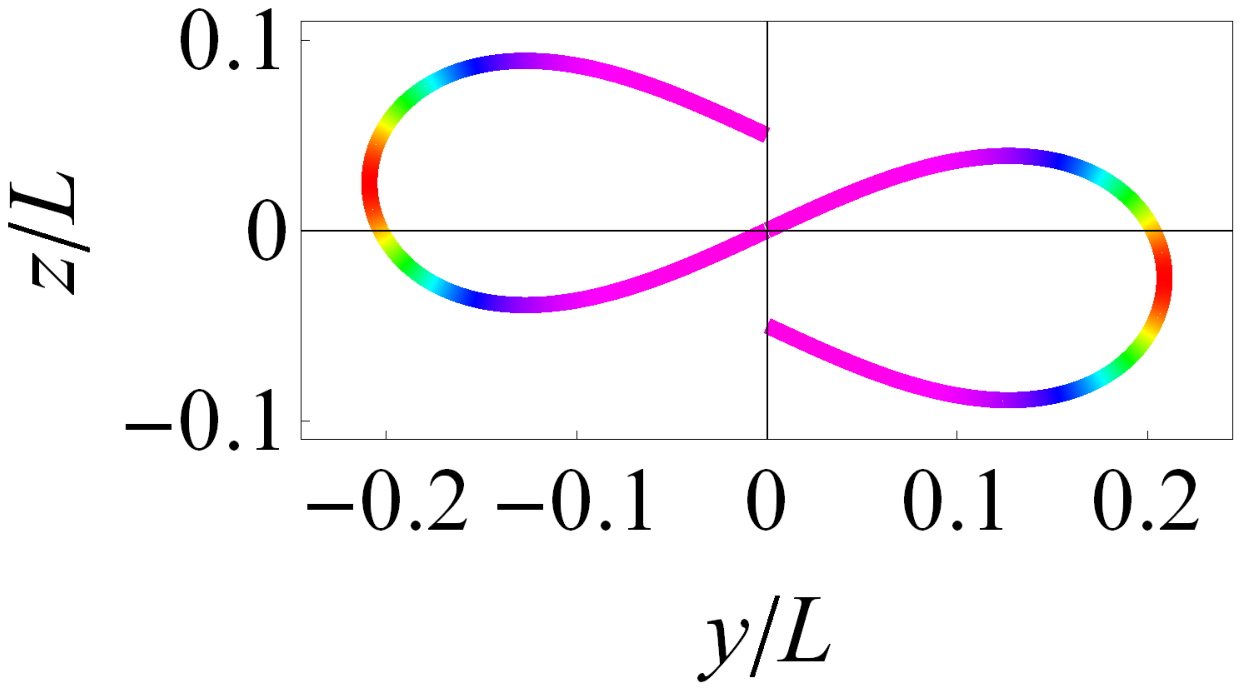} &
\includegraphics[scale=0.33]{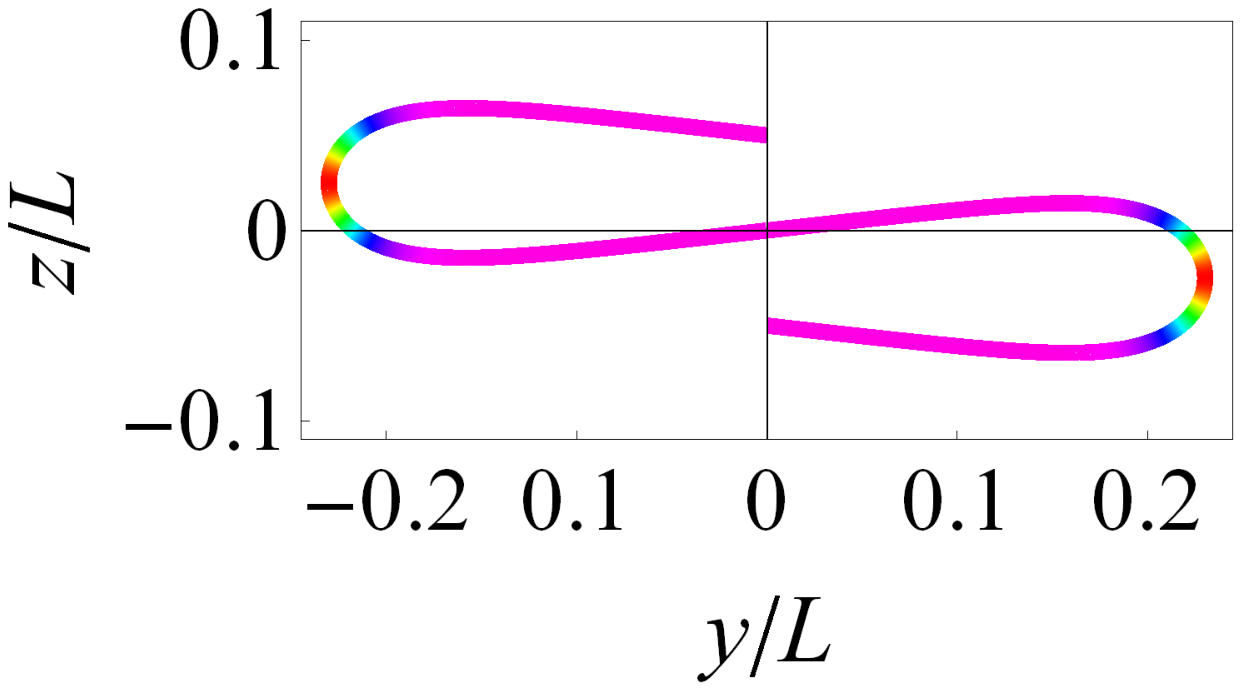} \\
  {$\gamma = 1$} & {$\gamma = 10$} &  {$\gamma = 100$} & {$\gamma=1000$}
\end{tabular}
\includegraphics[scale=0.15]{Fig5m}
 \caption{(Color online) Planar configurations with $n=2$ in regime $II$ ($\calM < 0$) for different values of the separation between boundaries: $\xi = 0.8$ (top row), $\xi=0.457$ (middle row), and $\xi=0.1$ (bottom row). The local energy density is color-coded in these figures.} \label{fig:VertFilsn2Mneg}
\end{figure}

\begin{figure}[htbp]
\centering
\begin{tabular}{cc}
\subfigure[]{\includegraphics[width=0.4\textwidth]{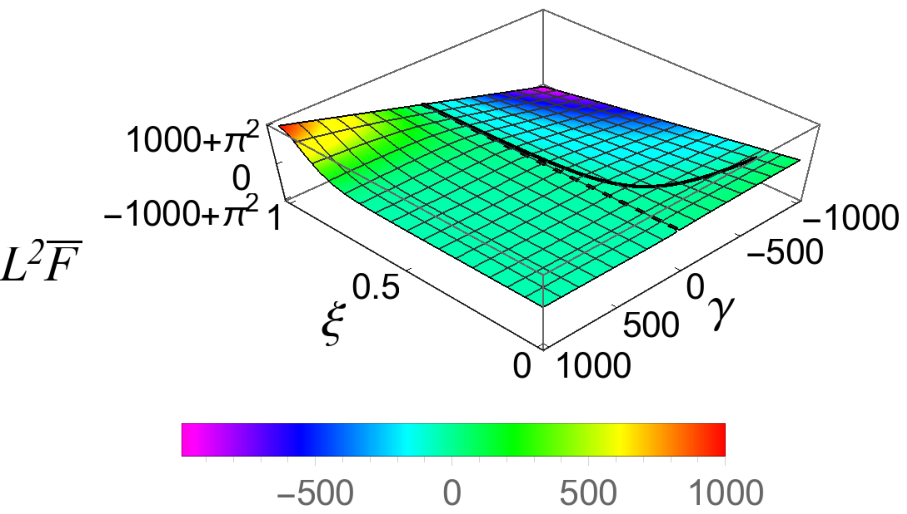}} &
\subfigure[]{\includegraphics[width=0.4\textwidth]{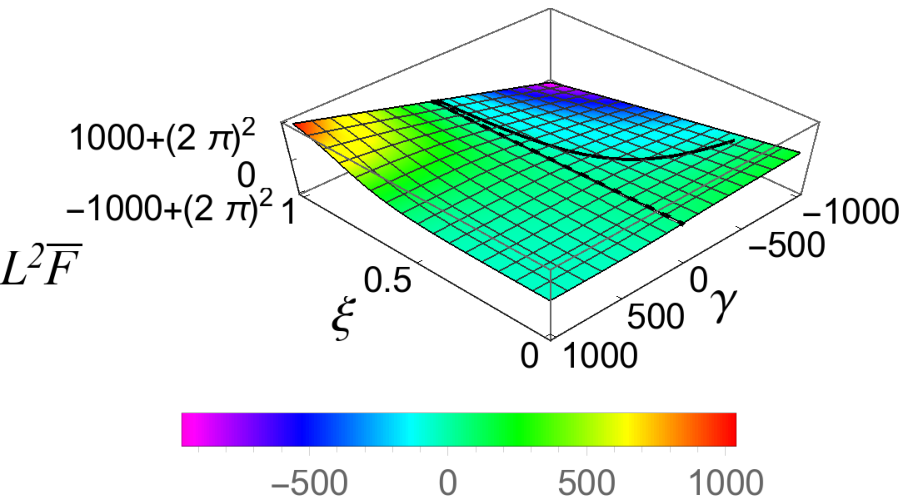}} 
\end{tabular}
\caption{(Color online) Magnitude of the scaled force $\bar{F}$ for the first two planar states (a) $n=1$ and (b) $n=2$. Configurations with vanishing force $F=0$ are represented with a solid black line; elastic curves with $\gamma=0$ are represented with a dashed black line.} \label{fig:Fn1n2}
\end{figure}

\begin{figure}[htbp]
\centering
\begin{tabular}{ccc}
\subfigure[]{\includegraphics[width=0.325\textwidth]{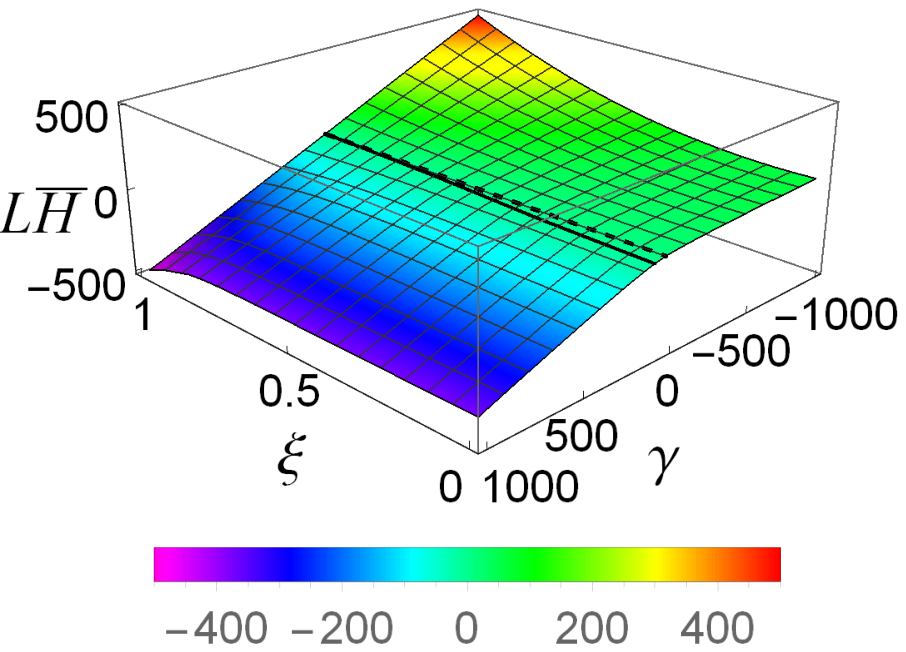}} &
\subfigure[]{\includegraphics[width=0.325\textwidth]{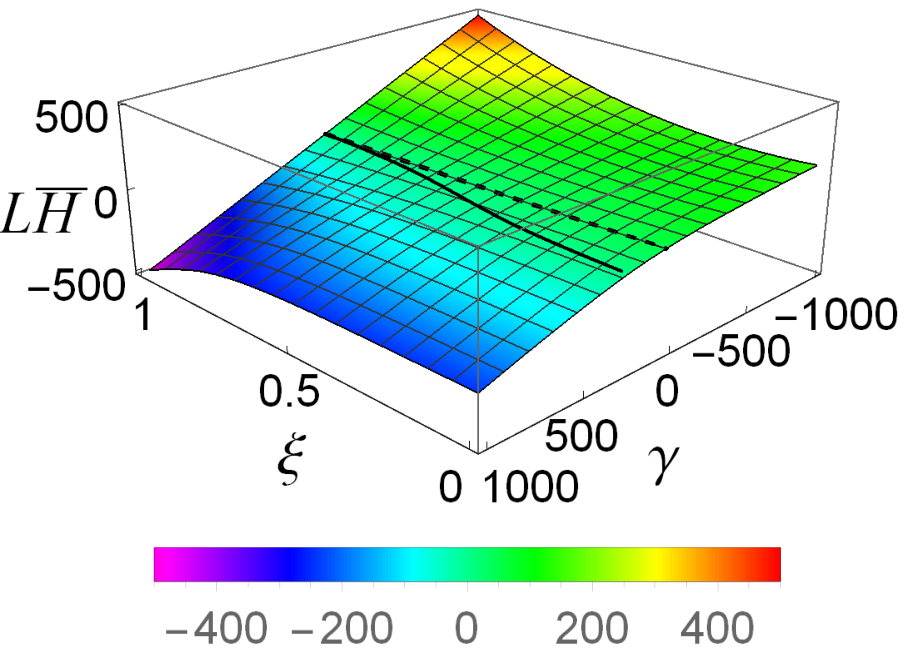}} &
\subfigure[]{\includegraphics[width=0.325\textwidth]{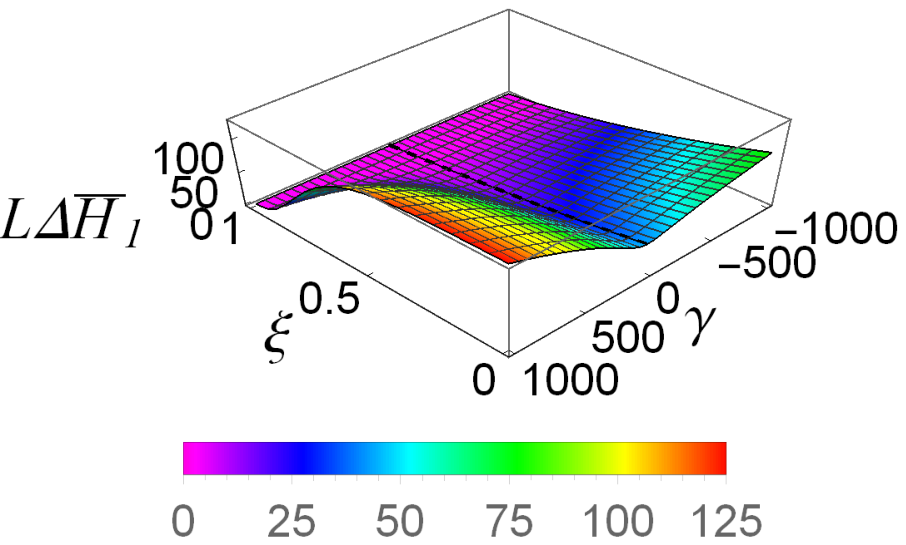}} 
\end{tabular}
\caption{(Color online) Scaled total energy $\bar{H}$ for the first two planar states (a) $n=1$ and (b) $n=2$, the solid black line represents configurations with $H=0$ and the dashed line elastic curves with $\gamma=0$. (c) Difference between the total energies, $\Delta H_{1}= H_{n=2} - H_{n=1}$, as a function of the boundary separation $\xi$ and the magnetoelastic parameter $\gamma$. $\Delta H_{1} \geq 0$ everywhere.} \label{fig:FigHTn1n2}
\end{figure}

\noindent
Now, we describe the behavior of the filaments in the non-linear regime, i.e. deformations far from straight configurations under the influence of a magnetic field (for comparison purposes, the case of elastic curves, $\calM=0$ is reviewed in Appendix \ref{sec:vertEulerElastica}). If $\calM \neq 0$ ($\vartheta \neq \vartheta_m$), the quadrature (\ref{eq:quadpsi}) can be recast as
\begin{equation} \label{eq:quadroots}
\Theta' {}^2 = \bar{\calM} \left(\cos \Theta -\cos \Theta_M \right) \left( 2 \chi- \cos \Theta_M - \cos \Theta\right)\,, \quad \chi = \frac{F}{\calM} \,.
\end{equation}
Integrating the quadrature twice, we obtain the coordinates in terms of elliptic functions in each regime,
(details are provided in Appendix \ref{sec:vertEllipInts} for the interested reader):
\\
\begin{subequations} \label{eq:yzhor}
\begin{eqnarray}
y_I &=& \frac{1}{q} \sqrt{\frac{2(a_I-1)}{1-\eta_I}} \, \arctan\left(\sqrt{\frac{\eta_I}{1-\eta_I}} \cn(q (s-s_0)|m)\right)\,, \nonumber \\
z_{I} &=& a_I s-\frac{(a_I-1)}{q} \, \left( \Pi \left(\eta_I;\am(q (s-s_0)|m)|m\right) - \mathrm{mod}(n-1,2) \, \Pi(\eta_I|m)\right)\,;\\
y_{II} &=& \frac{1}{q} \sqrt{\frac{2 (a_{II}+1)}{1+\eta_{II}}} \, \arctanh\left(\sqrt{\frac{\eta_{II}}{1+\eta_{II}}} \cn(q (s-s_0)|m) \right) \,,  \nonumber\\
z_{II} &=&- a_{II} s+ \frac{a_{II}+1}{q} \, \left( \Pi \left(-\eta_{II};\am(q (s-s_0)|m)|m\right) - \mathrm{mod}(n-1,2) \, \Pi(-\eta_{II}|m)\right)\,;
\end{eqnarray}
\end{subequations}
where the constants $a$ and $\eta$ are defined by
\begin{subequations} \label{def:aeta}
\begin{align}
a_I &= \left(q \ell\right)^2 \pm \sqrt{\left(\left(q \ell\right)^2+1\right)^2 -4 m (q \ell)^2}\,,  &\quad \eta_I&= \frac{2\,m}{a_I+1}\, ; \\
a_{II} &= \left(q \ell\right)^2 \pm \sqrt{\left(\left(q \ell\right)^2-1\right)^2 + 4 m  (q \ell)^2} \,, &\quad \eta_{II} &= \frac{2 m}{a_{II}-1}\,;
\end{align} 
\end{subequations}
$\sn(u|m)$, $\cn(u|m)$ and $\dn(u|m)=\sqrt{1-m \, \sn^2(u|m)}$ are the sine, cosine, and delta Jacobi elliptic functions;  $\am(u|m)$ is the Jacobi amplitude; $\Pi(\eta,u,m)$ is the incomplete elliptic integral of the third kind \cite{Abramowitz1974, Gradshteyn2007}. Recall $\ell = \sqrt{\calB/\calM}$ is the buckling characteristic length.
\\
Like in the perturbative case, if the filament possesses $n$ half periods, the wave number is given by 
\begin{equation} \label{def:qvert}
q = \frac{2 n \rmK(m)}{L}\,, \quad \mbox{or} \quad q \ell = \frac{2 n \mathrm{K}(m)}{\sqrt{|\gamma|}}\,,
\end{equation}
where $\rmK(m)$ is the complete elliptic integral of the first kind \cite{Abramowitz1974, Gradshteyn2007}.
The last relation permits us to express constants $a$ and $\eta$, defined in Eq. (\ref{def:aeta}), in terms of the modulus $m$ and the magnetoelastic parameter $\gamma$.
\\
The FS curvature of the filament is given by
\begin{subequations} \label{eq:kappaplanar}
\begin{align}
\kappa_I &= \frac{\kappa_{M\,I} \, \cn(q (s-s_0)|m)}{1 - \eta_I \,\sn(q (s-s_0)|m)^2}\,, &\quad \kappa_{M I} &= q \sqrt{2 \eta_I \left(a_I-1\right)}\,;\\
\kappa_{II} &= \frac{\kappa_{M\,II} \, \cn(q (s-s_0)|m)}{1 + \eta_{II} \,\sn(q (s-s_0)|m)^2}\,, &\quad \kappa_{M II} &= q \sqrt{2 \eta_{II} \left(a_{II}+1\right)}\,;
\end{align}
\end{subequations}
where $\kappa_M$ is the maximum value of the curvature. The condition (\ref{def:BCfixedends}) of vanishing curvature at the boundaries determines, $s_0 = -\mathrm{mod} (n-1,2) L/(2 n)$.
\\
Evaluating expressions (\ref{eq:yzhor}) for $z$ at the boundaries and using the identities $\am(n \rmK(m)|m) = n \pi/2$ and $\Pi(\eta;n \frac{\pi}{2}|m) = n \Pi(\eta|m)$, we get the following equation for the scaled boundary separation $\xi$, defined in Eq. (\ref{def:xi})
\begin{subequations} \label{eq:nlreqm}
\begin{eqnarray}
\xi_I &=&  a_I - (a_I-1) \frac{\Pi \left(\eta_I| m \right)}{\rmK(m)}\,, \\
\xi_{II} &=&  - a_{II} + (a_{II}+1)\frac{\Pi \left(-\eta_{II}| m \right)}{\rmK(m)} \,.
\end{eqnarray}
\end{subequations}
To determine $m$, these equations are solved numerically for given values of $n$, $\gamma$, and $\xi$.\footnote{Alternatively, one could extend the method employed in Ref. \cite{Hu2013} for the purely elastic case, in which case $m$ would be expanded as a series in $\xi$ and $\gamma$ and the coefficients would be determined from Eq. (\ref{eq:nlreqm}).} This completes the determination of all parameters of the curve. 
\vskip0pc \noindent
States with $n=1,2$ in regime $I$ are plotted for different values of $\xi$ and $\gamma$ in Figs. (\ref{fig:VertFilsn1Mpos}) and (\ref{fig:VertFilsn2Mpos}). Corresponding states with $n=1,2$ in regime $II$ are plotted in Figs (\ref{fig:VertFilsn1Mneg}) and (\ref{fig:VertFilsn2Mneg}). In these sequences we choose initial states with different boundary angles $\Theta_b= \Theta(s_b)$, specifically $\Theta_b < \pi /2$ (top rows with $\xi = 0.8$), $\Theta_b =\pi/2$ (middle rows with $\xi = 0.457$) and $\Theta_b >\pi/2$ (bottom rows with $\xi = 0.1$).\footnote{For smaller separations $\xi$, the boundaries, and also upper and bottom segments of the filaments for strong magnetic fields, get close and in consequence our nearest neighbors approximation is no longer valid.} In both regimes we observe that for relative small absolute values of the magnetoelastic parameter, $|\gamma|= 1, 10$, the filaments behave mainly as elastic curves (shown with dashed black lines in the plots with $\gamma=1$), adopting $U$ and $S$ shapes for $n=1$ and $n=2$, respectively. As $|\gamma|$ is increased, we observe deviations from elastic behavior depending on the regime: in regime $I$, at $\gamma=100$, filaments begin to elongate along the precession axis and squeezing inwards along the orthogonal direction, and for a large value, $\gamma=1000$, they form thin vertical hairpins connected by straight segments aligned with the precession axis; in regime $II$ the converse behavior is observed, at $\gamma = - 100$ they begin to stretch outwards and orthogonally to the precession axis, and at $\gamma=-1000$ they are mostly straightened and with the filament's horizontal extremum farthest from the precession axis.
\vskip0pc \noindent
The magnitude of the forces in these planar curves is given by\footnote{It can be positive or negative depending on the sign of the constant $a$ satisfying Eq. (\ref{eq:nlreqm}).}
\begin{subequations}
 \begin{eqnarray}
 L^2 \, \bar{F}_I &=& \pm \gamma \sqrt{\left(\left(q \ell\right)^2+1\right)^2 -4 m (q \ell)^2} \,,\\
 L^2 \, \bar{F}_{II} &=& \mp \gamma \sqrt{\left(\left(q \ell\right)^2-1\right)^2 + 4 m (q \ell)^2}\,.
\end{eqnarray}
\end{subequations}
$F$ is plotted for states $n=1,2$ in Fig \ref{fig:Fn1n2}. For vertical curves with $\xi =1$, the force is linear in the magnetoelastic parameter, $F = \gamma + (n \pi)^2$, as found in the perturbative analysis. We see that in regime $I$ $F$ is positive for all values of $\xi$ and $\gamma$, indicating that the filaments are under compression, as is usual for elastic curves bent under compression. By contrast, there are regions in regime $II$ where $F$ becomes negative in which case filaments are under tension, reflecting the fact that they tend to lie orthogonally to the precession axis.
\vskip0pc \noindent
The bending and magnetic energy densities in terms of arc length read 
\begin{subequations}
\begin{align} \label{def:HBHMplanarcurves}
\mathcal{H}_{BI} &= q^2 \eta_I (a_I-1)  \frac{\cn^2 (q (s-s_0)|m)}{\left(1-\eta_I \sn^2(q (s-s_0)|m)\right)^2} \,, &\quad \mathcal{H}_{MI} &= -\frac{\calM}{2} \left(a_I-\frac{a_I-1}{1-\eta_I \sn^2 (q (s-s_0)|m)}\right)^2\,; \\
\mathcal{H}_{BII} &= q^2 \eta_{II} (a_{II}+1)  \frac{\cn^2 (q (s-s_0)|m)}{\left(1+\eta_{II} \sn^2(q (s-s_0)|m)\right)^2} \,, &\quad \mathcal{H}_{MII} &= -\frac{\calM}{2} \left(a_{II}-\frac{a_{II}+1}{1+\eta_{II} \sn^2 (q (s-s_0)|m)}\right)^2\,.
\end{align}
\end{subequations}
The bending energy is positive, increasing with $q^2$ ($\propto n^2$); the magnetic energy is positive in regime $I$ and negative in regime $I$, thus the total energy density $\calH = \mathcal{H}_B + \mathcal{H}_M $, can be positive or negative in regime $I$, but it is strictly positive in regime $II$. $\calH$ is shown with a color scale for states $n=1,2$ in Figs. (\ref{fig:VertFilsn1Mpos})-(\ref{fig:VertFilsn2Mneg}). For filaments in regime $I$, we see that initially $\calH$ is concentrated in the extremum and low in the boundaries, but as $\gamma$ is increased, the high-energy regions migrates towards the hairpins near the boundaries and low-energy regions move to the extremum where straight segments (minimizing both energies) are developed. In regime $II$, high-energy regions always occur at the extremum where the curvature concentrates, whereas low-energy regions correspond to the straight segments near to the boundaries. Moreover, the former regions become more localized and the latter regions more spread as $\gamma$ is increased.
\\
In the calculation of the total energy, although integration of $\calH_M$ is simple, integration of $\calH_B$ is rather complicated because it involves $\kappa^2 = (\Theta')^2$. However, we can integrate expression (\ref{def:endensimp}) for the total energy density, where $\kappa^2$ was replaced in favor of $\rmtz=\cos \Theta$ by means of the quadrature, (\ref{eq:FprojsT}), obtaining the following expressions of the total energy for each case (details are presented in Appendix \ref{sec:vertEllipInts}):
\begin{subequations} \label{def:totenplanar}
\begin{eqnarray} 
\bar{H}_I &=& \frac{\gamma}{2 L} \left[ \frac{a_I-1}{1-\eta_I} \left( 2 \frac{\rmE(m)}{\rmK(m)} - 1\right) +\left(\frac{a_I -1}{1-\eta_I}-a_{II}+b_I\right) \xi_I - a_I b_I\right]\,,\\
\bar{H}_{II} &=& \frac{|\gamma|}{2 L} \left[ \frac{a_{II}+1}{1+\eta_{II}}  \left(2 \frac{\rmE(m)}{\rmK(m)} -1\right) + \left(\frac{a_{II}+1}{1+\eta_I}-a_{II}-b_I\right) \xi_{II} - a_{II} b_{II}\right] \,, 
\end{eqnarray}
\end{subequations}
where the constant $b$ is defined by
\begin{subequations}
\begin{eqnarray}
b_I &=& -\left(q \ell\right)^2 \pm \sqrt{\left(\left(q \ell\right)^2+1\right)^2 -4 m (q \ell)^2} \,,\\
b_{II} & = & \left(q \ell\right)^2 \mp \sqrt{\left(\left(q \ell\right)^2-1\right)^2 + 4 m (q \ell)^2}\,.
\end{eqnarray}
\end{subequations}
The total energy $H$ of states $n=1,2$ are plotted in Figs. \ref{fig:FigHTn1n2}(a) and \ref{fig:FigHTn1n2}(b). As found in the perturbative regime, regardless of $n$, straight lines with $\xi =1$ have scaled total energy $L \bar{H} = -\gamma /2$.  We see that $H$ is negative almost everywhere (except in a small fringe of values in the vicinity of $\gamma=0$) in regime $I$ and it is positive everywhere in regime $II$. Values of $\xi$ and $\gamma$ for which $H=0$ are shown with a solid black line, and the energies of elastic curves ($\gamma=0$) are shown with a dashed black line. 
\\
The total energy of the filaments increases as $n$ augments for any value of $\xi$ and $\gamma$. To show this, in Fig. \ref{fig:FigHTn1n2}(c), we plot the energy difference between states $n=1$ and $n=2$, $\Delta H_{1} = H_{n=2} - H_{n=1}$, where we see that $\Delta H_{1} > 0$ everywhere on the parameter space $\xi-\gamma$, result that can be verified for states with higher $n$, i.e, $H_{n} < H_{n+1}$. Thus $n=1$ is the ground state among planar configurations for all parameter values. However, as we will see below this does not hold in general when non-planar configurations are considered, in particular the energy may be lowered for some values of $\xi$ and $\gamma$ if filaments adopt helical configurations, which we examine in the next section.

\section{Helices} \label{sec:Helices}
\begin{figure}[htbp]
\begin{center}
\includegraphics[height=5cm]{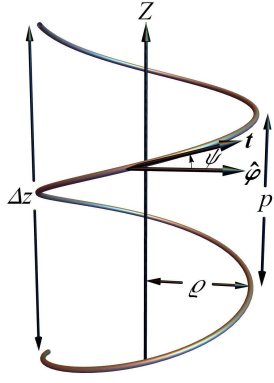}
\end{center}
\caption{(Color online) The helix is characterized by its radius $\varrho$ and pitch $p= 2 \pi \varrho \tan \psi$, with $\psi$ the angle formed by $\bft$ and $\hat{\bm \varphi}$.}\label{Fig:helix}
\end{figure}

\begin{figure}[htbp]
\centering
\begin{tabular}{cc}
\subfigure[]{\includegraphics[width=0.4\textwidth]{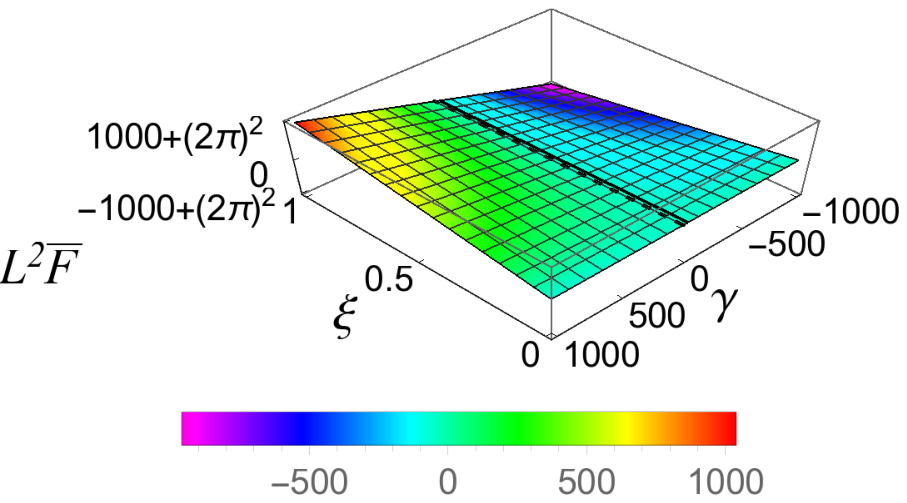}} &
\subfigure[]{\includegraphics[width=0.4\textwidth]{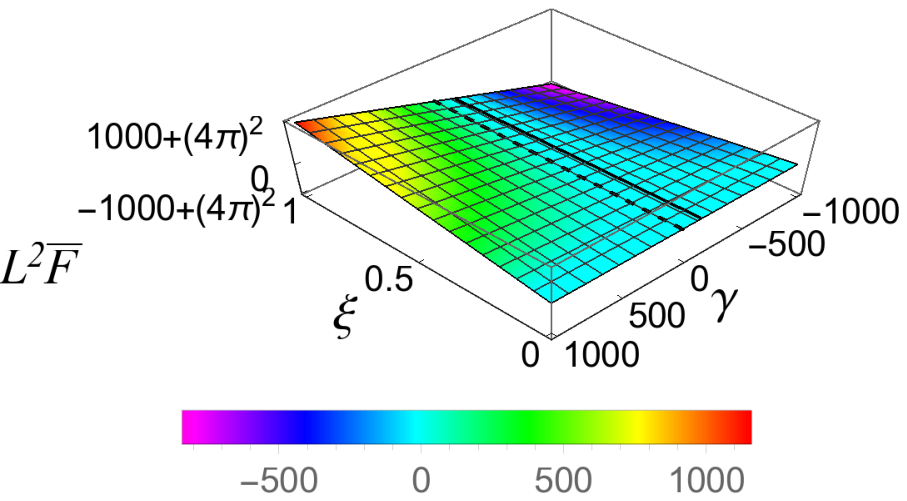}} 
\end{tabular}
\caption{(Color online) Magnitude of the scaled force $\bar{F}$ for the first two helical states (a) $n=1$ and (b) $n=2$. The solid black line represents configurations with $F=0$, the dashed black line elastic curves with $\gamma=0$.} \label{fig:Fn1n2helix}
\end{figure}

\begin{figure}[htbp]
\centering
\begin{tabular}{ccc}
\subfigure[]{\includegraphics[width=0.325\textwidth]{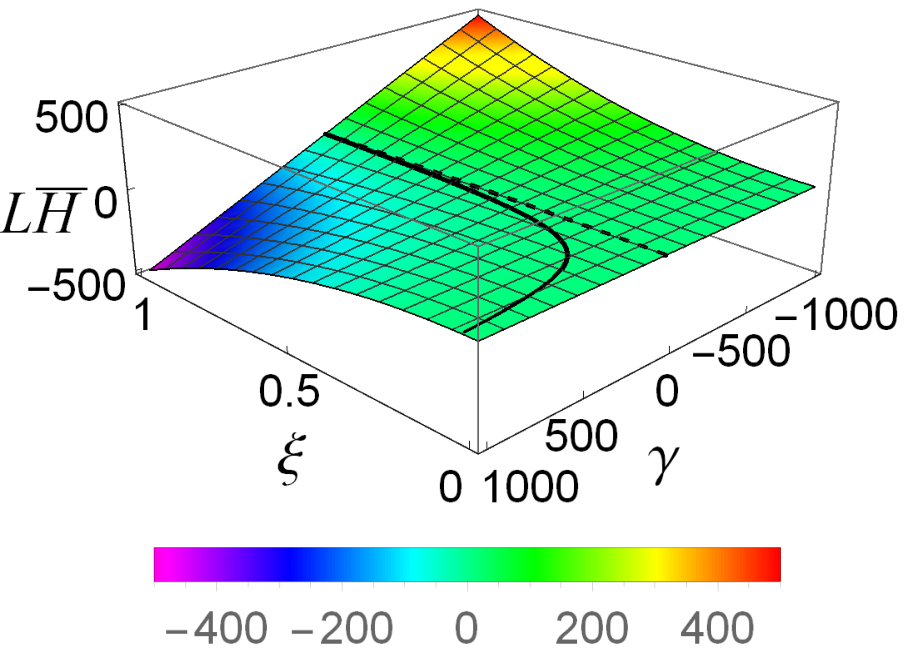}} &
\subfigure[]{\includegraphics[width=0.325\textwidth]{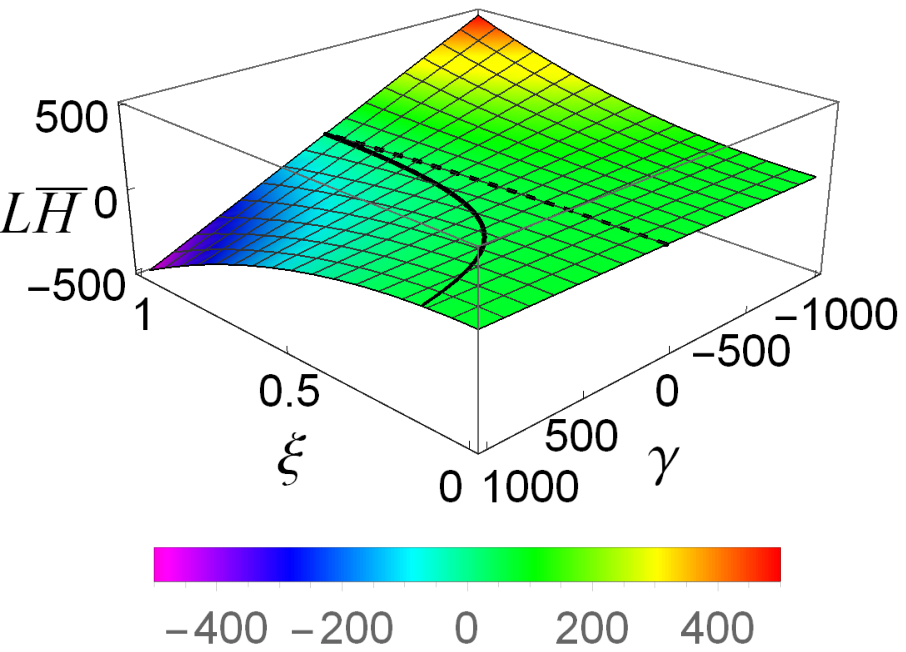}} &
\subfigure[]{\includegraphics[width=0.325\textwidth]{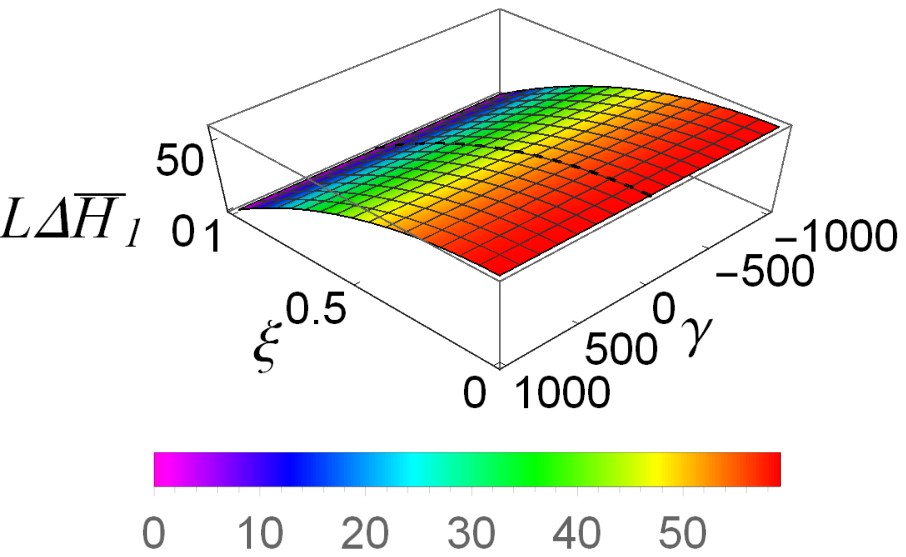}} 
\end{tabular}
\caption{(Color online) Total energy of helices with (a) $n=1$ and (b) $n=2$, the solid black line represents configurations with $H=0$ and the dashed line elastic curves with $\gamma=0$. (c) Difference between the total energies, $\Delta H_{1} = H_{n=2} - H_{n=1}$, as a function of the boundary separation $\xi$ and the magnetoelastic parameter $\gamma$. $\Delta H_{1} \geq 0$ everywhere.} \label{fig:FigHhelix}
\end{figure}

\noindent
Here we demonstrate that helices are also critical points of the total energy. Recall that a helix is characterized by its radius $\varrho$ and pitch $p = 2 \pi \varrho \tan \psi$, with $\psi$ the pitch angle defined by $\cos \psi = \bft \cdot \hat{\bm \varphi}$ (see Fig. \ref{Fig:helix}).
The helix can be parametrized in cylindrical coordinates by the azimuthal angle $\varphi$ as,
\begin{equation} \label{eq:helix}
\bfY = \varrho \hat{\bm \varrho} + \frac{p}{2 \pi} \, \varphi \, \hat{\bf z}\,.
\end{equation}
A helical segment is specified by the total azimuthal angle $\Phi$. We consider helices completing $n$ full turns, so $\Phi = 2 \pi n$ and the distance between the end points is $\Delta z = n p$. Arc length is proportional to $\varphi$, $s = \sqrt{\varrho^2 + \left(\frac{p}{2 \pi}\right)^2} \varphi$, so total length is proportional to $\Phi$, $L =\sqrt{\varrho^2 + \left(\frac{p}{2 \pi}\right)^2} \, \Phi$. From these relations follows that $\xi = \Delta z /L = \sin \psi $.
Inverting to get $\varrho$ and $p$ in terms of $L$, $\xi$, and $\Phi$, we get
\begin{equation}
\varrho = \frac{L}{\Phi} \cos \psi\,, \quad \frac{p}{2 \pi} = \frac{L}{\Phi} \sin \psi \,. 
\end{equation}
The FS basis adapted to the helix is
\begin{equation} \label{eq:FSframehelix}
\bft = \cos \psi \hat{\bm \varphi} + \sin \psi \hat{\bf z} \,, 
\qquad
\bfn = -\hat{\bm \varrho} \,, 
\qquad
\bfb = -\sin \psi \hat{\bm \varphi} + \cos \psi \hat{\bf z} \,;
\end{equation}
whereas the FS curvature and torsion are given by
\begin{equation} \label{eq:FSkappatau}
\kappa = \frac{\Phi}{L} \cos \psi \,, 
\quad 
\tau = \frac{\Phi}{L} \sin \psi \,.
\end{equation}
The sign of the torsion determines the chirality of the helix, $\psi>0$ ($\psi<0$) 
corresponds to right (left) handed helices. The degenerate cases $\psi \rightarrow 
0$ and $\psi \rightarrow \pi/2$ correspond to circles on the plane $X$-$Y$ and to vertical lines, respectively. For helices, the Darboux vector is along the helical axis $\bfD = \Phi/L \hat{\bf z}$.
\\
Since $\kappa$ and $\tau$ are constant and $\rmnz=0$, the EL Eq. (\ref{eq:ELN}) is satisfied if $\Lambda$ is constant, taking the value
\begin{equation} \label{eq:muhelix}
 \bar{\Lambda} = \frac{1}{2 L^2} \left[ (1-3 \xi^2) \, \Phi^2 - \xi^2 \, \gamma \right]\,,
\end{equation}
and the EL Eq. (\ref{eq:ELB}) vanishes identically. Hence, helices satisfying Eq. (\ref{eq:muhelix})  minimize the total energy $H$.
\vskip0pc \noindent
Using these expressions for $\kappa$, $\tau$, and  $\Lambda$ in Eq. (\ref{def:FBFM}) for $\bfF$, we find that the scaled force required to hold the helix is linear in the separation of the end points and directed along the helical axis,
\begin{equation} \label{eq:FMhelix}
L^2 \bar{\bfF} = (\gamma + \Phi^2) \xi \, \hat{\bf z}\,,
\end{equation}
The magnitude of the force is plotted for states $n=1,2$ as a function of $\xi$ and $\gamma$ in Fig. \ref{fig:Fn1n2helix}. In this plot the line $L^2 \bar{F} = \gamma + \Phi^2$ over $\xi=1$ ($\psi=\pi/2$) represents the scaled force required in the Euler buckling instability of a straight line, with the elastic term $\Phi^2=(2n\pi)^2$ four times larger as compared with the scaled force required in the planar case, $(n\pi)^2$. Like the case of planar curves, the force can be tensile or compressive depending on value of the magnetoelastic parameter relative to the total azimuthal angle: if $\gamma > -\Phi^2$ ($\gamma < -\Phi^2$) the magnitude of the axial force is positive (negative), $F>0$ ($F<0$), and the helix is under compression (tension). For circles with $\xi=0$ ($\psi=0$) and configurations with $\gamma = -\Phi^2$, there is no vertical force, $F=0$.
\\
The difference of the force between successive states $n$ and $n+1$,
\begin{equation}
 L^2 \Delta \bar{F} = (2 \pi)^2 (2 n +1 ) \xi\,,
\end{equation}
is independent of $\gamma$ and positive for any value of $\xi$, and it increases with $n$.
\vskip0pc \noindent
The torque vector has two components, one introduced by the magnetic field and another one of elastic character
\begin{equation} \label{eq:Mhelix}
L \, \bar{\bfM} =  \cos \psi \left( \sin \psi \frac{\gamma}{\Phi}  \, \hat{\bm \varphi} + 
\cos \psi \, \Phi \,  \hbfz \right)\,.
\end{equation}
The magnitude of the azimuthal torque is linear in the magnetoelastic parameter, so it vanishes for elastic curves with $\gamma=0$ and its direction is reversed when changing from regime $I$ to regime $II$. For a given value of $\gamma$ it vanishes for circles and lines with $\xi=0,1$ $(\psi=0,\pi/2)$, respectively, and is maximum for helices of maximum torsion with $\psi=\pi/4$. The magnitude of the axial torque is proportional to the total azimuthal angle and increases as the boundary points are approached, so it is maximal for circles and vanishing for vertical filaments.
\\
The total scaled energy of the helices is harmonic in the separation of the end points \cite{Dempster2017}
\begin{equation} \label{eq:Ehelix}
L \bar{H} = -\frac{1}{2} (\gamma + \Phi^2)\, \xi^2 +  \frac{\Phi^2}{2} \,.
\end{equation}
The scaled total energy of helical states with $n=1$ and $n=2$ are plotted in Figs. \ref{fig:FigHhelix}(a) and \ref{fig:FigHhelix}(b) as a function of $\xi$ and $\gamma$. Configurations with $H=0$ are shown with a solid black line, and the energies of elastic curves ($\gamma=0$) are shown with a dashed black line.
Like planar curves the scaled total energy of straight lines with $\xi =1$ is independent of $n$, $L \bar{H} = -\gamma /2$. By contrast, in the idealistic limit $\xi \rightarrow 0$, we would have an $n$ covering of a circle whose total energy is $L \bar{H} =\Phi^2/2$, so it scales with $n^2$. Thus, for a given $n$, in regime $I$, the total energy increases from negative values towards positive values as the pitch of the helices (or $\xi$) decreases, whereas in regime $II$, in general $H$ decreases monotonically with the pitch.
\\
The difference between the total energy of two successive states, $\Delta H_{n}= H_{n+1} -H_n$, reads
\begin{equation}
 L \Delta \bar{H}_n = 2 \pi ^2 (2 n+1) \left(1-\xi ^2\right) \,.
\end{equation}
It is independent of $\gamma$ and positive for any $n$, as exemplified for $n=1$ in Fig. \ref{fig:FigHhelix}(c). Since the total energy of excited states increases with $n$, we have that $n=1$ corresponds to the ground state. 
\vskip0pc \noindent
Note that the magnitude of axial force is given by the derivative of $H$ with respect to $\xi$, i.e. $F = -\partial H/ \partial \Delta z$. Moreover, $\partial^2 H/ \partial \xi^2 \propto -(\gamma + \Phi^2)$, so $\partial^2 H/ \partial \xi^2\geq 0$ if $\gamma \leq - \Phi^2 < 0$. This suggests that helices would be stable only in regime $II$ and if the absolute value of the magnetoelastic parameter is greater than the squared total azimuthal angle, in which case they are stress free or under tension. Although this criterion is not precise, in the next section, by comparing their energies, we argue that planar curves are unstable to decay into helices with same parameters (length, boundary separation, bending rigidity, and magnetic field) in a domain where the magnetoelastic parameter satisfies such inequality.

\section{Comparison of total energies} \label{sec:Hcomp}
\begin{figure}[htbp]
\centering
\begin{tabular}{cc}
\subfigure[]{\includegraphics[width=0.4\textwidth]{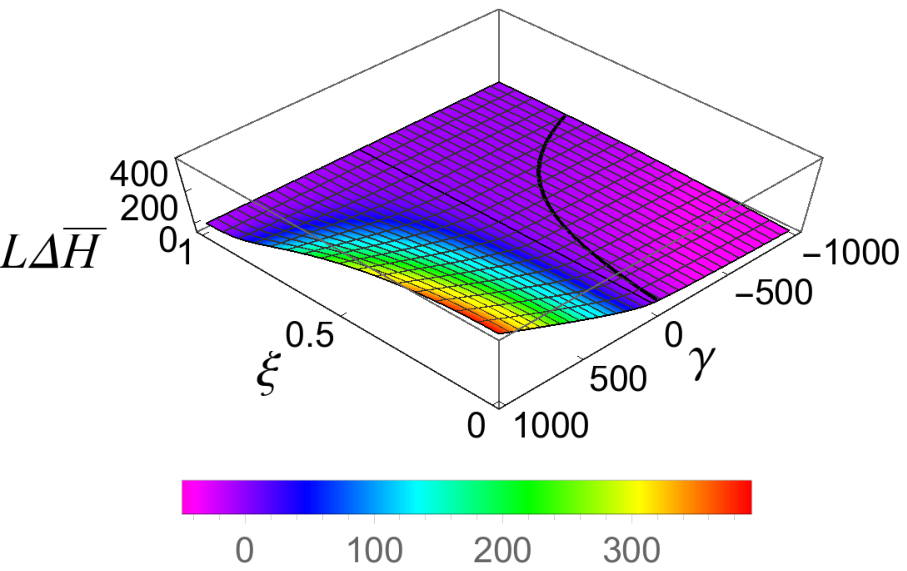}} & 
\subfigure[]{\includegraphics[width=0.35\textwidth]{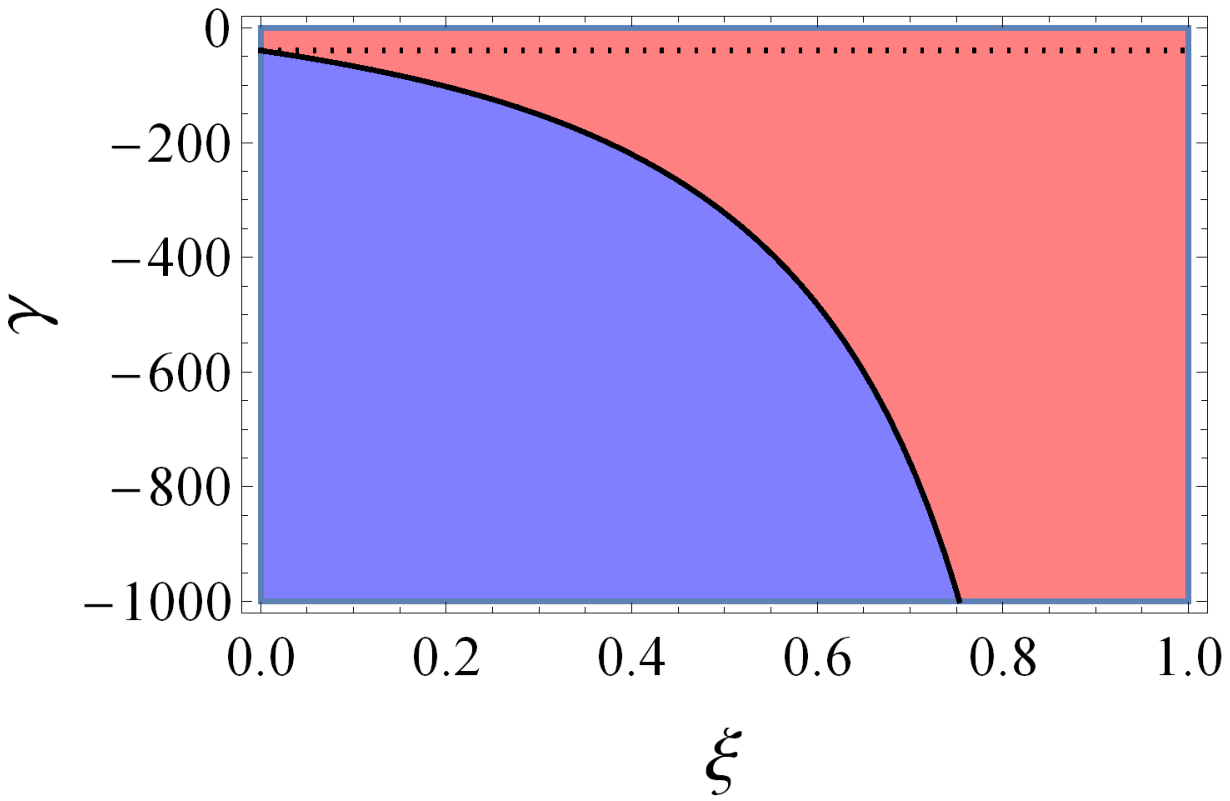}} 
\end{tabular}
\caption{(Color online) (a) Difference between the total energy of the helices and planar curves with $n=1$ and same values of $\xi$ and $\gamma$, the black line indicates points where $\Delta H = 0$ and the dotted line $\gamma = -\Phi^2$. (b) Zoom of the region where $\Delta H$ becomes negative, red regions correspond to $\Delta H>0$ (planar curves have lower energy) and blue regions correspond to $\Delta H<0$, (helices have lower energy).} \label{fig:FigdifHhlxpln}
\end{figure}

\begin{figure}[htbp]
\centering
\begin{tabular}{cc}
\subfigure[]{\includegraphics[height=5cm]{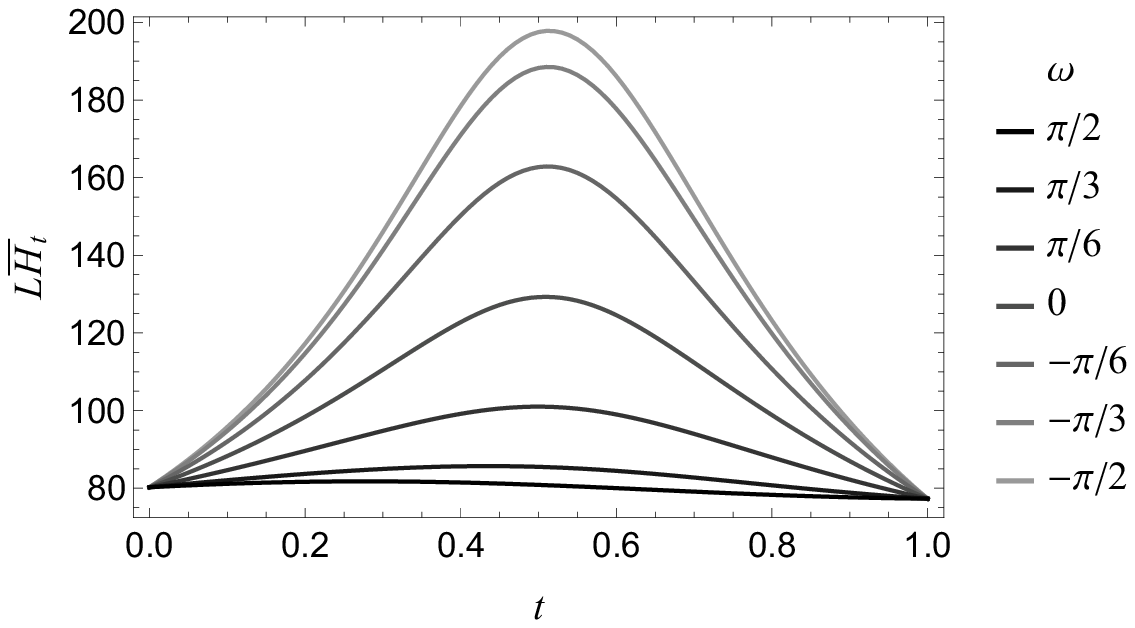}} & 
\subfigure[]{\includegraphics[height=5cm]{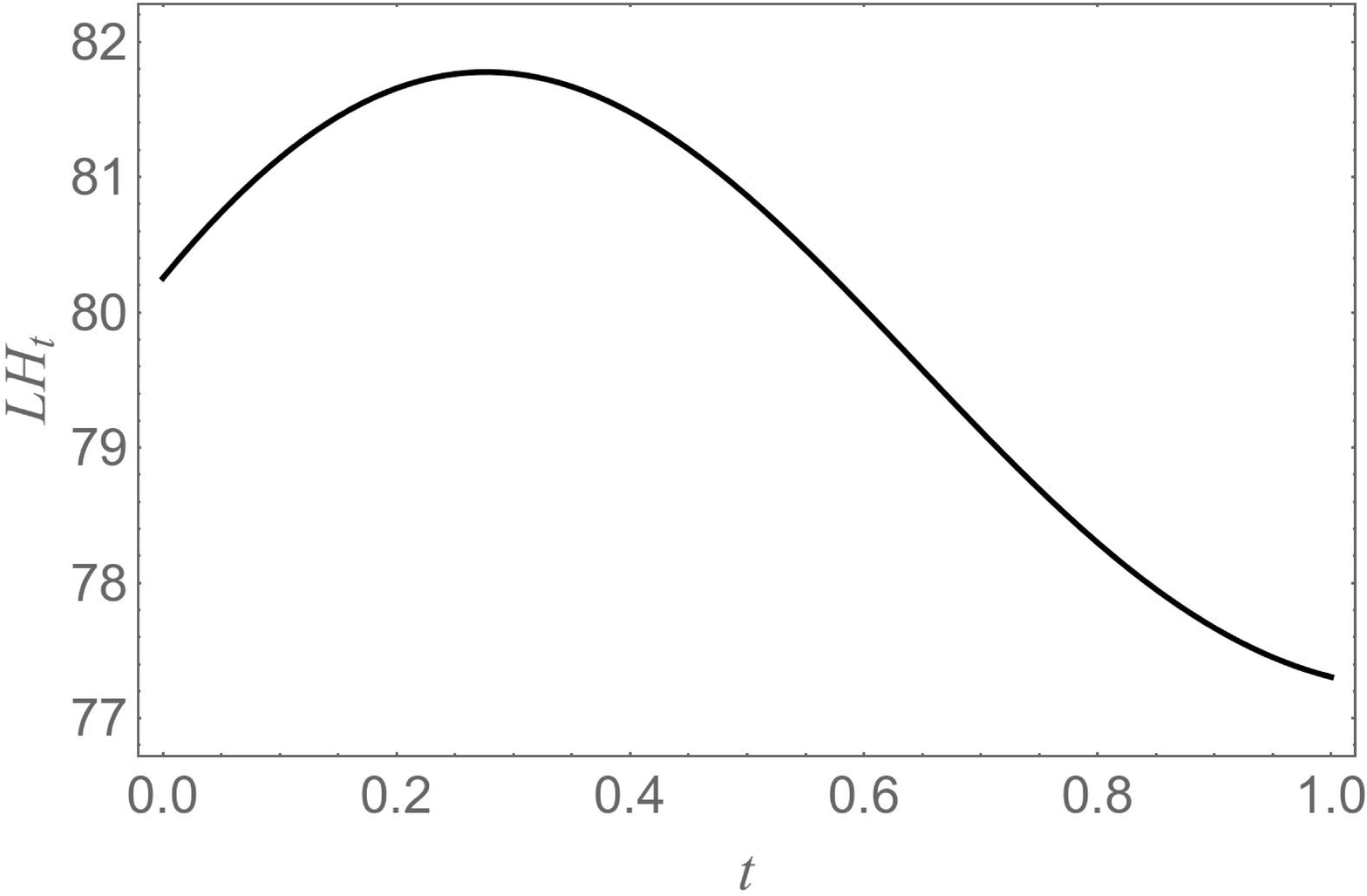}} 
\end{tabular}
\caption{(a) Scaled total energy as a function of $t$ of the sequence of homotopic configurations interpolating the planar and helical states with $n=1$, and parameter values $\xi=0.5$ and $\gamma=-500$. Sequence with $\omega = \pi/2$ ($-\pi/2$) exhibits the smallest (largest) amplitude (b) Zoom of the total energy of sequence with $\omega = \pi/2$.} \label{fig:Htothomotopy}
\end{figure}

\begin{figure}[htbp]
\begin{tabular}{cccccc}
$\vcenter{\hbox{\includegraphics[height=5cm]{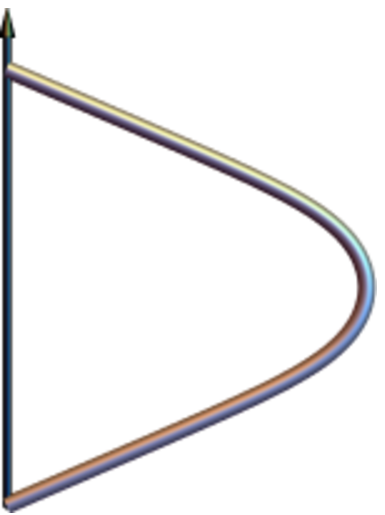}}}$ \hspace{0.5cm}&
$\vcenter{\hbox{\includegraphics[height=5cm]{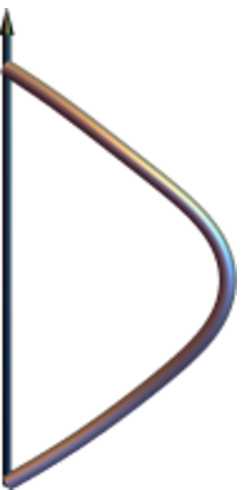}}}$ \hspace{0.4cm} &
$\vcenter{\hbox{\includegraphics[height=5cm]{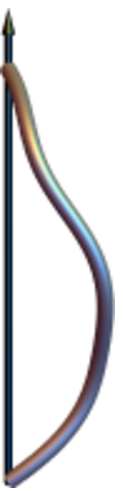}}}$ \hspace{0.8cm} &
$\vcenter{\hbox{\includegraphics[height=5cm]{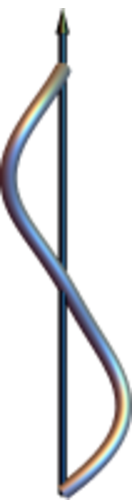}}}$ \hspace{0.5cm} &
$\vcenter{\hbox{\includegraphics[height=5cm]{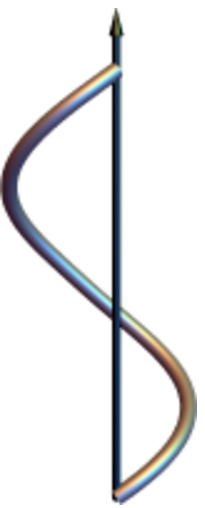}}}$ \hspace{0.5cm} &
$\vcenter{\hbox{\includegraphics[height=5cm]{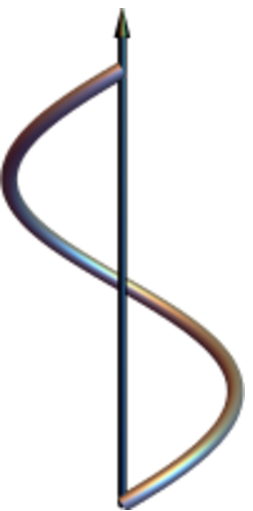}}}$ \\
\footnotesize{(a) $t=0$} & \footnotesize{(b) $t=0.2$} \hspace{0.5cm} & \footnotesize{(c) $t=0.4$} \hspace{0.8cm} & \footnotesize{(d) $t=0.6$} & \footnotesize{(e) $t=0.8$} & \footnotesize{(f) $t=1$}
\end{tabular}
 \caption{(Color online) Sequence of homotopic configurations starting with (a) the planar curve and ending with (f) the helix, both with $n=1$ and parameters $\xi = 0.5$ and $\gamma=-500$.} \label{fig:homotopy}
\end{figure}

\noindent
In order to compare the energies of planar curves and helices, we calculated the difference of their energies $\Delta H = H_{Helix} - H_{Planar}$ as a function of $\xi$ and $\gamma$, and plotted it in Fig. \ref{fig:FigdifHhlxpln}(a).
We observe that there is a region, satisfying the inequality $\gamma \leq -\Phi^2$, where $\Delta H<0$ [shown in blue in Fig. \ref{fig:FigdifHhlxpln}(b)], so the helices have lower energy and are energetically favorable to occur. The drop in the energy of helices below that of their planar counterparts may occur in two ways: the magnitude of the magnetic field should be high enough for weakly deformed filaments (ends not very close), whereas in weak magnetic fields it suffices to bring the ends of the filament close enough. In the other region where $\Delta H>0$ [shown in red in Fig. \ref{fig:FigdifHhlxpln}(b)], helices, in particular purely elastic ones with $\gamma =0$, have higher energy than their planar counterparts, so the latter ones would be realized.
\\
Although it is not viable to obtain an analytical expression of threshold separating the two regions, we can determine numerically the values of $\xi$ and $\gamma$ for which $\Delta H=0$, and fit them as a polynomial $\gamma = a_i \xi^i$, where the first 10 coefficients are given in Table \ref{Table1}.
\begin{table}
 \begin{tabular}{||c|c|c|c|c|c|c|c|c|c|c||}
 \hline \hline
 $i$ & 0 & 1 & 2 & 3 & 4 & 5 & 6 & 7 & 8 & 9 \\
 \hline
 $a_i$ & -39.4902 & -234.104  & -328.414 & -1027.81 & 11873  & -75436.9 & 230163 &  -385608 & 337094 & -123100 \\
\hline \hline
 \end{tabular}
 \caption{Coefficients of the magnetoelastic parameter polynomial in the ends separation $\xi$ for which the energy of the planar curves equals the energy of the helices.} \label{Table1}
\end{table}
\vskip0pc \noindent
To support our stability argument and to illustrate the decay of a planar curve into a helix, we now look at the homotopy connecting planar and helical states $n=1$ of the same length, boundary separation  and with a magnetoelastic parameter
\begin{equation}
\bfY_t = (1-t) \bfY_0  + t \mathrm{R}(\omega) (\bfY_1 + \rho \hat{\bfx}) \,,
\end{equation}
where the initial configuration $\bfY_0$ is the planar curve and the final state $\bfY_1$ is the helical state. We have translated the final state to make the ends of the two states coincide and also introduced a rotation $\mathrm{R}$ about the precession axis by an angle $\omega$ to account for the rotational freedom in the orientation of the final state relative to the initial state. Due to the symmetry with respect to the plane of the initial state the periodicity of the rotation gets reduced to $\pi$, so we choose $-\pi/2\leq \omega \leq \pi/2$. We find that, regardless of the values of $\xi$ and $\gamma$, the total energy $H_t$ of the sequence exhibits a barrier whose amplitude is maximized for $\omega = - \pi/2$ and minimized for $\omega = \pi/2$, as illustrated in Fig. \ref{fig:Htothomotopy}(a) for the case $\xi=0.5$ and $\gamma = -500$. The homotopic transition corresponding to these values and $\omega = \pi/2$ is shown in Fig. \ref{fig:homotopy}. Moreover, in general, we see that for the sequence of lowest energy, $\omega=\pi/2$, there is a small asymmetric barrier of the order $L \bar{H} \approx 1-10$, exemplified in Fig. \ref{fig:Htothomotopy}(b) for $\xi=0.5$ and $\gamma = -500$. 
This means that for a filament of bending rigidity $\calB \approx 10^{-25} \mathrm{Jm}$ and length $L \approx 100 \upmu \mathrm{m}$, we have the required energy is $H \approx 1 \, kT$, so it is plausible that under the effect of thermal fluctuations the planar state decays into the helical state.\footnote{On account of the asymmetry of the barrier, the converse transition would require approximately twice the energy.} Although this result is not a rigorous proof that the helical state is stable in this regime, it hints that it is the stable ground state.

\section{Discussion and conclusions} \label{sec:Conc}
\noindent
We have shown how the competition of the elasticity and magnetic properties of paramagnetic filaments can determine their  behavior. In particular, by controlling the precession angle, filaments lying on plane may behave as if they were pushed or pulled in the direction orthogonal to the precession axis, tending to get as close or as far as possible to or from the precession axis by changing the magnitude of the magnetic field. 
\\
We found that ground states for planar curves and helices correspond to filaments developing one undulation and one period respectively. Moreover, a stability analysis in the perturbative regime indicated that the planar ground state is the only stable state. Although excited states with a higher number of undulations are possible if the magnetoelastic parameter is increased, the presence of some external agent is required to render them stable (for instance, a surrounding polymer gel hindering local rotations of the filament \cite{Huang2016}), otherwise they would decay to the ground state. The same considerations might apply for helices in viscous medium, so it is conceivable that, for high values of the magnetoelastic parameter and under appropriate conditions, helical filaments with higher number of periods could be observed experimentally.
\\
The evaluation of the stability of planar states in the non-linear regime would be more involved, because deformations out of plane should be considered in the calculation of the second variation of the energy, which complicates the derivation of the differential operators associated with normal deformations, as well as the determination of their eigenvalues. Needless to say, a rigorous stability analysis of the transition from planar curves to helices in the non-linear regime would be a complex task. Despite of this fact, the comparison of the energies of planar curves and helices allowed us to identify the threshold separating the regions in the parameter space of scaled boundary separation and the magnetoelastic number where the occurrence of each family is propitious. Additionally, the calculation of the total energy of a homotopic sequence, representing a possible transition from planar to helical families, permitted us to estimate the energy required to drive such a transition, finding that it is of the order of a few $kT$, barrier easily surmounted under the influence of thermal fluctuations. Although this result is not a direct proof, in the sense that intermediate states in the homotopy are not equilibrium states, and that we cannot rule out the possibility that one intermediate state has lower energy than the planar or helical states, this bolsters our claim that each family will be realized in the regions of the parameter space we identified from their energetic comparison.
\\
Qualitatively speaking, we found that helical configurations may appear only for values of the magnetoelastic parameter smaller than the negative square of total azimuthal angle (which requires the precession angle to be greater than the critical angle). Moreover, the required values of the magnetoelastic parameter become more negative as the pitch is increased. On account of this, allowable helices only exert tensile forces. In contrast, planar curves may occur in both precession regimes, smaller or bigger than the critical precession angle, so they can exert contractile or tensile forces. Stress-free configurations of both families are possible for precession angles greater that the critical angle, when $\gamma=-(n\pi)^2$ for planar curves and $\gamma = - (2 n \pi)^2$ for helices. These features confer these filaments the ability of actuation, tunable though the parameters of the magnetic field.
\\
Although we have employed some idealizations, the model we have employed captures the main features of the superparamagnetic filaments and describes to a good approximation configurations obtained by molecular dynamics simulations. One could explore the implications of considering generalized models including additional degrees of freedom of the filament. For instance, in our model chirality of helices would occur with the same probability, but one could consider a twist degree of freedom along with an spontaneous twist in order to break the chiral symmetry \cite{Goriely2000, Belovs2009a, Belovs2009b}. Such helical filaments might be good candidates for swimmers with a given chirality, property capable of influencing their motility \cite{Tjhung2017}.
\\
One direction of future research would be the extension of this framework to the study of equilibrium conformations of membranes composed by magnetic beads. It will be interesting to probe how a precessing magnetic field influences the geometry of such magnetic membranes, as well as the stresses on them, as compared with their elastic analogues. 

\section*{Acknowledgments}
\noindent
We have benefited from conversations with Profs. Michael Cates and Jemal Guven. This work was supported by the Center for Bio-Inspired Energy Science (CBES), which is an Energy Frontier Research Center funded by the U.S. Department of Energy, Office of Science, Office of Basic Energy Sciences under Award No. DE-SC0000989.

\begin{appendix}

\section{Derivation of the magnetic energy density} \label{sec:MagEn}
\noindent
The interaction energy of two magnetic dipoles with dipole moments $\bm{\mu}_i$ and $\bm{\mu}_j$ and at positions $\bfx_i$ and $\bfx_j$ is \cite{GriffithsBook, JacksonBook}
\begin{equation} \label{eq:dipEn}
U_{ij} = \frac{\mu_0}{4 \pi \, |\bfx_{ij}|^3} \left(\bm{\mu}_i \cdot \bm{\mu}_j 
- 3 (\bm{\mu}_i \cdot \hat{\bfx}_{ij}) \, (\bm{\mu}_j \cdot \hat{\bfx}_{ij}) 
\right)\,, 
\quad 
\bfx_{ij} = \bfx_j - \bfx_i\,.
\end{equation}
We first consider a homogeneous strand of isotropic paramagnetic beads of radius $a$ and magnetic susceptibility $\chi$, connected by elastic linkers in an external magnetic field $\bfH = \mathrm{H} \hat{\bfH}$. In the regime of high temperature and magnetic saturation, the induced dipole moment of each bead is aligned with the magnetic field, so $\bm{\mu}_i = V \chi \bfH$, where $V$ is the volume of the bead, so its magnitude is $\mu_i = 4/3 \pi a^3 \chi H$ and its direction $\hat{\bm \mu}_i = \hat{\bfH}$. Moreover, the separation between the centers of neighboring beads is approximately equal, $\Delta \bfx = \bfx_{i+1} -\bfx_{i}$. The magnitude of the separation of the centers of the beads $\Delta l = | \Delta \bfx|$ is bounded from below by the diameter $\Delta l \geq 2a$.
\vskip0pc \noindent
The leading term of the energy per unit length of the strand, $u$, is given by the dipolar interaction \cite{Zhang1995}, which in this case reads
\begin{equation} \label{eq:diplinEnDen}
u = \frac{\mu_0}{4 \pi} \left(\frac{\mu}{\Delta l^2}\right)^2 \left(1 - 3 \left(\frac{\Delta \bfx}{\Delta l} \cdot \hat{\bm \mu}\right)^2  \right) \,.
\end{equation}
In order to describe the strand by a continuum model, we assume that the separation between the center of the beads is much smaller than the total length of the filament,\footnote{In experiments, the beads separation is two orders of magnitude smaller than the total length, $\Delta l \approx 1 \upmu \mathrm{m}$ and $L \approx 100 \upmu \mathrm{m}$ \cite{Biswal2003, Goubalt2003, Biswal2004}.} $\Delta l/L \ll 1$. In consequence, we can consider the beads separation as the line element along the strand, $\Delta l \rightarrow \rmd s$, so the limit curve $\Gamma$ passing through the center of the beads can be parametrized by arc length $s$ by the embedding $\Gamma: s \rightarrow \bfY(s) \in \mathbb{E}^3$. Thus, in this limit, we have $\frac{\Delta \bfx}{\Delta l} \rightarrow \bft := \frac{\rmd \bfY}{\rmd s}$, the unit tangent vector along $\Gamma$. Moreover, taking into account that the bead separation is bounded from below by the diameter, we have that of the magnitude of the dipole moment to the squared bead separation scales is bounded from above by the product of the radius of the beads and the magnitude of the magnetic field, $\mu/\Delta l^2 \leq \pi/3 a \chi \mathrm{H}$. Thus, although $\Delta l$ is very small, for constant $\mathrm{H}$, $\mu/\Delta l^2$ converges to some constant finite value (with units of electric current, i.e. amperes).
\vskip0pc \noindent
If the magnetic field is precessing about the $Z$ axis at an angle $\vartheta$ and with frequency $\omega$, then in the quasi-static regime (fast precession frequency) the magnetic dipole direction will be $\hat{\bm \mu} = (\sin \vartheta \, \cos \omega t , \sin \vartheta \, \sin \omega t, \cos \vartheta)$ and the tangent vector $\bft = (t^x,t^y,\rmtz)$ can be regarded as constant. In one period the average of the squared scalar product of $\hat{\bm \mu}$ and $\bft$ is\footnote{The time average is invariant under a change $\omega \rightarrow -\omega$, so the sense of the precession is immaterial.}
\begin{eqnarray} \label{eq:avmdotT}
\left\langle (\hat{\bm \mu} \cdot \bft)^2 \right\rangle &=& \frac{1}{2} \sin^2 \vartheta (t^x{}^2 + t^y{}^2) + \cos^2 \vartheta \, \rmtz{}^2 \nonumber\\
&=& \frac{1}{2} \left(\sin^2 \vartheta - (1-3 \cos^2 \vartheta)\rmtz{}^2\right)\,,
\end{eqnarray}
so the averaged linear energy density is
\begin{eqnarray} \label{eq:avlindensen}
\left \langle h_M \right\rangle &=& \frac{\mu_0}{4 \pi} \left(\frac{\mu}{\Delta l^2}\right)^2 \left(1 - \frac{3}{2} \left(\sin^2 \vartheta - (1-3 \cos^2 \vartheta)\rmtz{}^2\right)  \right) \nonumber \\
&=& -\frac{\mu_0}{8 \pi} \left(\frac{\mu}{\Delta l^2}\right)^2 (1- 3 \cos^2 \vartheta) (1- 3 \, \rmtz{}^2)\,.
\end{eqnarray}
Defining the magnetic modulus as in Eq. (\ref{def:magmodprec}), the energy density is given by
\begin{equation} \label{eq:lindensmagen}
h_M = \frac{\calM}{2} \left(\frac{1}{3} - \rmtz{}^2  \right)\,.
\end{equation}
The constant term of the magnetic dipolar energy adds to a constant $\lambda$ implementing the inextensibility of the filament, yielding an effective line tension
\begin{equation} \label{def:eflinten}
\Lambda = \lambda + \frac{\calM}{6}\,.
\end{equation}
Thus, we see that the magnetic field renormalizes the intrinsic line tension, and we end with the magnetic energy density defined in Eq. (\ref{eq:lindensen}).
\\
In the case of beads of anisotropic magnetic susceptibility, the magnetic energy density has approximately the same dependence on the projections of the tangent vector but the magnetic module gets modified. Consider a prolate spheroidal paramagnetic particle of long and short axis lengths $\mathrm{a}$ and $\mathrm{b}$, with magnetic susceptibilities $\chi_\parallel$ and $\chi_\perp$ along such directions. Since the magnetization of the particles and alignment with the magnetic field occur mainly along the long axis, the tangential direction of the filament lies approximately along such symmetry axis \cite{Martin2000, Tierno2014}. Thus the induced magnetic moment is now given by \cite{Tierno2009, Cimurs2013}
\begin{equation}
 \bm{\mu}_i = V \mathrm{H} \left(\chi_\perp \hat{\bm \mu} + \left(\chi_\parallel - \chi_\perp  \right) \left(\bft_i \cdot \hat{\bm \mu}\right) \bft_i \right)\,,
\end{equation}
where $V= 4\pi/3 \mathrm{a} \mathrm{b}^2$ is the volume of the particle. Moreover, the radius vector between the centers of two neighboring particles is
\begin{equation}
 \Delta \bfx = a (\bft_i + \bft_{i+1}) \,.
\end{equation}
Its magnitude is given by $\Delta l = \lvert \Delta \bfx\rvert = 2 \mathrm{a} \sin^2 \alpha/2$, with $\cos \alpha = \bft_i \cdot \bft_{i+1}$. Assuming $\alpha$ is small, we have $\bft_{i+1}  \approx \bft_i + \kappa \bfn_i \Delta l$, so $\Delta \hat{\bfx} \approx \bft_i + a \kappa \bfn_i$. Thus, at the lowest order, the normalized distance between the centers of neighboring particles is still given by the tangent vector $\Delta \bfx / \Delta l \approx \bft_i$. Taking this into account, we have that the anisotropic magnetic dipolar energy per unit length is
\begin{equation}
u = \frac{\mu_0}{4 \pi} \left(\frac{V\mathrm{H}}{\Delta l^2}\right)^2 \left( \chi_\perp^2 - \left(\chi_\perp^2 +2 \chi_\parallel^2\right) (\hat{\bm \mu} \cdot \bft)^2 \right)\,.
\end{equation}
Taking the time-average of this energy per unit length and using Eq. (\ref{eq:avmdotT}), we obtain the magnetic energy density
\begin{equation}
 \langle h_M \rangle  = \frac{\mu_0}{4 \pi} \left(\frac{V\mathrm{H}}{\Delta l^2}\right)^2 \left(\chi_\perp^2 - \frac{\chi_\perp^2 +2 \chi_\parallel^2}{2} \left(\sin^2 \vartheta -(1-3 \cos^2 \vartheta) \rmtz{}^2 \right)  \right)\,.
\end{equation}
Therefore, in this case the magnetic modulus is defined by
\begin{equation}
 \calM = \frac{3 \mu_0}{4 \pi} \left(\frac{V\mathrm{H}}{\Delta l^2}\right)^2 \left(\chi_\perp^2 + 2 \chi_\parallel \right) \left( \cos^2 \vartheta -\frac{1}{3}\right)\,,
\end{equation}
and the constant terms contribute to the effective line tension
\begin{equation}
 \Lambda = \lambda + \frac{\calM}{3 \cos^2 \vartheta -1} \left( \frac{\chi_\perp^2}{\chi_\perp^2 + 2 \chi_\parallel } - \frac{\sin^2 \vartheta}{2}\right)\,,
\end{equation}
so to lowest order the magnetic energy density is given by Eq. (\ref{eq:lindensen}). These expressions reduce to the isotropic case for which $\chi_\perp = \chi_\parallel = \chi$.

\section{Hamiltonian formalism for planar curves} \label{sec:Hamform}
\noindent
The Lagrangian density for a planar magnetic filament reads
\begin{equation} \label{eq:LPlanar}
\mathscr{L} = \frac{\calB}{2}  \Theta'{}^2 - \frac{\calM}{2} \cos^2 \Theta  + F_y \left(\sin \Theta - y' \right) + F_z \left(\cos \Theta - z' \right)  \,,
\end{equation}
where the first two terms represent the bending and magnetic energy densities, whereas the last two terms implement the definition of the tangent vector in terms of the angle $\Theta$ as the derivative of the coordinates $y$ and $z$. Regarding $\Theta$, $y$ and $z$ as generalized coordinates, the conjugate momenta $P_i = \partial \mathscr{L} /\partial q^i{}'$, are
\begin{equation} \label{def:Pi}
P_\Theta =  \calB \Theta'\,, \quad P_y  = - F_y \,, \quad P_z = - F_z\,.
\end{equation}
We see that the conjugate momenta to the coordinates correspond to the force on the curve . The Hamiltonian density $\mathscr{H} = q^i{}' P_i  -\mathscr{L}$, is 
\begin{equation} \label{eq:Hamiltonian}
\mathscr{H} = \frac{P_\Theta^2}{2 \calB} + \frac{\calM}{2} \cos^2 \Theta  + P_y \sin \Theta + P_z \cos \Theta \,.
\end{equation}
Since there is not explicit dependence of $s$ in the Lagrangian, the Hamiltonian is constant. Identifying $\Lambda = \mathscr{H}$, we see that Eq. (\ref{eq:Hamiltonian}) corresponds to the second integral (\ref{eq:quadpsi}), whereas the equation of motion for $\Theta$,
\begin{equation} \label{eq:EqmotHam}
P_\Theta '= - \frac{\partial \mathscr{H}}{\partial \Theta} =   \frac{\calM}{2} \sin 2 \Theta - P_y \cos \Theta + P_z \sin 
\Theta\,,
\end{equation}
corresponds to the first integral (\ref{eq:motion}).

\section{Second variation of the energy of planar curves} \label{Sect:app2ndvar}
\noindent
Under a deformation of the curve $\bfY \rightarrow \bfY + \delta \bfY$, the second order variation of the energy of planar curves, required to analyze their stability, is given by
\begin{equation} \label{eq:delta2H}
\delta^2 H = \int \rmd s \delta \phi \delta \varepsilon_\bfn + \int \rmd s \delta (\delta Q')\,,
\end{equation}
where $\varepsilon_\bfn$ is given by Eq. (\ref{eq:ELPlanar}), so\footnote{The deformation of the curve does not affect the magnetic modulus, $\delta \calM = 0$, but the constant $\Lambda$ fixing length might have a first-order variation.}
\begin{equation} \label{eq:deltaELn}
 \delta \varepsilon_\bfn = \delta (\kappa'') + \left( \frac{3}{2} \kappa^2 -\bar{\calM} \left(\frac{\rmtz{}^2}{2} - \rmnz{}^2 \right) -\bar{\Lambda} \right) \delta \kappa - \bar{\calM} \kappa \left(\rmtz \delta \rmtz -2 \, \rmnz \delta \rmnz \right) -\kappa \delta \bar{\Lambda}\,.
\end{equation}
We need to calculate how the FS basis and the curvature change. First, we decompose the deformation in tangent and normal components as $\delta \bfY = \delta \psi \, \bft + \delta \phi \, \bfn $. The inextensibility of the filament implies that the variation and differentiation with respect to arc length commute, so 
\begin{equation}
 \delta \bft = (\delta \bfY)' = \left( \delta \phi' + \kappa \delta \psi \right) \bfn\,.
\end{equation}
In this calculation we used the FS equations and the fact that $\bft$ is a unit vector, so $\delta \bft \cdot \bft = 0$, from which follows that the derivative of the tangential component to the normal component are related: 
\begin{equation} \label{eq:isomcond}
\delta \psi ' = \kappa \delta \phi\,.
\end{equation}
Also, the variation of the normal vector follows from the orthogonality of the FS basis, $\delta \bfn = - (\bfn \cdot \delta \bft) \bfn$. Thus the required projections are
\begin{equation} \label{eq:deltatznz}
 \delta \rmtz =  \left( \delta \phi' + \kappa \delta \psi \right) \rmnz\,, \qquad \delta \rmnz = - \left( \delta \phi' + \kappa \delta \psi \right) \rmtz\,.
\end{equation}
The variation of the curvature is obtained from the variation of the FS equations, $\delta (\bft') = \delta \kappa \bfn + \kappa \delta \bfn$, so 
\begin{equation} \label{eq:deltakappa}
\delta \kappa = \bfn \cdot (\delta \bft)' =  \left( \delta \phi' + \kappa \delta \psi \right) ' = \delta \phi '' + \kappa^2 \delta \phi + \kappa' \delta \psi\,,
\end{equation}
where in the second identity we used the relation (\ref{eq:isomcond}). Likewise, its second derivative is
\begin{equation} \label{eq:deltakappapp}
\delta \kappa'' = \delta \phi^{(4)} + \kappa^2 \delta \phi'' + \frac{5}{2} (\kappa^2)' \delta \phi' + \left(\frac{3}{2} (\kappa^2)'' + \kappa \kappa''\right) \delta \phi  + \kappa''' \delta \psi\,.
\end{equation}
Using the EL derivative, Eq. (\ref{eq:ELPlanar}), and its arc length derivative in this expression, it is possible to replace $\kappa''$ and $\kappa'''$ in favor of $\kappa$. Substituting Eqs. (\ref{eq:deltatznz})-(\ref{eq:deltakappapp}), Eq. (\ref{eq:deltaELn}) can be recast in terms of the components of the deformation
\begin{equation} \label{eq:deltaELnsimp}
 \delta \varepsilon_\bfn = \delta \phi^{(4)} + \mathcal{V} \delta \phi '' + \mathcal{V}' \delta \phi '+  \left(\mathcal{W} + 4 \kappa  \varepsilon_\bfn \right) \delta \phi + \varepsilon _\bfn' \, \delta \psi - \kappa \delta \bar{\Lambda}\,,
\end{equation}
where
\begin{subequations} \label{eq:defVW}
\begin{eqnarray}
\mathcal{V} &=& \frac{5}{2} \kappa^2 - \bar{\calM} \left(\frac{\rmtz{}^2}{2}-\rmnz{}^2\right) -\bar{\Lambda}\,, \\
\mathcal{W} &=& 3 \kappa'{}^2-\frac{\kappa^4}{2} + 3 \kappa^2 \left( \bar{\calM} \left(\frac{\rmtz{}^2}{2}-\rmnz{}^2\right) + \bar{\Lambda}\right) \,.
\end{eqnarray}
\end{subequations}
By virtue of the isometry condition (\ref{eq:isomcond}), the last term in Eq. (\ref{eq:deltaELnsimp}) multiplied by $\delta \phi$ can be recast as a total derivative, $\delta \phi \kappa \delta \Lambda = (\delta \Lambda \delta \psi)'$, so it only contributes to boundary terms, which vanish if the boundaries are fixed (or if the curve is closed or periodic).
\\
Hence, in equilibrium the second variation of the energy, Eq. (\ref{eq:delta2H}), can be recast in terms of the normal deformation\footnote{The commutation of variation and differentiation implies that the second term in Eq. (\ref{eq:delta2H}) is again a total derivative, so it does not contribute to the variation in the bulk.}
\begin{equation} \label{eq:2ndvar}
\delta^2 \bar{H} = \int \rmd s \, \delta \phi \mathcal{L} \delta \phi \,,
\end{equation}
where $\mathcal{L}$ is a fourth order self-adjoint differential operator defined by
\begin{equation} \label{eq:defLdiff}
\mathcal{L} = \frac{\partial^4}{\partial s^4} + \frac{\partial}{\partial s} \mathcal{V} 
\frac{\partial}{\partial s} + \mathcal{W}\,.
\end{equation}
The normal deformation modes $\delta \phi$ can be spanned in terms of the eigenfunctions of the operator $\mathcal{L}$, so for a given deformation mode $\delta \phi_m$ satisfying $\mathcal{L} \delta \phi_m = e_m \delta \phi_m$, the corresponding second order variation of the energy is $\delta^2 H = e_m \int \rmd s \delta \phi_m{}^2$. Thus the sign of the eigenvalues $e_m$ determine the stability of the equilibrium configurations, if positive (negative) the deformation mode increases (decreases) the energy,\footnote{Recall that in equilibrium the first order change in the energy, proportional to $\varepsilon_\bfn$, vanishes.} indicating they are stable (unstable).

\section{Vertical Planar Euler elastica} \label{sec:vertEulerElastica}
\begin{figure}[htbp]
\centering
\begin{tabular}{cc}
\subfigure[]{\includegraphics[width=0.49\textwidth]{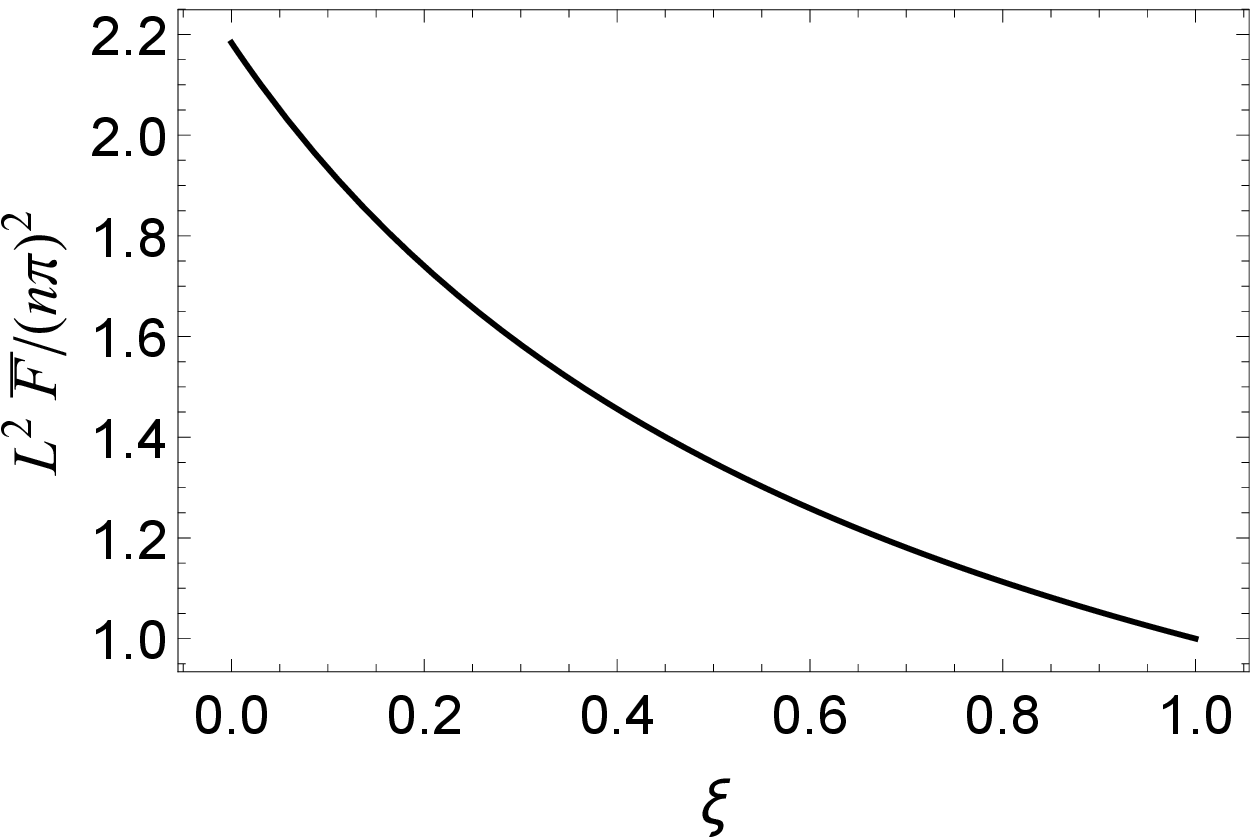}} &
\subfigure[]{\includegraphics[width=0.49\textwidth]{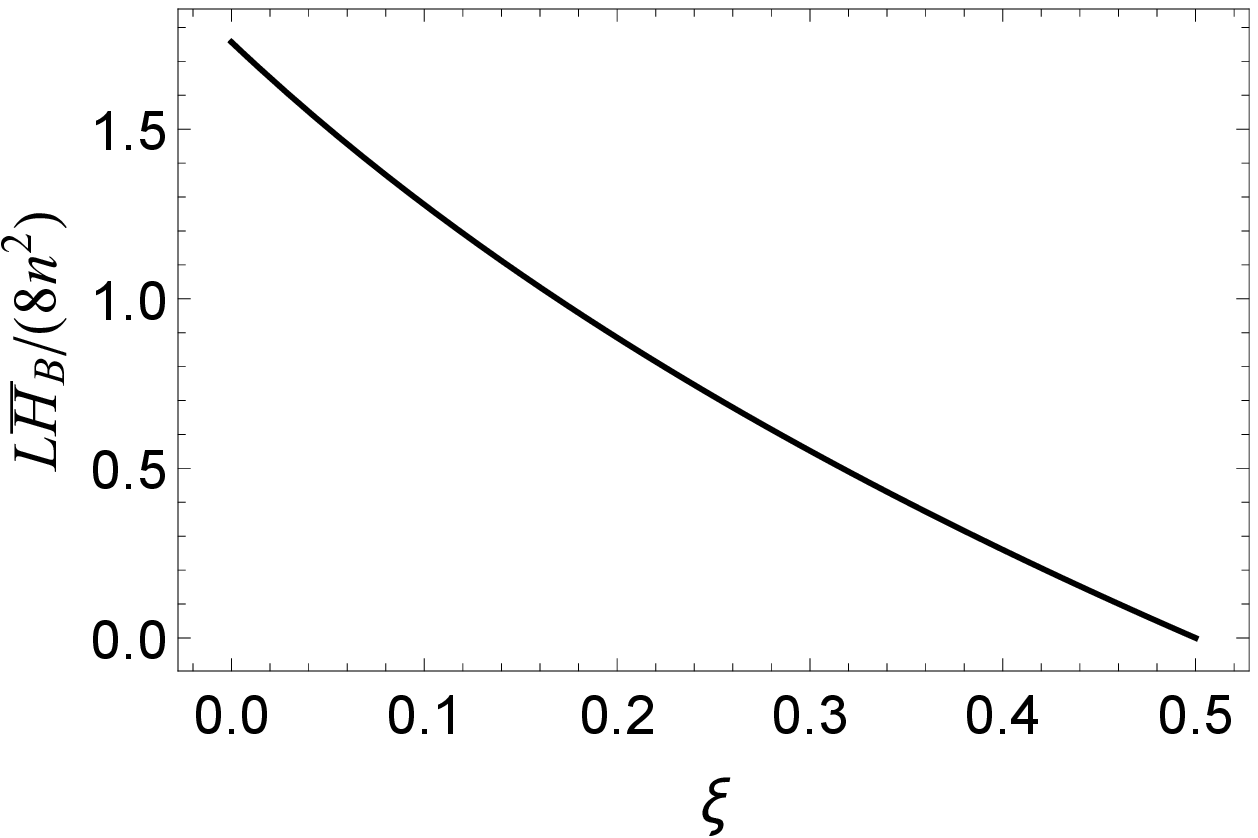}}
\end{tabular}
\caption{Force and total bending energy of planar Euler elastica as a function 
of the separation of the ends, $\xi$, both increase monotonically as $\xi$ 
decreases.} 
\label{fig:FigFHPEE}
\end{figure}

\noindent
For $\vartheta = \vartheta_m$, the magnetic modulus vanishes, $\calM = 0$, so in this regime the filaments behave as the classical Euler elastica. We consider the compression of a vertical rod lying along the $Z$ axis, with edges at $z=\pm z_b$. We are interested in solutions oscillating symmetrically about the vertical ($\Theta=0$), so the potential $V$ has to be convex and $F>0$ (compression). Thus, the potential reduces to $V=-\bar{F} \cos \Theta$, which has a global minimum at $\Theta = 0$ and maxima at $\Theta = \pm \pi$. The maximum angle the curve develops, occurring at inflexion points with 
$\kappa = 0$. Defining $v = \cos \Theta$ and $v_M = \cos \Theta_M$, the quadrature 
can be written as the elliptic integral
\begin{equation} \label{eq:quadvertPEE}
\int_{v}^1 \frac{\rmd \mathrm{v}}{\sqrt{(1-\mathrm{v}^2) (\mathrm{v} - 
\mathrm{v}_M)}} = \sqrt{2 \bar{F}} \int_0^s \rmd S\,, \quad 1>v>v_M>-1\,,
\end{equation}
which after integrating\footnote{Using for instance formulas (2.131) ($5$)-($6$) of 
Ref. \cite{Gradshteyn2007}.} and solving for $y'=\sqrt{1-v^2}$ 
and $z'=v$ we get
\begin{equation} \label{eq:ypzpvertPEE}
y' = 2 \sqrt{m} \, \sn(q (s-s_0)|m) \, \dn(q (s-s_0)|m) \,, \quad z' = 2 \, 
\dn^2(q (s-s_0)| m) -1 \,,
\end{equation}
where the wave number and parameter are given by the square root of the scaled 
magnitude of the force and the squared sine of half the maximum angle:
\begin{equation} \label{eq:qmEulerElast}
q = \sqrt{\bar{F}} \,, \quad m =\sin^2 \frac{\Theta_M}{2}\,.
\end{equation}
The curvature is $\kappa = -\Theta' = -2 \sqrt{m} \, q \, \cn(q (s-s_0)| m)$. 
Integrating once 
more we get
\begin{equation}  \label{yzvertPEE}
y(s) = -\frac{2}{q} \, \sqrt{m} \, \cn(q (s-s_0)|m) + y_0\,, \quad z(s) =  
\frac{2}{q}\,\mathrm{E}(\am(q (s-s_0)|m)| m) - s +z_0\ \,, \\
\end{equation}
$q$ and $m$ are determined by specifying the total length of the curve $L$ and 
the separation between the two edges.
If the filament develops $n$ oscillations, each oscillation of length $L/n$ 
corresponds to half period of the Jacobi elliptic functions, which is $2 
\rmK(m)$, then we have
\begin{equation} \label{def:qs0vertPEE}
q = \frac{2 n \rmK (m)}{L}\,, 
\end{equation}
where $\mathrm{K}[m]$ is the complete elliptic function of the first kind. This 
relation determines the wave number in terms of $L$ and $m$. 
From the boundary conditions requiring that the curvature vanished at the end 
points at $s=\pm s_b$, $s_b =L/2$, we obtain $s_0 = -\mathrm{mod}(n-1,2) L/(2 n)$, $y_0 = 0$ and $z_0 = -2\mathrm{mod}(n-1,2) \mathrm{E}(m)/q$, where $\mathrm{mod}(a,b)$ stands for $a$ modulo $b$. Using these results, and imposing the condition  $z\left(\pm s_b \right) =\pm z_b$ in expressions (\ref{yzvertPEE}) for $y$, we obtain the following equation for the ratio $\xi = 2 z_b/L \leq 1$ ($2 z_b \leq L$):
\begin{equation} \label{eq:halfyb}
\xi = 2 \frac{\mathrm{E}(m)}{\mathrm{K}(m)} - 1\,, 
\end{equation}
This equation allow us to numerically determine $m$ for a given $\xi$. States with $n=1,2$ are shown for different values of $\xi$ in Figs. \ref{fig:VertFilsn1Mpos}-\ref{fig:VertFilsn2Mneg} with a dashed line. $\xi = 1$ ($m=0$) corresponds to a vertical line, and as $\xi$ decreases the curves buckle, developing a horizontal tangent at $\xi = 0.4568$ ($m=1/2$). 
At $\xi = 0.1947$ ($m=0.7013$) the curve reaches it maximum height $z=0.4031$ 
and it touches itself at $\xi=0$ ($m=0.8261$). 
\\
The magnitude of the scaled force, given by the squared wave number $\bar{F} = q^2$, is shown in Fig. \ref{fig:FigFHPEE}(a) [also with dashed lines in Figs. \ref{fig:Fn1n2}(a) and \ref{fig:Fn1n2}(b)],  where we see that there is an initial force required to drive an Euler buckling instability $\bar{F}_0 = (n \pi/L)^2$,\footnote{This agrees with the perturbative analysis about the vertical line.} after which the force increases monotonically as the edges are brought together ($\xi \rightarrow 0$). 
\\
The total bending energy of the curves is proportional to $n^2$,
\begin{equation} \label{eq:EPEE}
\bar{H} = \frac{8 n^2}{L} \, \mathrm{K}[m] \left(\mathrm{E}[m] + (m-1) 
\mathrm{K}[m]\right)\,.
\end{equation}
$H$ is plotted in Fig. \ref{fig:FigFHPEE}(b) [also with dashed lines in Figs. \ref{fig:FigHTn1n2}(a) and \ref{fig:FigHTn1n2}(b)], where we observe that it increases monotonically as the ends are joined. Expression (\ref{eq:EPEE}) for the energy can be written as $L\bar{H} = (q L)^2 \left(\xi + 2 m  -1 \right)$, so it is proportional to the magnitude of the force.\footnote{It can be checked that the force is given by the derivative of the energy with respect to the edges separation, i.e. $F = \partial H_B /\partial \Delta z$.}

\section{Jacobi Integrals for vertical filaments} \label{sec:vertEllipInts}
\begin{figure}[htbp]
\centering
\begin{tabular}{cc}
\subfigure[$\calM >0$]{\includegraphics[height=5cm]{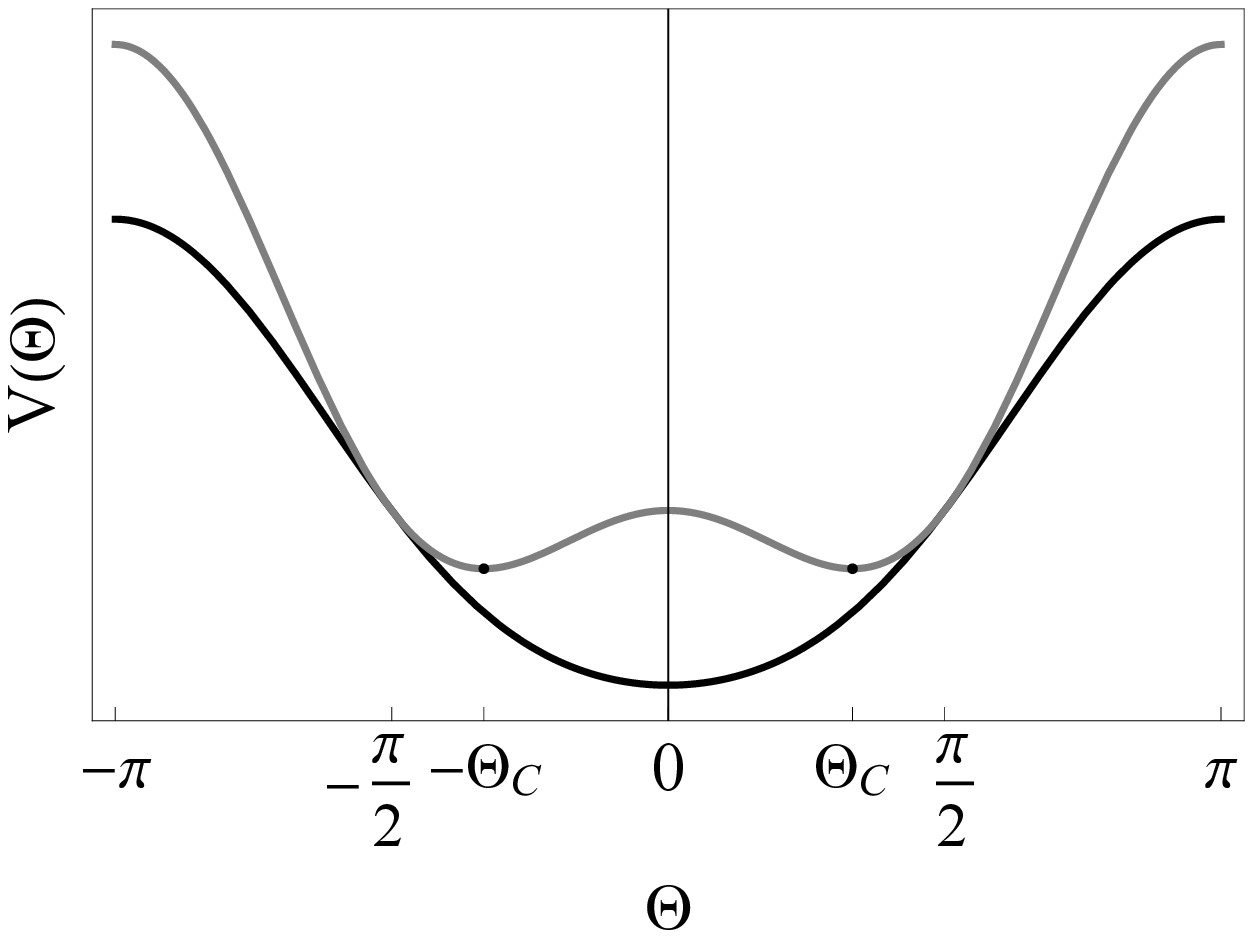}} &
\subfigure[$\calM <0$]{\includegraphics[height=5cm]{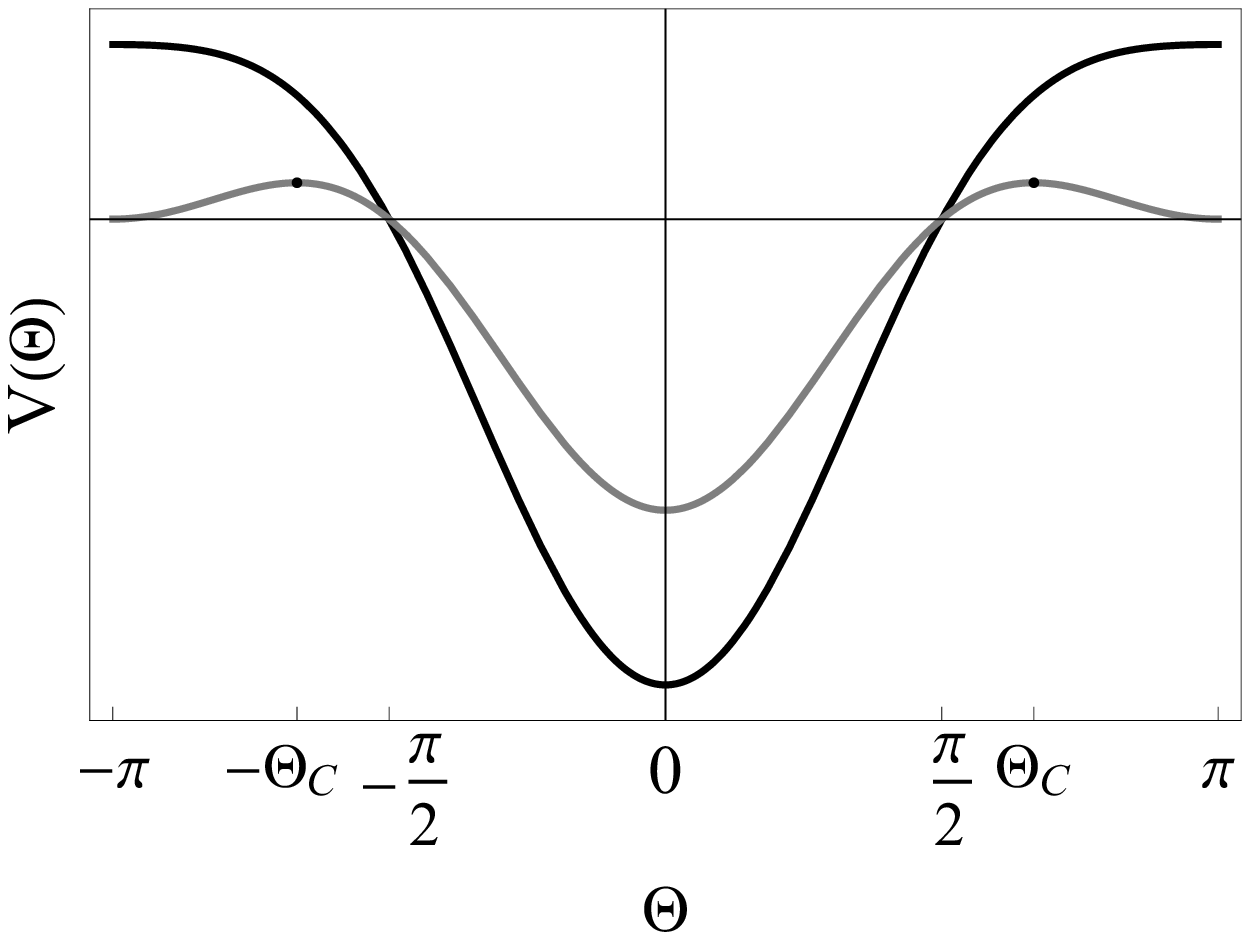}}
\end{tabular}
\caption{Potential for $\calM$ (a) positive and (b) negative. For $\chi>1$, there is only one minimum (black lines) and for $\chi<1$ two additional critical points are developed (black dots on gray lines).} \label{fig:FigVbeta}
\end{figure}

\noindent
In order to integrate the quadrature (\ref{eq:quadroots}) in the non-linear regime we need to determine how the  critical points of the potential $V$ depend on the ratio $\chi = F/\calM $. As found in the perturbative analysis, in regime $I$, we have that only for $0< \chi < \infty$ (compression) $V$ is a convex function. For $0 <\chi< 1$, $V$ possesses five critical points, two global maxima at $\Theta = \pm \pi$, one local maximum at $\Theta = 0$ and two local minima at $\Theta=\pm \Theta_C$, with $\Theta_C = \arccos \chi$ [black dots on gray line in Fig \ref{fig:FigVbeta}(a)], and since we consider filament with up-down symmetry and its end points aligned, in this case we choose $\Lambda > V(0)$.\footnote{If $\Lambda < V(0)$, then $\Theta$ oscillates only on one well of the potential with a definite sign, and in consequence the boundary points will not be aligned along the $Z$ axis.} If $\chi\geq1$, $V$ possesses only two global maxima at $\Theta = \pm \pi$ and one global minimum at $\Theta =0$ (see black curve in Fig. \ref{fig:FigVbeta}(a)).
\\
In regime $II$, $V$ is a convex function if $-\infty < \chi < 1$. For $|\chi| < 1$ (tension or compression) the two maxima of $V$ are located at $\Theta_C = \pm \arccos \chi$ [black dots on gray line in Fig. \ref{fig:FigVbeta}(b)], while for $\chi \leq -1$ (compression) the maxima occur at $\Theta = \pm \pi$ [see black curve in Fig. \ref{fig:FigVbeta}(b)]; in both cases there is global minimum at $\Theta=0$. 
\vskip0pc \noindent
With these considerations, we define the constants $a_I = -a_{II} = 2 \chi - b_I$ and $b_I = b_{II} = \cos \Theta_M$, such that under the change of variable $v = \cos \Theta$ ($v_M = b$) the quadrature (\ref{eq:quadroots}) can be expressed as an elliptic integral in each regime:
\begin{subequations}
\begin{align} \label{eq:ellintvertMpos}
\calM > 0: & \qquad \int_v^1 \frac{\rmd \mathrm{v}}{\sqrt{(a_I-\mathrm{v})(1-\mathrm{v})(\mathrm{v}-b_I)(\mathrm{v}+1)}} = \frac{(s-s_0)}{\ell} \,, \qquad 
a_I >1>v>b_I>-1 \,,\\ 
\calM < 0: & \qquad \int_v^1 \frac{\rmd \mathrm{v}}{\sqrt{(1-\mathrm{v})(\mathrm{v}-b_{II})(\mathrm{v}+1) (\mathrm{v}+a_{II})}} = \frac{(s-s_0)}{\ell} \,, \qquad 
1>v>b_{II}>-1>-a_{II} \,.
\end{align}
\end{subequations}
Integrating we get $\rmF(\alpha|m) = q (s-s_0)$, where $\rmF(u,m)$ is the incomplete elliptic function of the first kind \cite{Abramowitz1974, Gradshteyn2007}. The amplitude $\alpha$, the parameter $m$, and wave number $q$ are given by
\begin{subequations}
\begin{align} \label{eq:alphamqvert}
\calM > 0: \quad \sin \alpha &= \sqrt{\frac{(a_I-b_I)(1-v)}{(1-b_I) (a_I-v)}}\,, \quad m = \frac{(1+a_I)(1-b_I)}{2 (a_I-b_I)}\,, \quad q \ell = \sqrt{ \frac{a_I-b_I}{2}}\,,\\
\calM < 0: \quad \sin \alpha &= \sqrt{\frac{(a_{II}+b_{II})(1-v)}{(1-b_{II}) (a_{II}+v)}}\,, \quad m = \frac{(a_{II}-1)(1-b_{II})}{2 (a_{II}+b_{II})}\,, \quad q \ell = \sqrt{\frac{(a_{II}+b_{II})}{2}}\,.
\end{align}
\end{subequations}
Inequalities for $a_i$, $i=I,II$ and $b$ imply that $m<1$. $a$ and $b$ can be expressed in terms of $q$ and $m$\footnote{The second  pair of solutions with minus the squared root does not satisfy the inequality $a>1>b$, so is not considered.}:
\begin{subequations}
\begin{align} \label{eq:abABvert}
a_I &= \left(q \ell\right)^2 \pm \sqrt{\left(\left(q \ell\right)^2+1\right)^2 -4 m (q \ell)^2}\,, &\quad b_I &= -\left(q \ell\right)^2 \pm \sqrt{\left(\left(q \ell\right)^2+1\right)^2 -4 m (q \ell)^2} \,,\\
a_{II} &= \left(q \ell\right)^2 \pm \sqrt{\left(\left(q \ell\right)^2-1\right)^2 + 4 m (q \ell)^2}\,, &\quad b_{II} & = \left(q \ell\right)^2 \mp \sqrt{\left(\left(q \ell\right)^2-1\right)^2 + 4 m (q \ell)^2}\,.
\end{align}
\end{subequations}
Solving Eq. (\ref{eq:alphamqvert}) for $v=z'$ and $\sqrt{1-v^2} = y'$, we get
\begin{subequations} \label{eq:ypzpevodvert}
\begin{align}
y_I' &= \frac{\sqrt{2 \eta_I (a_I-1)} \,\sn(q(s-s_0)|m) \, \dn(q(s-s_0)|m) }{1- \eta_I \, \sn^2(q(s-s_0)|m)}  \,, &\quad
z_{I}' &= -\frac{a_I-1}{1- \eta_{I} \, \sn^2 (q (s-s_0)|m)}+a_I\,,\\
y_{II}' &= \frac{\sqrt{2 \eta_I (a_{II}+1)} \,\sn(q(s-s_0)|m) \, \dn(q(s-s_0)|m) }{1+ \eta_I \, \sn^2(q(s-s_0)|m)}   \,, &\quad
z_{II}' &= \frac{a_{II}+1}{1 + \eta_{II} \, \sn^2 (q (s-s_0)|m)} -a_{II}\,.
\end{align}
\end{subequations}
where $\eta_I$, $i=I,II$ are defined in Eq. (\ref{def:aeta}). Finally, integrating again Eqs. (\ref{eq:ypzpevodvert}) we obtain expressions for the coordinates, Eq. (\ref{eq:yzhor}).\footnote{Alternatively, expressions of the coordinates in terms of $\Theta$ could be obtained by combining the quadrature (\ref{eq:quadpsi}) with the expressions (\ref{eq:ypzpvert}) for the arc length derivatives of the coordinates. From them we obtain
\begin{equation} \label{def:intsyz}
y-y_0 = \int\limits_0^\Theta \frac{\rmd \theta \sin \theta}{\sqrt{2 \left(V(\theta_M)-V(\theta)\right)}}\,, \quad  z-z_0 = \int\limits_0^\Theta \frac{\rmd \theta \cos \theta}{\sqrt{2 \left(V(\theta_M)-V(\theta)\right)}}\,,
\end{equation}
However, to reproduce expressions (\ref{eq:yzhor}) in terms of $s$, one has to use expressions (\ref{eq:ypzpevodvert}) for $\cos \Theta= z'(s)$ after integration of Eq. \ref{def:intsyz}.
}
Expressions (\ref{eq:kappaplanar}) of the curvature are obtained by using Eq. (\ref{eq:ypzpevodvert}) in the identity $y'' = - \kappa z'$.

\subsection{Energy}
\noindent
In each case, the total bending and magnetic energy are
\begin{equation} \label{eq:totbenmagenint}
\bar{H}_B = \frac{1}{2}\int\limits_{-L/2}^{L/2} \rmd s \kappa^2  =   \bar{F} \int\limits_{-L/2}^{L/2} \rmd s \, \rmtz  + \bar{H}_M + \bar{\Lambda} L\,, \qquad
\bar{H}_M = - \frac{\bar{\calM}}{2}\int\limits_{-L/2}^{L/2} \rmd s \rmtz{}^2 \,.
\end{equation}
Since $\rmtz= z'$, the first integral is just give the separation of the boundaries $\int\limits_{-L/2}^{L/2} \rmd s \, \rmtz = 2 z_b = \xi L$.
Using expressions (\ref{eq:ypzpevodvert}) for $z'$, along with relations (\ref{eq:nlreqm}) for $\xi$, we get
\begin{subequations}
\begin{eqnarray}
\calM >0: \quad \int\limits_{-L/2}^{L/2} \rmd s \, \rmtz{}^2 &=& -L \left[ \left (\frac{a-1}{1-\eta} \right) \left( \frac{\rmE(m)}{\rmK(m)} -\frac{1}{2}\right) + \left(\frac{a-1}{1-\eta}-2 a\right) \frac{\xi}{2} \right]	\,,\\
\calM < 0: \quad \int\limits_{-L/2}^{L/2} \rmd s \, \rmtz{}^2 &=& L \left[ \left (\frac{a+1}{1+\eta} \right) \left(\frac{\rmE(m)}{\rmK(m)} - \frac{1}{2} \right) + \left( \frac{a+1}{1+\eta} - 2 a\right) \frac{\xi}{2} \right]\,.
\end{eqnarray}
\end{subequations}
The magnitude of the force $F$ and the constant $\bar{\Lambda}$ can be expressed also in terms of $a$ and $b$ as 
\begin{subequations}
\begin{align}
\bar{F}_I &= \left(\frac{a_I+b_I}{2}\right) \frac{\gamma}{L^2} \,, & \bar{\Lambda}_I &= -\frac{a_I b_I}{2} \, \frac{\gamma}{L^2}\,,  \\
\bar{F}_{II} &= -\left(\frac{a_{II}-b_{II}}{2}\right) \frac{\gamma}{L^2} \,, & \bar{\Lambda}_{II} &= \frac{a_{II} b_{II}}{2} \, \frac{\gamma}{L^2}\,.
\end{align}
\end{subequations}
Substituting these expressions and simplifying we obtain the total energy giving by Eq. (\ref{def:totenplanar}). 

\end{appendix}

\bibliographystyle{apsrev4-1}
\bibliography{BibMagPol}

\end{document}